\documentclass[prd,floats,floatfix, nofootinbib]{revtex4}
\usepackage[utf8]{inputenc}

\usepackage{color}
\usepackage{float}

\usepackage{graphicx}
\usepackage{mathtools, makecell}
\usepackage{subcaption}

\usepackage{hyperref}
\usepackage{cleveref} 
\usepackage{url}

\newcommand{\scriptR}{\mathcal{R}}

\begin{document}

\title{Singularity-free cosmology from interactions in the dark sector}

\author{Molly Burkmar${{^a}}$\footnote{molly.burkmar@port.ac.uk}, Marco 
Bruni${{^{a,b}}}$\footnote{marco.bruni@port.ac.uk} and Thomas Waters${{^c}}$\footnote{thomas.waters@port.ac.uk}} 
\vspace{0.5cm}

\affiliation{${{^a}}$Institute of Cosmology {\rm \&} Gravitation, University of Portsmouth, Dennis Sciama Building, Burnaby Road, Portsmouth, PO1 3FX, United Kingdom\\
${{^b}}$INFN Sezione di Trieste, Via Valerio 2, 34127 Trieste, Italy \\
${{^c}}$School of Mathematics and Physics, 
University of Portsmouth,
PO1 3HF, 
United Kingdom}

\date{\today}

\begin{abstract}
    
    We study the dynamics of Friedmann-Lema{\^i}tre-Robertson-Walker models where a dark energy component with a quadratic equation of state (EoS) nonlinearly interacts with cold dark matter. Thus, two energy scales naturally come into play: $\rho_*$ is the scale at which the nonlinearity of the EoS becomes relevant; $\rho_i$ is the energy scale around which the interaction starts to play an important role in  the dynamics. Our focus is to understand whether there are parameter ranges for this system that can produce nonsingular bouncing and emergent cosmologies for any initial condition. We complete a dynamical systems analysis, and find the parameter range such that trajectories always expand from a high energy nonsingular de Sitter state. For flat and negative curvature models this de Sitter state is represented by a fixed point,  the asymptotic past from which the Universe emerges. We find a subset of  positive curvature models that during contraction get arbitrarily close to the de Sitter state, thus having a quasi-de Sitter bounce, then emerge from the bounce and expand, evolving toward spatial flatness. We find that the dimensionless parameter $q\equiv \rho_*/\rho_i$, which measures the relative strength of the nonlinear terms in the system, plays a crucial role in the topology of the phase space. When $q < 3$, some trajectories expand toward a singularity, while others evolve toward a  low energy cosmological constant at late times, with a subset going through a decelerated matter dominate era before the final acceleration. When $q > 3$, all trajectories are nonsingular, and evolve toward a late-time cosmological constant. We find a subclass in this case in which all trajectories have at least one decelerated matter dominated phase, and accelerate at late times. Therefore, the nonlinear interacting cosmology presented here allows for a subclass of models that are singularity-free and qualitatively realistic for any initial condition. 

\end{abstract}

\maketitle

\onecolumngrid

\section{Introduction}

The $\Lambda$ cold dark matter ($\Lambda$CDM) model plus inflation is our current standard model of cosmology, as it has been the most consistent theory, agreeing with a wealth of observations \cite{Planck2018I, Planck2018VI}. However, there are theoretical problems which require addressing. One problem is that of singularities. Assuming the strong energy condition (SEC) holds, cosmological singularities are inevitable in general relativity (GR), as shown by Hawking and Penrose \cite{Penrose1965,Hawking1965,Hawking1966I, Hawking1966II, Hawking1967III,Hawking1970, Hawking1973}. However, their current interpretation is that they represent points where GR breaks down \cite{Joshi1993,Hawking1979,Gibbons2003,Ashtekar2015}. There are different ways in which we can tackle this problem: we can develop a quantum theory of gravity to replace GR at Planck energies and above, or we can take a semiclassical approach; for example, running vacuum models in quantum field theory (QFT) can expand from a nonsingular state \cite{SolaPeracaula2020,Sol_Peracaula_2025}. Another possibility is to avoid a singularity by developing a nonsingular classical framework for high but sub-Planckian energies. It is this approach that we take in this paper. One alternative to an initial singularity are bouncing cosmologies, which transition from an initially contracting period to an expanding phase \cite{Gasperini1993,Khoury2001,Kallosh2001,Steinhardt2002,Novello2008,Wands2008,Maier2012,Maier2013,Battefeld2014,Cai2017i,Brandenberger2017,Cai2017ii,Ijjas2018,Ijjas2019,Ganguly2019,Ilyas2020,Battista2021,Zhu2021}. An emergent universe is another alternative, originally conceived as initially quasistatic, expanding from a nonsingular Einstein universe state with positive curvature \cite{Ellis2003,Ellis2004,Mukherjee2006,Verlinde2017,Ellis2016,Parisi2007,Mulryne2005,Bonanno2017,Gionti2017,Sengupta2024,Barrow2003}. The problem with this scenario is that it is not self-consistent, at least classically, as it relies on some previously existing mechanism to establish such very special initial conditions, which otherwise is a set of measure zero in phase space \cite{Burkmar2024}. In Ref. \cite{Burkmar2024}  we contended that a different emergent scenario can be built, where all trajectories in phase space representing expanding models evolve from a past nonsingular de Sitter phase. A dark energy with a quadratic equation of state (EoS), which evolves between two effective cosmological constants \cite{Ananda2006}, is an example of a noninteracting model which can give rise to nonsingular bouncing and emergent cosmologies.

Another problem facing the standard model of cosmology is that of dark energy. The cosmological constant $\Lambda$ is the simplest form of dark energy that we have to explain the accelerated expansion of our Universe \cite{Riess1998,Perlmutter1998}. However, the observed value of $\Lambda$ is at odds by up to 120 orders of magnitude with theoretical estimates of the contributions to the effective cosmological constant expected from QFT \cite{Weinberg1989, Straumann1999, Ellis2012}. Various attempts have been made to resolve this problem within the Friedmann-Lema{\^i}tre-Robertson-Walker (FLRW) model; for example, certain running vacuum models in QFT may be able to alleviate the fine-tuning, see e.g. Ref. \cite{Sol2022} and references therein.

It has also been debated as to whether $\Lambda$ is a true constant, or an approximation to a more complex dynamic dark energy. Beyond a cosmological constant, an interacting vacuum has been widely studied \cite{Wands2012,De-Santiago2012,Hogg2021,Kaeonikhom2020,Quercellini2008}, where energy-momentum is transferred between the vacuum and other matter fields. It has been shown that certain running vacuum models are actually favored by observations over $\Lambda$CDM \cite{Sol__2015,Sol__2017,Sol__2017_2,Sol_Peracaula_2018,Sol_Peracaula_2018_2}, and the recent Dark Energy Spectroscopic Instrument (DESI) results hint towards a dynamic dark energy \cite{DESIDR2II}. The coincidence problem also needs addressing: that is, the energy densities of cold dark matter and dark energy are of the same order of magnitude. Running vacuum models have been found to help alleviate the coincidence problem \cite{Grande2006,Grande2007}, as well as models with interactions in the dark sector \cite{Lip2010}. An interacting vacuum model has also been studied in the context of nonsingular cosmologies \cite{Bruni2022}.



In our previous work, which we will refer to as Paper I \cite{Burkmar2023}, we extended Refs. \cite{Ananda2006,Ananda20062}, and studied a dark energy with a nonlinear quadratic EoS. This dark energy was past asymptotic to a high energy effective cosmological constant during expansion, and evolved towards a low energy effective cosmological constant at late times. We included dark matter and radiation in the setup, without any interaction term between the components. Nonsingular models were possible; however, when we quantitatively set the energy scales for the dark energy, such that the high energy effective cosmological constant was between Planck and inflationary energy scales, and the low energy effective cosmological constant was close to the observed value of $\Lambda$, we lost a decelerated period where large scale structure could form. In this paper, we want to understand whether including an interaction between the dark components would allow for qualitatively realistic nonsingular bouncing and emergent models, which could be consistent with observations, with quantitatively realistic energy scales. We introduce an interaction term which is quadratic and proportional to the product of the dark matter and dark energy densities, such that only the high energy dynamics are modified where we want to avoid singularities. We expect a realistic model to bounce or emerge between Planck and inflationary scales, and have a low energy, dark energy dominated acceleration phase at late times. At high energy, we would need to constrain the model with current bounds on primordial gravitational waves. If we linearize our equations, we obtain a low energy model which has already been tested against, and is consistent with data from the Wilkinson Microwave Anisotropy Probe (WMAP), the Sloan Digital Sky Survey (SDSS) and the Hubble Space Telescope \cite{Balbi2007,Pietrobon2008}. In addition, models with a linear interaction term have been tested against supernovae data \cite{Quercellini2008}, therefore we expect the late-time behavior of our model to be consistent with these observations. In this paper, we would like to understand if there are parameter ranges in which nonsingular models emerge between Planck and inflationary scales, have a decelerated expansion phase, and asymptotically tend towards a low energy cosmological constant for any initial condition. We note that we are really interested in the qualitative behavior of the model we present in this paper. Here our aim is to understand whether a singularity can be avoided at high energy, with late-time behavior that is qualitatively approximate to the observed universe, such that we obtain a low energy, accelerating phase dominated by dark energy. However, in light of the DESI results \cite{DESIDR2II}, we appreciate that the late-time behavior of the model we consider may need modifying in order to accommodate a dark energy that is dynamical at low energy, rather than asymptotically constant. The analysis for this work is available through GitHub, provided the reader has a \texttt{Mathematica} license \cite{BurkmarIntDEnb}.


This paper is organized as follows: in Sec. \ref{sec:Equations}, we present the system of equations for the dark energy, dark matter and the Hubble expansion function in terms of dimensionless variables. In Sec. \ref{sec:param_space}, we explore the parameter space, and set parameters such that the topology of the phase space is qualitatively meaningful.  With this, here and in the following, we mean that  expanding models go through a matter dominated decelerated phase before the late-time acceleration of the expansion, thereby -- at least qualitatively -- corresponding to the observed universe. In Sec. \ref{sec:qLT3}, we study the dynamics of the phase space when 
$q$, which measures the relative strength of the nonlinear terms in the system, satisfies the condition $q < 3$, and in Sec. \ref{sec:qGT3} we study the dynamics when $q > 3$, showing that in this case all universe models are singularity-free. We present our conclusions in Sec. \ref{sec:Conclusions}. In this paper, we employ natural units such that $8\pi G = c = 1$.

\section{FLRW Dynamics} \label{sec:Equations}
\subsection{Physical variables}
The evolution of FLRW models is described by a system of ordinary differential equations (ODEs), which consist of the energy conservation equations for each component, as well as the Raychaudhuri equation which describes the expansion scalar $H$, also known as the Hubble function. The Friedmann equation (the Hamiltonian constraints) relates $H$ to the energy densities of the components and to the three-curvature of the constant time hypersurfaces.

In this paper we consider dark matter and dark energy with a nongravitational interaction term. We take dark matter to be cold (CDM) and pressureless. Then, its energy conservation equation is expressed as a sum of the standard matter part and the interaction,

\begin{equation}
    \dot{\rho}_m=-3H\rho_m + \frac{H\rho_x\rho_m}{\rho_{i}} \,,
    \label{eqn:rhomdot_interacting}
\end{equation}

\noindent where $\rho_m$ is the dark matter energy density, $\rho_x$ is the dark energy density, and $\rho_i$ is the energy scale characterizing the interaction, i.e. the scale at which the interaction term becomes significant in the dynamics; overdots are derivatives with respect to time $t$. An interaction term which is proportional to the product of the energy densities of the dark components has been studied before, and has been found to help alleviate the coincidence problem \cite{Lip2010}. This type of model can also admit solutions where the density parameters oscillate, such that the dark energy and dark matter repeatedly switch between being the dominant component, and their energy densities do not become infinite \cite{Arevalo2012,Perez2014,Aydiner2018}. In this paper, we implement a quadratic interaction term with the aim of only modifying the dynamics at high energy, to understand if the energy densities of the dark components remain finite, such that we can avoid singularities.


The dark energy has a self-interaction term with a quadratic EoS, and the pressure of the dark energy, $P_x$, is expressed as \footnote{The parametrization here is different from Paper I \cite{Burkmar2023}.}

\begin{equation}
    P_x = -\rho_\Lambda(1+w_x) + \rho_x\left(w_x-\epsilon\frac{\rho_\Lambda}{\rho_*}\right) + \epsilon \frac{\rho_x^2}{\rho_*}\,,
\label{eqn:DE_EoS_interacting}
\end{equation}

\noindent where $w_x$ sets its equation of state at low energy; $\epsilon$ sets the sign of the quadratic term; $\rho_\Lambda$ is the effective low energy cosmological constant, which as we will see in Sec. \ref{sec:param_space} is the low energy attractor; and $\rho_*$ is the characteristic energy scale of the dark energy, i.e. the scale at which the nonlinear terms in Eq. \eqref{eqn:DE_EoS_interacting} become significant. We note that although this model is phenomenological, it has been shown that a perfect fluid with affine barotropic EoS can represent a standard or k-essence scalar field \cite{Quercellini2007}. The energy conservation equation for the dark energy then becomes

\begin{equation}
    \dot{\rho_x}=-3H(\rho_x-\rho_\Lambda)\left(1 + w_x + \epsilon\frac{\rho_x}{\rho_*}\right)-\frac{H\rho_x\rho_m}{\rho_i}\,.
    \label{eqn:rhoxdot_interacting}
\end{equation}

\noindent The interaction term has been set such that if the initial conditions for the dark energy density satisfy $\rho_x > 0$ and $\dot{\rho_x} < 0$, then the dark energy density is always positive. This also implies that during expansion the energy flows from the dark energy to the dark matter. Depending on the values of parameters, it is possible for the dark energy to violate the null energy condition (NEC) during certain periods of its evolution. Therefore, the dark energy can be phantom, meaning $\rho_x$ increases with expansion during those periods.  
For $\rho_m=0$, it is clear from Eq. \eqref{eqn:rhoxdot_interacting} that $\rho_x=\rho_\Lambda$ represents a cosmological constant: as we show later, this is the low energy attractor of our dynamical system. We think of $\rho_\Lambda$ as being close to the observed dark energy density. Similarly, for $\rho_m=0$ the energy scale $\rho_*$ also gives a cosmological constant, albeit a high energy and unstable one. We assume that $\rho_*$ and $\rho_i$ are energy scales between the Planck scale and that of inflation. This is so our model does not interfere with particle production at lower energies, and so a decelerated matter dominated phase can follow. We want to understand if nonsingular models which admit a matter dominated decelerated phase are possible, particularly when we set realistic energy scales.

The evolution of the Hubble expansion function $H$ is given by the Raychaudhuri equation,

\begin{equation}
    \dot{H}=-H^2-\frac{1}{6}\left[\rho_m + \rho_x(1 + 3w_x - 3\epsilon\frac{\rho_\Lambda}{\rho_*})
    - 3\rho_\Lambda(1 + w_x) + 3\epsilon\frac{\rho_x^2}{\rho_*}\right]\,,
    \label{eqn:Hubble_rate_interacting}
\end{equation}

\noindent and

\begin{equation}
    \frac{\ddot{a}}{a} = \dot{H}+H^2
\end{equation}

\noindent is the related acceleration equation. Equation \eqref{eqn:Hubble_rate_interacting} admits a first integral, the Friedmann equation:

\begin{equation}
    H^2=\frac{\rho_m}{3}+\frac{\rho_x}{3}-\frac{k}{a^2}\,,
    \label{eqn:Friedmann_Dimensions_interacting}
\end{equation}

\noindent where $k$ is the spatial curvature and $a$ is the dimensionless scale factor, which is connected to the Hubble expansion function by $H = \dot{a}/a$. $k$ is an arbitrary constant with the same dimensions as $H^2$, and we set $a_0 = 1$; $k$ is positive for closed models, negative for open models, and zero for flat models.

\vspace{0.1cm}

\subsection{Dimensionless variables} \label{subsec:normalisation}

We normalize the above system of equations with respect to the characteristic energy scale of the dark energy, $\rho_*$, thereby introducing dimensionless variables and parameters. Following Paper I \cite{Burkmar2023} and Refs. \cite{Ananda2006,Ananda20062}, we define these variables as

\begin{equation}
    x=\frac{\rho_x}{\rho_*}, ~~ y=\frac{H}{\sqrt{\rho_*}}, ~~ z=\frac{\rho_m}{\rho_*}, ~~
    \eta=\sqrt{\rho_*}t.
    \label{eqn:dimensionless_variables}
\end{equation}

\noindent The variables $x$ and $z$ are the normalized dark energy density and the normalized dark matter energy density, respectively, $y$ is the normalized Hubble expansion function, and $\eta$ is the normalized time variable. In Paper I \cite{Burkmar2023}, we included radiation in the setup, which we do not do here, and the system of ODEs in Paper I could be solved analytically, which is not possible here. This means that unlike in Paper I, we cannot project the dynamics onto lower-dimensional phase spaces. In any case, including radiation would only give a transition regime between the high energies, dominated by the dark energy and the interaction, and the matter dominated era. We also introduce the dimensionless parameters,

\begin{equation}
  \scriptR=\frac{\rho_\Lambda}{\rho_*}, ~~   q= \frac{\rho_*}{\rho_i},
\end{equation}

\noindent where $\scriptR$ is the ratio of the low energy effective cosmological constant $\rho_\Lambda$ to the characteristic energy scale of the dark energy $\rho_*$, which takes a value in the range $0 < \scriptR < 1$, and $q$ is the strength of the interaction parameter. We assume that $\rho_\Lambda$ is close to the observed value of $\Lambda$, and $\rho_*$ is between Planck and inflationary energy scales, such that the evolution is completely classical without the need of a quantum gravity era at high energy. Observations show that the current vacuum energy density is $\rho_{obs} \sim 10^{-47}$GeV$^4$ \cite{Planck2018VI,Prat2022}. The Planck scale is approximately 120 orders of magnitude larger than $\rho_{obs}$, and if measurements of the tensor-to-scalar ratio fall within the range $0.001 < r < 0.1$, then it would point to a scale of inflation of $\rho_{_{inf}} \sim 10^{64}$GeV$^4$ \cite{Bass2015}. These measurements provide a realistic range of $\scriptR$ to be $10^{-120} < \scriptR < 10^{-111}$. In Sec. \ref{sec:qLT3}, we will use a larger value of $\scriptR$ purely for the purpose of producing readable phase spaces. In terms of dimensionless variables, Eq. \eqref{eqn:DE_EoS_interacting} becomes

\begin{equation}
    \frac{P_x}{\rho_*} = -\scriptR(1+w_x) + x(w_x - \epsilon \scriptR) + \epsilon x^2 \,,
    \label{eqn:P_x_dimensionless}
\end{equation}

\noindent and our system of equations in Eqs. \eqref{eqn:rhoxdot_interacting}, \eqref{eqn:rhomdot_interacting} and \eqref{eqn:Hubble_rate_interacting} become

\begin{equation}
    x' = -3y(x - \scriptR)(1 + w_x + \epsilon x) - qxyz \,,
    \label{eqn:x_prime}
\end{equation}

\begin{equation}
    z' = -3yz + qxyz \,,
    \label{eqn:z_prime}
\end{equation}

\begin{equation}
    y' = -y^2 - \frac{1}{6}\left[z + 
      x(1 + 3w_x - 3\epsilon\scriptR) - 3\scriptR(1 + w_x) + 3\epsilon x^2\right] \,,
      \label{eqn:y_prime}
\end{equation}

\noindent where primes indicate differentiation with respect to the time variable $\eta$. We note that models with affine EoS $P = P_0 + \alpha\rho$ have already been tested against, and were found to be consistent with, observational data from WMAP, SDSS and the Hubble Space Telescope \cite{Balbi2007,Pietrobon2008}. If we linearize the EoS in Eq. \eqref{eqn:P_x_dimensionless}, we obtain $P_x/ \rho_* = -\scriptR(1+w_x) + x\left(w_x - \epsilon \scriptR \right)$, and as the interaction term is quadratic, we can neglect it in Eq. \eqref{eqn:x_prime} at late times. Therefore, the effective dark energy that we observe at low energy depends on $w_x$ and $\epsilon$, as $\scriptR$ is already constrained to $10^{-120} < \scriptR < 10^{-111}$ in the model. We therefore expect that if we set the sign of $\epsilon$ and constrain $w_x$ with WMAP, SDSS and Hubble Space Telescope data, our model will fit observations, just as $\alpha$ was constrained in models with EoS $P = P_0 + \alpha\rho$ which were found to be consistent with data \cite{Balbi2007,Pietrobon2008}.

In dimensionless variables, the acceleration equation is

\begin{equation}
    \frac{a''}{a} = y' + y^2 \,,
      \label{eqn:accn}
\end{equation}

\noindent and the normalized Friedmann equation becomes

\begin{equation}
    y^2=\frac{x}{3}+\frac{z}{3}+\frac{r}{3}-\frac{k}{\rho_* a^2} \,,
    \label{eqn:Friedmann_dimensionless_Interacting}
\end{equation}

\noindent where $y$ can be connected to the scale factor $a$ through the expression $y = a'/a$. We only consider the region of phase space where $x, z > 0$, such that the energy densities of the dark energy and dark matter are always positive. From Eq. \eqref{eqn:z_prime}, it is clear that $z = 0$ is an invariant manifold that cannot be crossed. Although $x = 0$ is not an invariant manifold, from Eq. \eqref{eqn:x_prime} we see that along $x = 0$, $x' > 0$; although phantom trajectories ($x' > 0$) can cross from below $x = 0$, trajectories cannot cross $x = 0$ from above. Therefore, as long as initial conditions are chosen satisfying $x > 0$ and $\dot{x} < 0$, such that we are in nonphantom region, this is sufficient for trajectories to always satisfy $x > 0$, even if they become phantom at late times.


\section{Analysis of Parameters} \label{sec:param_space}

\subsection{Two-dimensional phase space}

For this system to be nonsingular, and to be compatible with observations, we require all trajectories to emerge from a high energy repellor during expansion, and tend toward an attractor at low energy. Therefore, we need to understand the parameter space that allows for this topology. To do this, we need to look at the two-dimensional phase space. Taking $y > 0$ ($H > 0$), and using the relation 

\begin{equation}
    y = \frac{a'}{a} = \frac{d}{d\eta}\ln{(a)}\,,
\end{equation}
\noindent we now change the time variable to $\ln(a)$. Our equations in two dimensions become 

\begin{equation}
    x' = -3(x - \scriptR)(1 + w_x + \epsilon x) - qxz \,,
    \label{eqn:x_prime_lna}
\end{equation}

\begin{equation}
    z' = -3z + qxz\,,
    \label{eqn:z_prime_lna}
\end{equation}

\noindent where now primes indicate differentiation with respect to $\ln({a})$. We then need to find the fixed points for the two-dimensional system, which are shown in Table \ref{tab:zx_Fixed_Points}, and their eigenvalues, which can be found in Table \ref{tab:Eigenvalues}. We classify all the fixed points in the above system as representing de Sitter models, because they have constant energy densities $x' = z' = 0$, where $y \neq 0$ and can vary for positively or negatively curved de Sitter models. In general, this system admits three fixed points. The fixed point $dS_{1+}$ comes from the interaction term, and $dS_{2+}$ and $dS_{3+}$ come from the nonlinear equation of state for the dark energy in Eq. \eqref{eqn:DE_EoS_interacting}. 

\begin{table}
\centering
 \begin{tabular}{||c | c c||} 
 \hline
 
 Name & $z$ & $x$\\ [0.5ex] 
 
\hline\hline

$dS_{1\pm}$ & $\frac{(q\scriptR-3)(q(1+w_x)+3\epsilon)}{q^2}$ & $\frac{3}{q}$ \\

$dS_{2\pm}$ & $0$ & $\scriptR$ \\

$dS_{3\pm}$ & $0$ & $-\frac{(1+w_x)}{\epsilon}$ \\
 
 \hline
 \end{tabular}
 \caption{\label{tab:zx_Fixed_Points} Fixed points of the $z$-$x$ system. $dS_{+}$ denotes an expanding (+) de Sitter universe ($x' = z' = 0$) when $y > 0$, and $dS_{-}$ denotes a contracting de Sitter universe when $y < 0$. In the two-dimensional phase spaces, only the $dS_{+}$ fixed points are plotted as we have fixed $y > 0$ so the trajectories are expanding.}
\end{table}

We want to find parameters such that when $y > 0$ ($H > 0$), the high energy fixed point $dS_{1+}$ is a repellor, so trajectories emerge from a nonsingular de Sitter fixed point during expansion. We also require that the low energy fixed point $dS_{2+}$ is an attractor, such that trajectories asymptotically approach a low energy cosmological constant to ensure the dynamics qualitatively matches our observed Universe. 
To understand whether this system allows for a high energy repellor and low energy attractor, we need to look at the eigenvalues of the fixed points, the sign of which depends on the specific values of $\epsilon$, $w_x$ and $q$. For $dS_{1+}$ to be a repellor, its eigenvalues must have positive real parts, and for $dS_{2+}$ to be an attractor its eigenvalues must have negative real parts.
Both of these fixed points must also exist at $x > 0$ and $z > 0$, so that the energy densities of the dark energy and dark matter are positive. In the following subsections, we analyze different parameter ranges to understand whether $dS_{1+}$ and $dS_{2+}$ can have the stability character we require.



\begin{table}[h]
\small
\centering
 \begin{tabular}{||c |c  c||} 
 \hline
 
 Name & $\lambda_1$ & $\lambda_2$\\ [0.5ex] 
 
\hline\hline

$dS_{1+}$ & $ -\frac{q^2\scriptR(1+w_x) + 9\epsilon + \sqrt{-12q(q+qw_x+3\epsilon)(q\scriptR - 3)+\left[q^2\scriptR(1+w_x)+9\epsilon\right]^2}}{2q}$ & $ -\frac{q^2\scriptR(1+w_x) + 9\epsilon - \sqrt{-12q(q+qw_x+3\epsilon)(q\scriptR - 3)+\left[q^2\scriptR(1+w_x)+9\epsilon\right]^2}}{2q}$ \\

$dS_{2+}$ & $ -3 + q\scriptR$ & $-3(1 + w_x + \epsilon\scriptR)$ \\
$dS_{3+}$ &  $-3 - \frac{q(1+w_x)}{\epsilon}$ & $3(1 + w_x + \epsilon \scriptR)$ \\
 
 \hline
 \end{tabular}
 \caption{\label{tab:Eigenvalues} Eigenvalues of the fixed points of the $z$-$x$ system. $dS_{+}$ denotes an expanding (+) de Sitter universe ($x' = z' = 0$). Note that for the contracting de Sitter fixed points $dS_{-}$, the eigenvalues change sign.}
\end{table}

\subsection{Parameter space for viable models}

In this subsection, we study how the fixed points and their eigenvalues change for $\epsilon = \pm 1$, and for $w_x > -1$ and $w_x < -1$. Notice that because of the interaction term, and because of the $\rho_\Lambda$ term in Eq. \eqref{eqn:DE_EoS_interacting}, the dark energy can be phantom, $x' > 0$, even for $w_x > -1$. We assume $q > 0$, meaning that during expansion energy transfers from the dark energy to the dark matter, and we take $0 < \scriptR < 1$.

\subsubsection{$w_x < -1$, $\epsilon = +1$}

For $dS_{2+}$ to have negative eigenvalues and be an attractor when $\epsilon = +1$, we require $-1 - \scriptR < w_x < -1$ and $q < 3/\scriptR$. However, for these parameter ranges, $dS_{1+}$ cannot have positive eigenvalues, and therefore cannot be a repellor. More generally, when $dS_{2+}$ is an attractor, $dS_{1+}$ can only exist at $z < 0$. We also note that for these parameter ranges, $dS_{3+}$ cannot be a repellor. Therefore, for $w_x < -1$ and $\epsilon = +1$, we cannot have a high energy repellor and a low energy attractor.

\subsubsection{$w_x < -1$, $\epsilon = -1$}

For the eigenvalues of $dS_{2+}$ to be negative when $\epsilon = -1$, we require $w_x > -1 + \scriptR$. Therefore, $dS_{2+}$ cannot be an attractor for $\epsilon = -1$ and $w_x < -1$, and so we do not have a late-time cosmological constant for these parameter values.

\subsubsection{$w_x > -1$, $\epsilon = +1$}

For the $dS_{2+}$ fixed point to have negative eigenvalues when $\epsilon = +1$, we require $w_x > -1$ and $q < 3/\scriptR$. However for these parameter values, $dS_{1+}$ cannot have eigenvalues with positive real parts and so cannot be a repellor. More generally, $dS_{1+}$ only exists at $z < 0$ for these parameters. The fixed point $dS_{3+}$ can be a repellor for these parameter ranges; however, it only exists at $x < 0$. Therefore, we cannot achieve our desired topology when $w_x > -1$ and $\epsilon = +1$.

\subsubsection{$w_x > -1$, $\epsilon = -1$}

Setting $\epsilon = -1$ and $0 < \scriptR < 1$, we find that $dS_{2+}$ is an attractor when $q < 3/ \scriptR$ and $w_x > -1 + |\epsilon|\scriptR$. These parameter constraints are sufficient for $dS_{1+}$ to exist at $z > 0$ and be a repellor. Depending on the specific value of $q$, the stability of $dS_{1+}$ can change. It is possible for the eigenvalues to be real and positive such that the fixed point is a repelling node, and it is also possible for the eigenvalues to be complex with positive real parts such that the fixed point is a spiral repellor. There also exits a limiting value between these two cases, such that the eigenvalues are real, positive and the same, meaning the repellor is an improper node. With these combinations of parameters, we find that $dS_{3+}$ exists at $x > 0$ and is a saddle fixed point, which always has a larger $x$ value than $dS_{2+}$, and always has a smaller $x$-value than $dS_{1+}$. For these parameters, the dark energy violates the NEC when $x < \scriptR$ and when $x > 1 + w_x$, and so has phantom behavior. Therefore, a high energy repellor and an attractor at low energy exist when $\epsilon = -1$ and $w_x > - 1$; however, we need to understand whether all trajectories emerge from $dS_{1+}$ during expansion and evolve towards $dS_{2+}$. From now on, we fix $\epsilon = -1$, $-1 + \scriptR < w_x < 0$ and $0 < q < 3/\scriptR$.

\subsection{Behavior at infinity} \label{subsec:compact}

Thus far, we have discussed the fixed points in the phase plane for $x$ and $z$; however, to understand the global behavior we introduce compactified variables. In the system above, $y$ takes values in the range $-\infty < y < \infty$, and we are interested in values of $x$ and $z$ in the range $0 < x, z < \infty$, such that the energy densities of the dark components are always positive. We note that in Paper I \cite{Burkmar2023}, we considered the dynamics in the range $\scriptR < x < 1$; if we had considered $x < \scriptR$ or $x > 1$, trajectories in these regions would have always been phantom, and therefore qualitatively unrealistic. At high energy, the interaction term decreases $x$, and is able to counter the phantom runaway behavior caused by the quadratic term in Eq. \eqref{eqn:P_x_dimensionless}, and at low energy, trajectories which were nonphantom in the past can be phantom at late times, unlike in Paper I \cite{Burkmar2023}. Therefore, in this paper we study $x$ in the range $0 < x < \infty$.

In order to understand the dynamics at infinity, we compactify our variables as follows:

\begin{equation}
    X = \frac{x}{1+x}\,,
\end{equation}

\begin{equation}
    Y = \frac{y}{\sqrt{1+y^2}}\,,
\end{equation}

\begin{equation}
    Z = \frac{z}{1+z}\,,
\end{equation}

\noindent where $X = Z = 1$ corresponds to $x, z \rightarrow \infty$, and $Y = \pm 1$ corresponds to $y \rightarrow \pm \infty$. For $z$, it is obvious that $z$ is always greater than zero from Eq. \eqref{eqn:z_prime}. Along $x = 0$, $x' > 0$, meaning for any positive initial $x$ value, $x$ will remain positive if we evolve the system forward in time. However, $x$ can be negative and cross $x = 0$ from below, therefore we need to set variable ranges such that we only consider trajectories where $x$ is always positive. We therefore analyze our system between $0 < X, Z < 1$, and $-1 < Y < 1$. The system of ODEs in Eqs. \eqref{eqn:x_prime} -- \eqref{eqn:y_prime} in compact variables becomes

\begin{equation}
    X' = -3\frac{Y}{(1-Y^2)^{1/2}}(1-X)^2\left(\frac{X}{1-X} - \scriptR \right) \left(1+w_x+\epsilon\frac{X}{1-X}\right) - q\frac{Y}{(1-Y^2)^{1/2}}X(1-X)\frac{Z}{1-Z}\,,
    \label{eqn:X'}
\end{equation}

\begin{equation}
    Z' = -3\frac{Y}{(1-Y^2)^{1/2}}Z(1-Z) + q\frac{Y}{(1-Y^2)^{1/2}}Z(1-Z)\frac{X}{1-X}\,,
    \label{eqn:Z'}
\end{equation}

\begin{equation}
    Y' = -Y^2(1-Y^2)^{1/2} - \frac{(1-Y^2)^{3/2}}{6} \left[\frac{Z}{1-Z} + \frac{X}{1-X} (1 + 3w_x -3\epsilon\scriptR) - 3\scriptR (1+w_x) + 3\epsilon\frac{X^2}{(1-X)^2} \right]\,,
    \label{eqn:Y'}
\end{equation}



\begin{figure}
    \centering
    \includegraphics[width=0.7\linewidth]{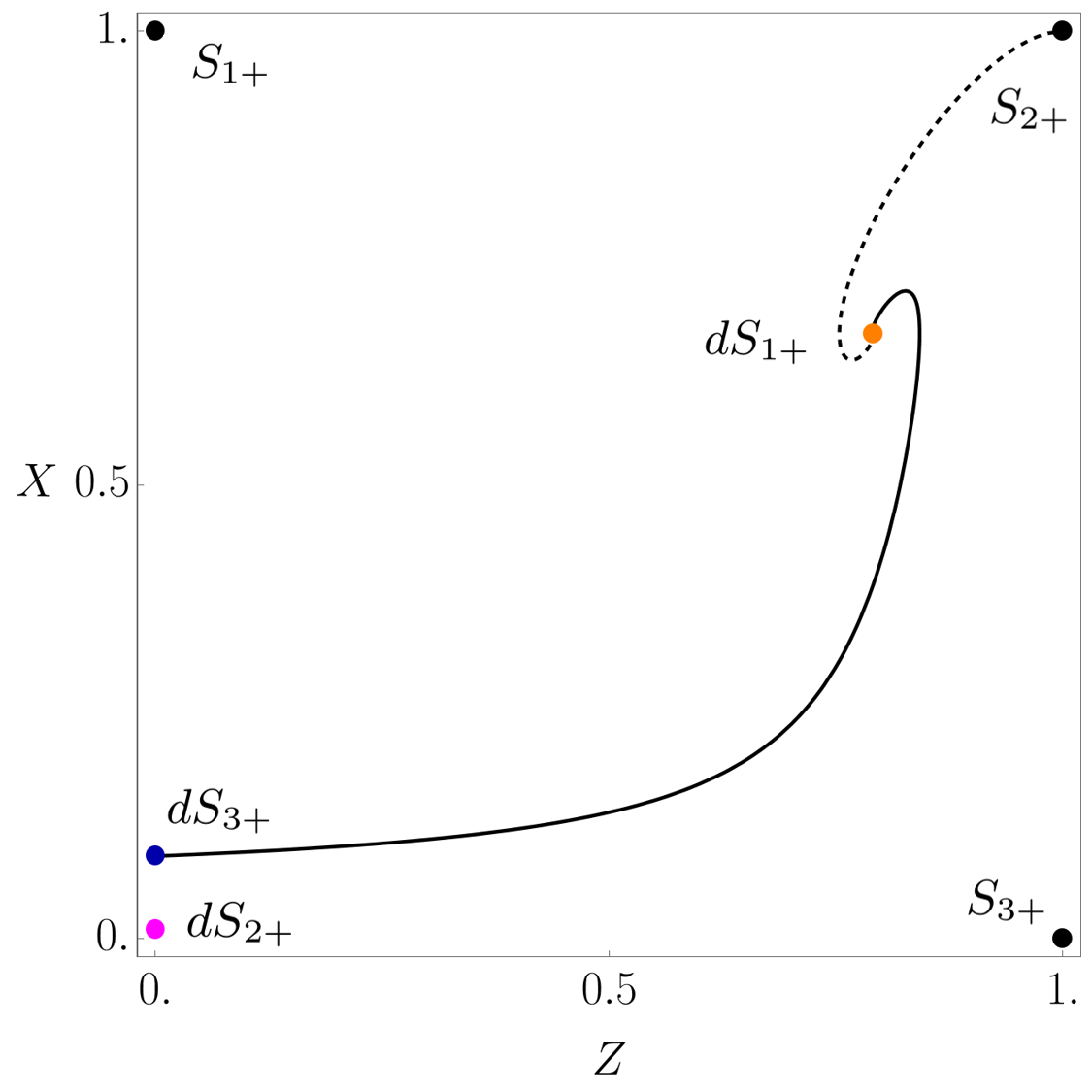}
    \caption{Schematic diagram of the fixed points and critical points in the two-dimensional $Z$-$X$ phase space. The solid black line shows the separatrix between $dS_{1+}$ and $dS_{3+}$ which exists for all values of $q$. The dashed black line shows the separatrix between $dS_{1+}$ and $S_{2+}$ which only exists for $q < 3$ (see Sec. \ref{subsec:compact} for an explanation).}
    \label{fig:FPs_Only}
\end{figure}

\noindent We can also write Eqs. \eqref{eqn:x_prime_lna} and \eqref{eqn:z_prime_lna} in compact form as

\begin{equation}
    X' = -3(1-X)^2\left(\frac{X}{1-X} - \scriptR \right)  \left(1+w_x-\frac{X}{1-X}\right) - qX(1-X)\frac{Z}{1-Z}\,,
    \label{eqn:X_Prime_lna}
\end{equation}

\begin{equation}
    Z' = -3Z(1-Z) + qZ(1-Z)\frac{X}{1-X}\,,
    \label{eqn:Z_Prime_lna}
\end{equation}

\noindent which have critical values/ singularities at $X=1$ and $Z=1$, corresponding to $x\to\infty$ and $z\to\infty$ respectively. These critical points are shown in Table \ref{tab:Singularities}, and a schematic diagram of the fixed points, critical points and separatrices in the two-dimensional system is shown in Fig. \ref{fig:FPs_Only}. It is especially important to understand the dynamics near $(Z,X)=(1,1)$, and in particular we want to know whether a separatrix exists between $dS_{1+}$ and $S_{2+}$. If such a separatrix does exist, then there will be trajectories trapped between it and the $X=1$ line that evolve to $S_{2+}$. On the other hand, if such a separatrix does not exist, then $S_{2+}$ will be a generalized saddle point and all trajectories will avoid the singularity and evolve to $dS_{2+}$.

\begin{table}
\centering
 \begin{tabular}{|| c | c | c ||} 
 \hline
 
 Name &  $Z$ & $X$ \\ [0.5ex] 
 
\hline\hline

$S_{1\pm}$ & $0$ & $1$ \\

$S_{2\pm}$ & $1$ & $1$ \\

$S_{3\pm}$ & $1$ & $0$ \\
 
 \hline
 \end{tabular}
     \caption{\label{tab:Singularities} Critical points that arise due to making the system compact. $S_{\pm}$ denotes a singularity with infinite expansion (+) or contraction (-). Note that in the $Z$-$X$ phase spaces, only the $S_{+}$ critical points are plotted as we have fixed $Y > 0$ so that the trajectories are expanding.}
\end{table}



First, we multiply Eqs. \eqref{eqn:X_Prime_lna} and \eqref{eqn:Z_Prime_lna} by $(1-X)(1-Z)$ which we can view as a rescaling of the vector field or a reparametrization, valid in the unit square $0<X,Z<1$. The singular point $(1,1)$ becomes a fixed point, and we can then Taylor expand around $(1,1)$; the linear part vanishes, and the equations to second order become

\begin{equation}
    X' = -q(1-X)^2 + 3(1-X)(1-Z),\qquad     Z' = q(1-Z)^2. 
    \label{eqn:TE_XZ_Prime}
\end{equation}

\noindent Note only the parameter $q$ remains. We find this system of equations admits the invariant line

\begin{equation}
    (X-1)=\frac{3-q}{q}(Z-1) \,,
\end{equation} 

\noindent through the point $(1,1)$ which naturally leads to two cases: when $q<3$, this invariant line lies in the unit square $0<X,Z<1$. The flow restricted to this line has the point $(1,1)$ as an attractor, and so there is a sector between this line and the $X=1$ line where all trajectories go onto $(Z,X)=(1,1)$ in the future. This invariant line, followed backward in time, goes to $dS_{1+}$. Therefore, when $q < 3$ a separatrix between $dS_{1+}$ and $S_{2+}$ exists [see Fig. \ref{fig:SR_No_Cross}(\subref{subfig:ZX_qLT3_CDNC}) for a representative example]. 

On the other hand, when $q > 3$ the invariant line does not enter the unit square and it appears all trajectories avoid the singularity at $(1,1)$, ultimately evolving towards $dS_{2+}$ (see Fig. \ref{fig:qGT3_SR_NoCross} for a representative example). We confirm this by writing $X=X(Z)$ and combining the equations in \eqref{eqn:TE_XZ_Prime} to give a homogeneous ODE for $dX/dZ$ with solution 

\begin{equation}
 X=1+\frac{a C(1-Z)^{a+1}}{1-C(1-Z)^a} \,,
 \label{eqn:X(Z)_TE}
 \end{equation}

\noindent where $C > 0$ is derived from the initial conditions and $a = \frac{3}{q}-1 < 0$ for $q > 3$. As $Z$ tends to 1 from below, the solution grows towards $X=1$; however, it reaches a maximum at $Z=1-\left(\tfrac{a+1}{C}\right)^{1/a}$ and then starts to decrease; as such no trajectories in the unit square go onto $(Z,X)=(1,1)$ when $q>3$. Alternatively we could consider Eqs. \eqref{eqn:x_prime_lna} and \eqref{eqn:z_prime_lna} in the ``high energy regime'' where we keep only the quadratic terms \begin{equation}
    x' = 3x^2 - qxz,\qquad  z' = qxz \,,
    \label{eqn:x_prime_HE}
\end{equation} whose quotient again leads to a homogeneous equation $\frac{dx}{dz} = \frac{3x}{qz} - 1$ 
with solution

\begin{equation}
    x(z) = \frac{1}{a} (z - cz^{a+1})
    \label{eqn:x{z}_HE}
\end{equation} with $c > 0$. This solution also admits a single maximum and so $x\not\to+\infty$ as $z\to+\infty$.

\section{Dynamics with $q < 3$} \label{sec:qLT3}

In this section, we analyze the dynamics when $q < 3$. For $q < 3$, there are models that become singular; however, there are trajectories that are nonsingular and qualitatively interesting. We fix $\scriptR = 0.01$; however, for realistic models we would expect $\scriptR$ to be much smaller if $\rho_*$ is between Planck and inflationary energy scales, and $\rho_\Lambda$ is close to the observed value of $\Lambda$. In Sec. \ref{sec:Equations}, we found a realistic range of $\scriptR$ to be $10^{-120} < \scriptR < 10^{-111}$. Ultimately, the dynamics are not affected by how small $\scriptR$ is; as $\scriptR \rightarrow 0$, $dS_{2+}$ just moves closer to the origin. In reality we could fix $w_x$ such that the system depends on one parameter and only change $q$ to show the full range of dynamics. However, we change both $w_x$ and $q$ to ensure the phase spaces are readable, while maintaining the same topology.

It is clear from Table \ref{tab:Eigenvalues} that there are three possible types of stability character for the repellor $dS_{1+}$: a repelling node, an improper node or a spiral repellor. In all three cases the dynamics is the same. In the following, we show the phase spaces where $dS_{1+}$ is a spiral repellor; however, we present examples of the improper node and repelling node cases in the Appendix.

\begin{table}
\centering
 \begin{tabular}{|| c | c | c ||} 
 \hline
 
 Name & Stability & Color \\ [0.5ex] 
 
\hline\hline

$dS_{1+}$ & Repellor & Orange \\

$dS_{2+}$ & Attractor & Magenta \\

$dS_{3+}$ & Saddle & Dark blue \\

$S_{1+}$ & Generalized saddle & Black \\

$S_{2+}$ & Nonsimple ($q<3$) or generalized saddle ($q>3$) & Black \\

$S_{3+}$ & Generalized saddle & Black \\
 
 \hline
 \end{tabular}
 \caption{\label{tab:ZX_FPs_Stability_Colour} Stability and color of the fixed points and critical points of the $Z$-$X$ system in the expanding case ($Y > 0$), with $\epsilon = -1$, $-1 + \scriptR < w_x < 0$ and $0 < \scriptR < 1$. $dS_{+}$ denotes an expanding (+) de Sitter universe ($x' = z' = 0$), and $S_{+}$ denotes a singularity with infinite expansion (+).}
\end{table}

\begin{table}
\centering
 \begin{tabular}{|| c | c | c ||} 
 \hline
 
 Name & Color \\ [0.5ex] 
 
\hline\hline

Separatrices & Black curves \\

Zero acceleration ($a'' = 0$) curve & Red \\

$Z' = 0$ curve &  Dark blue \\

$X' = 0$ curve & Green \\
 
 \hline
 \end{tabular}
 \caption{Colors of different features in the two-dimensional $Z$-$X$ phase space.}
 \label{tab:2-D_features}
\end{table}

 
 





 

In the $Z$-$X$ phase spaces two separatrices exist (see Fig. \ref{fig:FPs_Only}). The first is the \textit{repellor-saddle separatrix}, which joins the $dS_{1+}$ repellor and the $dS_{3+}$ saddle, and the second is the \textit{repellor-singularity separatrix}, which joins $dS_{1+}$ and the singularity $S_{2+}$. These separatrices separate two types of trajectories. The first evolve to the right of these separatrices, which are nonsingular models. They emerge during expansion from the nonsingular de Sitter fixed point $dS_{1+}$ and expand toward the late-time attractor $dS_{2+}$, where the dark energy asymptotically approaches a cosmological constant and the dark matter energy density tends to zero. Some trajectories have decreasing dark energy density ($\dot{X} < 0$) as they approach $dS_{2+}$, and some have phantom behavior with increasing dark energy density ($\dot{X} > 0$), which can be seen in Fig. \ref{fig:SR_No_Cross}(\subref{subfig:ZX_low_energy}). Trajectories to the left of the separatrices also expand from $dS_{1+}$, but evolve towards $S_{2+}$. 

There are three subcases for the $Z$-$X$ plane which we present. The topology of the phase space is the same, but they represent physically different models. The difference in each case is whether the repellor-saddle separatrix and the zero acceleration $a'' = 0$ curve intersect twice, touch or do not intersect. The zero acceleration curve is found by setting Eq. \eqref{eqn:accn} equal to zero. In compact variables, this equation is 

\begin{equation}
    \frac{Z}{1-Z} + \frac{X}{1-X} (1 + 3w_x -3\epsilon\scriptR) - 3\scriptR (1+w_x) + 3\epsilon\frac{X^2}{(1-X)^2} = 0 \,,
    \label{eqn:zero_accn}
\end{equation}

\noindent which we plot in red in the two-dimensional phase spaces. In all cases, either the $a'' = 0$ curve crosses the $Z = 0$ axes below the $dS_{2+}$ attractor, or it does not cross the $X$-axis at all, therefore all trajectories have late-time acceleration. In the full three-dimensional system, where we include the equation for $Y$, Eq. \eqref{eqn:zero_accn} represents a curve of static Einstein models along $Y = 0$. In Sec. \ref{sec:3D_qLT3}, we only plot first integral surfaces in the $X$-$Y$-$Z$ phase spaces, set by a specific $Z_0$ and $X_0$, which correspond to a single trajectory in the two-dimensional phase space. If the trajectory touches or intersects the $a'' = 0$ curve in the $Z$-$X$ phase space, then the first integral surface touches or intersects the zero acceleration surface in the corresponding three-dimensional phase space. Einstein fixed points exist in the three-dimensional phase spaces where these two surfaces touch or intersect at $Y = 0$. The physically interesting models are the ones which have a decelerated period, which are represented by the trajectories which intersect the $a'' = 0$ curve twice in the two-dimensional phase space, and correspond to a first integral surface in three dimensions which have two Einstein fixed points along $Y = 0$.

In the following, we first present the three cases for the two-dimensional phase spaces. The stability of the fixed points and critical points in the expanding ($Y > 0$) $Z$-$X$ phase spaces, along with their color in the plots, are given in Table \ref{tab:ZX_FPs_Stability_Colour}, and the color scheme for the different types of curves are given in Table \ref{tab:2-D_features}.

\subsection{$Z$-$X$ phase spaces}

\subsubsection{The repellor-saddle separatrix and the $a'' = 0$ curve do not intersect} \label{sec:qLT3_CDNC}

\begin{figure}
\begin{subfigure}[h]{0.47\linewidth}
\includegraphics[width=\linewidth]{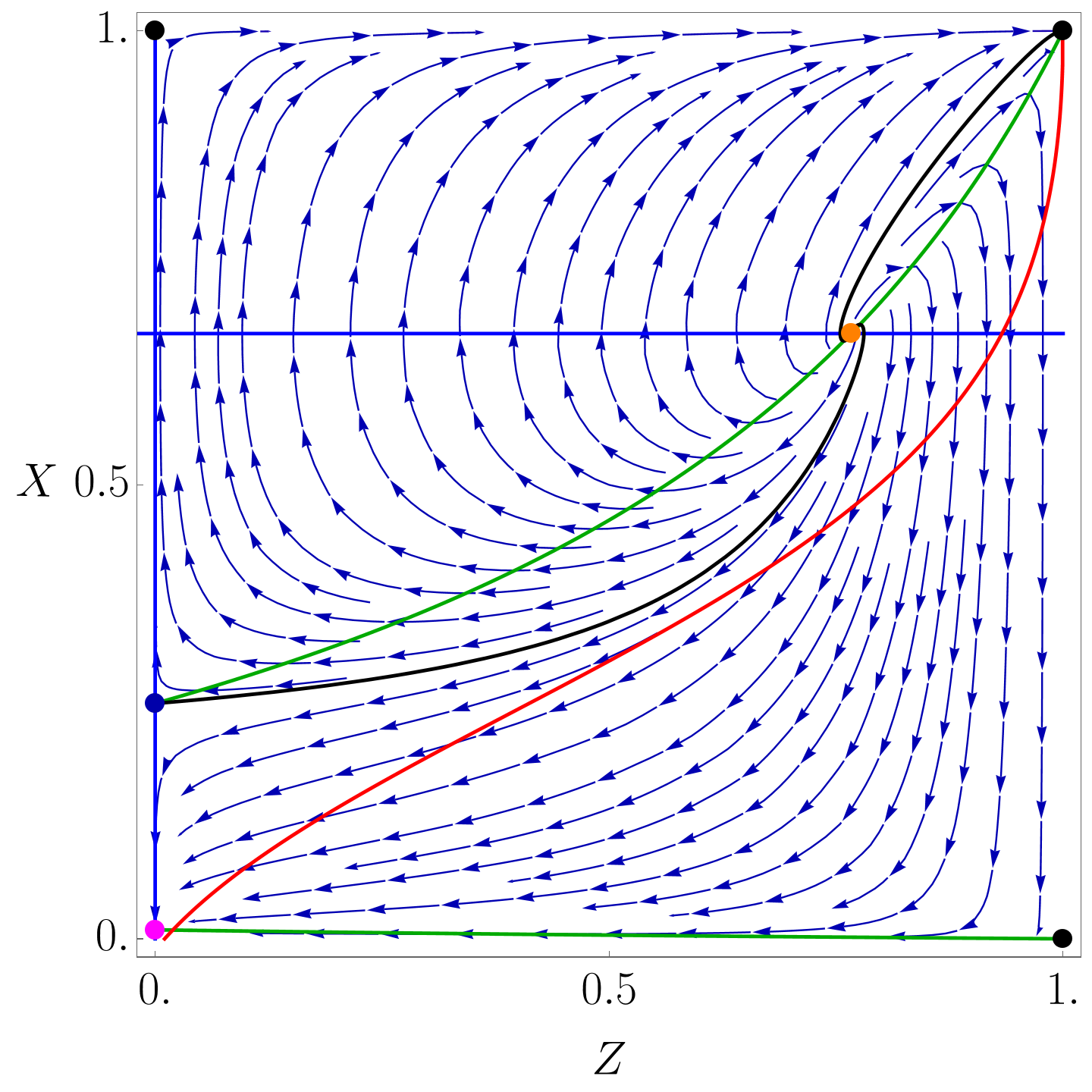}
\caption{Full phase space.}
\label{subfig:ZX_qLT3_CDNC}
\end{subfigure}
\hfill
\begin{subfigure}[h]{0.47\linewidth}
\includegraphics[width=\linewidth]{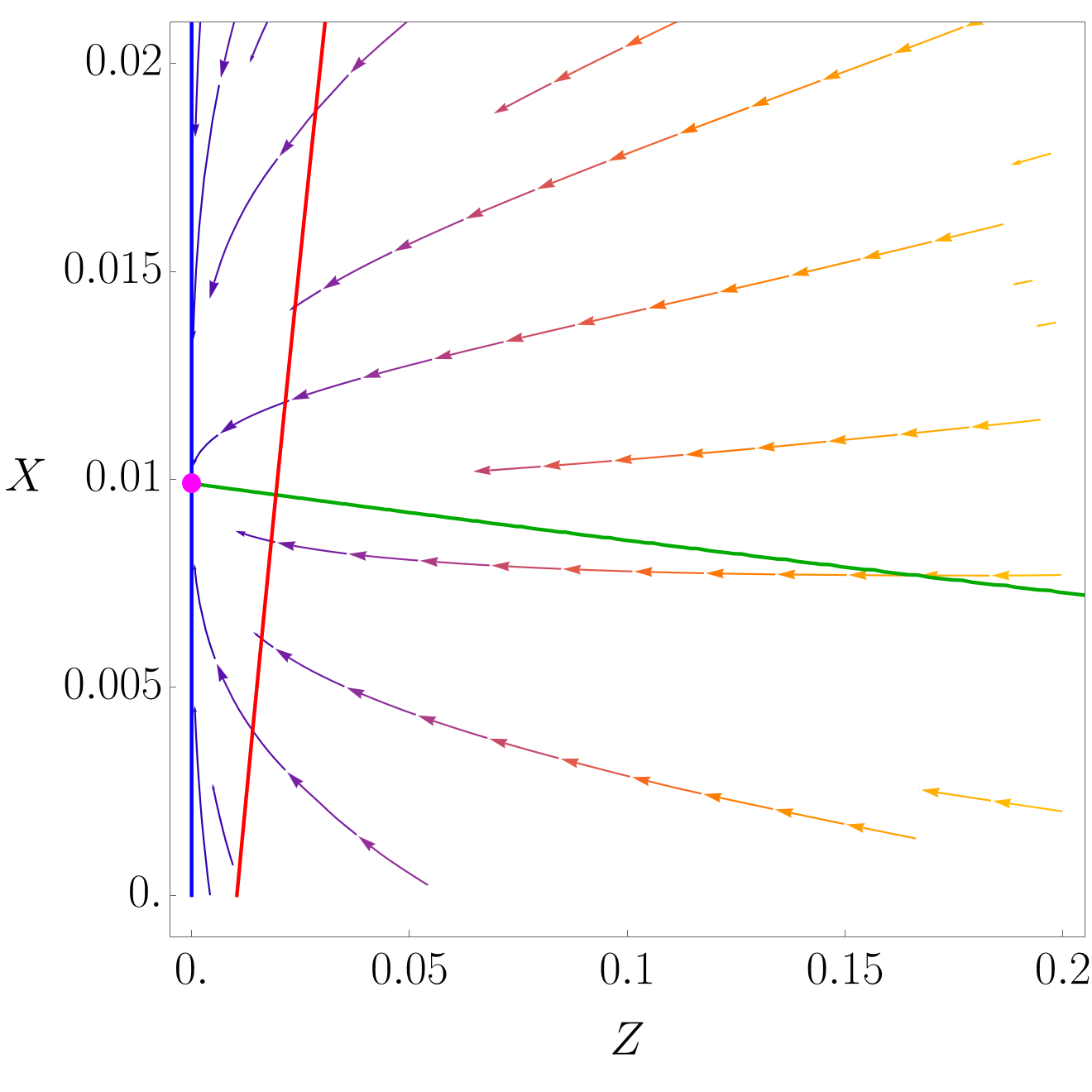}
\caption{The behavior of the trajectories that approach $dS_{2+}$.}
\label{subfig:ZX_low_energy}
\end{subfigure}
\caption{$Z$-$X$ phase space referred to in Sec. \ref{sec:qLT3_CDNC}, where parameters are set to $w_x=-0.65$ and $q=1.5$. (\subref{subfig:ZX_qLT3_CDNC}) shows the full phase space. Trajectories that evolve to the left of the two separatrices expand to the critical point $S_{2+}$ at $(1,1)$ representing a singularity, and those that evolve to the right expand towards the attractor fixed point $dS_{2+}$ (magenta). The zero acceleration curve does not intersect the repellor-saddle separatrix, therefore all trajectories that evolve to the left of the separatrices, and those that evolve to the right of the separatrices that do not intersect the $a'' = 0$ curve, always accelerate. Trajectories that intersect the $a'' = 0$ curve twice have a decelerated period, followed by a late-time acceleration. The trajectory that touches the zero acceleration curve reaches a point where there is zero acceleration, but accelerates otherwise. (\subref{subfig:ZX_low_energy}) is an enlargement of panel (\subref{subfig:ZX_qLT3_CDNC}) around the origin, showing the behavior of the trajectories that evolve to $dS_{2+}$. Those that evolve to $dS_{2+}$ from above the green $X' = 0$ curve have nonphantom behavior, with decreasing dark energy density. Some trajectories have late-time phantom behavior, and cross below the green $X' = 0$ curve and approach $dS_{2+}$ with increasing dark energy density. All trajectories accelerate as they approach $dS_{2+}$, as they are to the left of the red $a'' = 0$ curve. 
}
\label{fig:SR_No_Cross}
\end{figure}

The first subcase when $dS_{1+}$ is a spiral repellor is shown in Fig. \ref{fig:SR_No_Cross}.\footnote{We plot $Z$ on the horizontal axis and $X$ on the vertical axis so the two-dimensional phase spaces correspond more clearly to the three-dimensional phase spaces, where $X$ is plotted on the vertical axis and $Z$ is plotted on the depth axis.} Here, the $a'' = 0$ curve and repellor-saddle separatrix do not intersect, and not all trajectories have a decelerated period. Those trajectories that do not intersect the zero acceleration curve correspond to first integral surfaces in the $X$-$Y$-$Z$ phase space that have no Einstein fixed points. Some of these trajectories expand to the left of the two separatrices and evolve towards $S_{2+}$, and some evolve to the right of the two separatrices and asymptotically expand towards the late-time attractor $dS_{2+}$. Both of these sets of trajectories are always accelerating, and so are not of interest. One trajectory in the phase space touches the $a'' = 0$ curve, and corresponds to a first integral surface in three dimensions that has one Einstein fixed point. This trajectory evolves toward $dS_{2+}$, and will have zero acceleration where it touches the zero acceleration curve, but does not have a decelerated period so is not qualitatively interesting. The rest of the trajectories intersect the $a'' = 0$ curve twice, which correspond to first integral surfaces in three dimensions with two Einstein fixed points. These trajectories initially accelerate, then cross below the $a'' = 0$ curve and decelerate, and finally cross back above $a'' = 0$ and accelerate as they asymptotically approach the late-time attractor $dS_{2+}$. These trajectories are of interest as qualitatively they have a decelerated period where large-scale structure could form, followed by a late-time accelerated expansion that tends toward a cosmological constant.


\subsubsection{The repellor-saddle separatrix and the $a'' = 0$ curve touch} \label{sec:qLT3_CT}

\begin{figure}
    \centering
    \includegraphics[width=\linewidth]{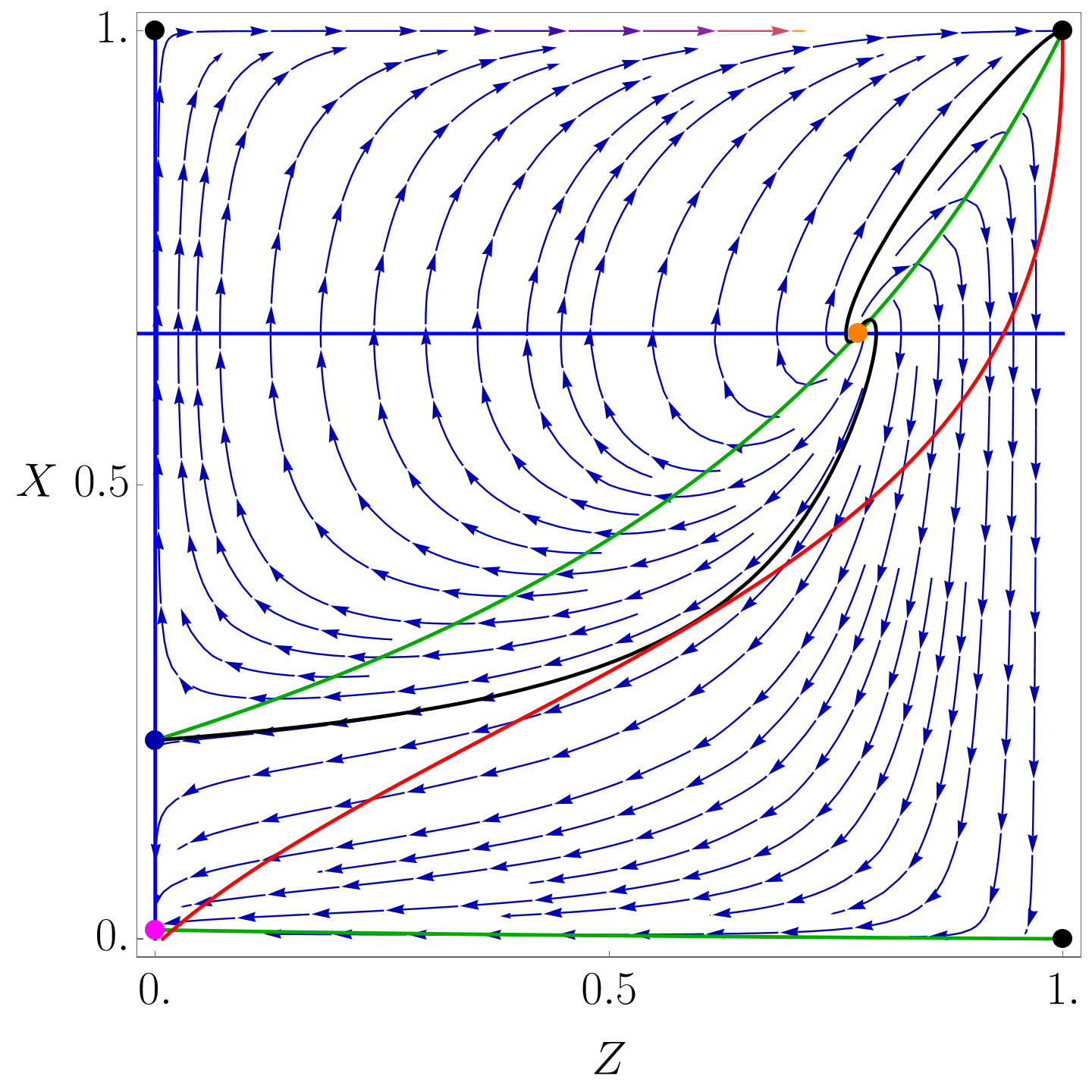}
    \caption{$Z$-$X$ phase space referred to in Sec. \ref{sec:qLT3_CT}. Here the parameters are set to $w_x=-0.72$ and $q=1.5$. Trajectories that evolve to the left of the two separatrices expand to the critical point $S_{2+}$ at $(1,1)$ representing a singularity, and those that evolve to the right expand towards the attractor fixed point $dS_{2+}$ (magenta). In this case, the zero acceleration curve touches the repellor-saddle separatrix. Trajectories that evolve to the left of the separatrices therefore always accelerate, and those that evolve to the right all have a decelerated period. The repellor-saddle separatrix is a trajectory that has a point of zero acceleration, but accelerates otherwise.}
    \label{fig:SR_qLT3_Touch}
\end{figure}

Figure \ref{fig:SR_qLT3_Touch} shows the subcase where the repellor-saddle separatrix and zero acceleration curve touch. The repellor-saddle separatrix itself corresponds to a first integral surface in the $X$-$Y$-$Z$ phase space with one Einstein fixed point. This trajectory initially accelerates, reaches a point with zero acceleration and then accelerates again, and tends towards the saddle fixed point $dS_{3+}$. As this trajectory has no decelerated period, it is not of interest. Trajectories that evolve to the left of the two separatrices correspond to first integral surfaces in three dimensions that have zero Einstein fixed points, and never decelerate. These trajectories also evolve to a singularity, therefore they are not of interest. The rest of the trajectories evolve to the right of the separatrices, and intersect the $a'' = 0$ curve twice, meaning they have a decelerated period. They also all tend towards $dS_{2+}$ at late times, and are therefore the trajectories of interest.

\subsubsection{The repellor-saddle separatrix and the $a'' = 0$ curve intersect} \label{sec:qLT3_CC}

\begin{figure}
    \centering
    \includegraphics[width=\linewidth]{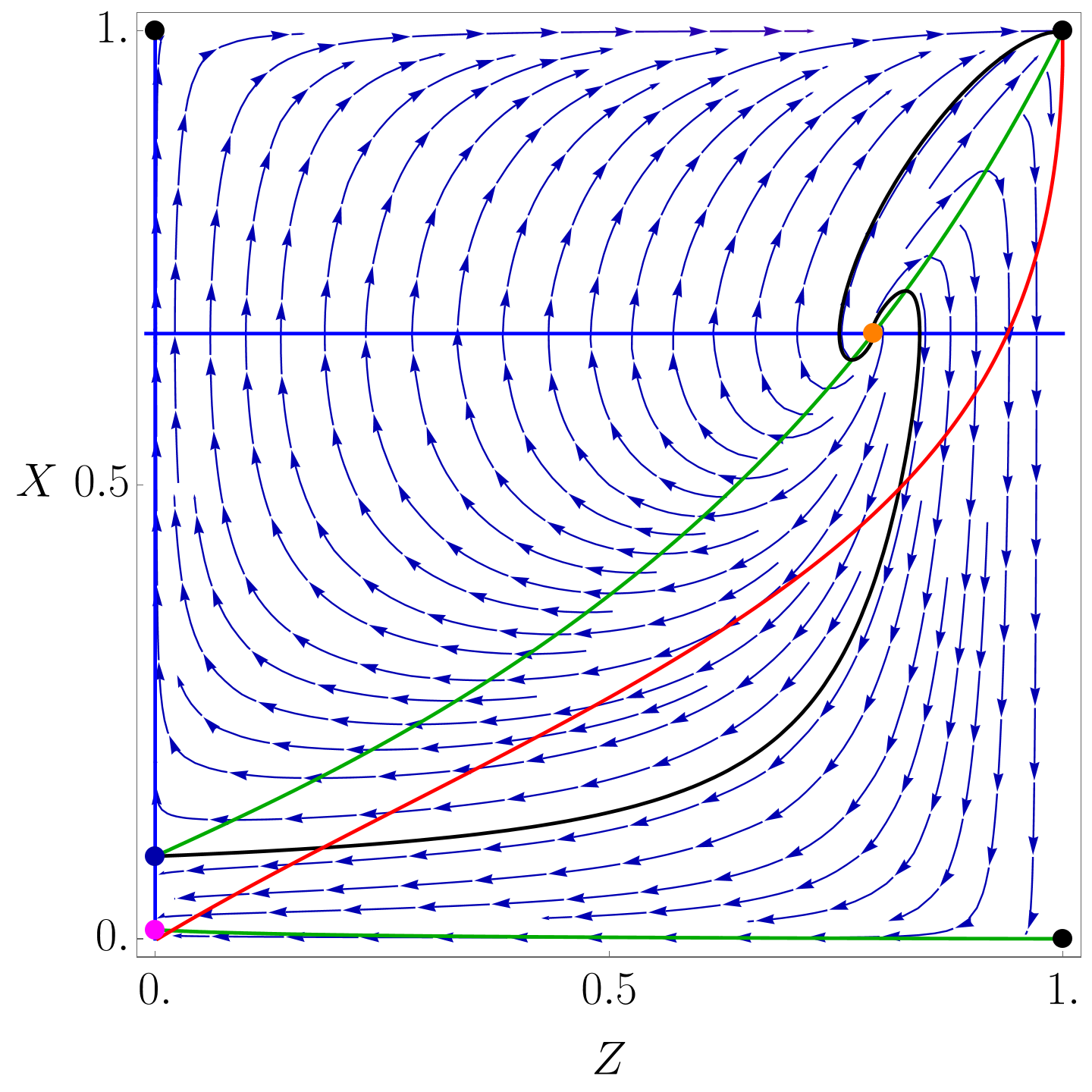}
    \caption{$Z$-$X$ phase space referred to in Sec. \ref{sec:qLT3_CC}. Here, the parameters are set to $w_x=-0.9$ and $q=1.5$. Trajectories that evolve to the left of the two separatrices expand to the critical point $S_{2+}$ at $(1,1)$ representing a singularity, and those that evolve to the right expand towards the attractor fixed point $dS_{2+}$ (magenta). In this case, the zero acceleration curve intersects the repellor-saddle separatrix twice, therefore trajectories that evolve to the right of the separatrices all have a decelerated phase followed by late-time acceleration.}
    \label{fig:SR_qLT3_Cross}
\end{figure}

The final subcase is where the repellor-saddle separatrix intersects the $a'' = 0$ curve twice, as shown in Fig. \ref{fig:SR_qLT3_Cross}. The trajectories that evolve to the left of the two separatrices all tend toward $S_{2+}$; those that intersect the $a'' = 0$ curve twice have a period of deceleration, and those that do not intersect $a'' = 0$ always accelerate. Regardless, these trajectories are not interesting with respect to our analysis as they evolve towards a singularity. Trajectories to the right of the separatrices all have a decelerated period followed by a late-time acceleration, where they tend toward $dS_{2+}$. These trajectories are all of interest, therefore this is the subcase that we focus our analysis in three dimensions on in the following section.


\subsection{Three-dimensional phase space for viable models} \label{sec:3D_qLT3}

\begin{table}
\centering
 \begin{tabular}{||c | c c c||} 
 \hline
 
 Name & $x$ & $y$ & $z$\\ [0.5ex] 
 
\hline\hline

 $E$ & $x$ & $0$ & $-x(3\scriptR+1+3w_x) + 3\scriptR(1+w_x) + 3x^2$ \\

$dS_{1\pm}$ & $\frac{3}{q}$ & $\pm \sqrt{\frac{\frac{1}{3}q^2\scriptR(1+w_x) - q(\scriptR + w_x) + 3}{q}}$ & $\frac{(q\scriptR-3)\left[q(1+w_x)-3\right]}{q^2}$ \\

$dS_{2\pm}$ & $\scriptR$ & $\pm\sqrt{\frac{\scriptR}{3}}$ & $0$ \\

$dS_{3\pm}$ & $1+w_x$ & $\pm \sqrt{\frac{(1+w_x)}{3}}$ & $0$ \\

 \hline
 \end{tabular}
 \caption{\label{tab:Fixed_Points_xyz} Fixed points of the full three-dimensional system in Eqs. \eqref{eqn:x_prime} -- \eqref{eqn:y_prime} with $\epsilon = -1$. $E$ denotes a static Einstein universe ($y' = y = 0$) and $dS_{\pm}$ an expanding (+) or contracting (-) de Sitter universe ($x' = z' = 0$).}
\end{table}

\begin{table}
\centering
 \begin{tabular}{||c | c c c ||} 
 \hline
 
 Name & $X$ & $Y$ & $Z$ \\ [0.5ex] 
 
\hline\hline

 $dS_{4\pm}$ & $\frac{3}{3+q}$ & $\pm 1$ & $\frac{(-3+q\scriptR)(q+qw_x-3)}{-3q+q^2-3qw_x + q^2 \scriptR+q^2w_x \scriptR -3q\scriptR +9}$ \\

 $dS_{5\pm}$ & $\frac{\scriptR}{1+\scriptR}$ & $\pm 1$ & $0$ \\

 $dS_{6\pm}$ & $\frac{1+w_x}{2+w_x}$ & $\pm 1$ & $0$ \\

 $S_{1\pm}$ & $1$ & $\pm 1$ & $0$ \\

 $S_{2\pm}$ & $1$ & $\pm 1$ & $1$ \\

 $S_{3\pm}$ & $0$ & $\pm 1$ & $1$ \\
 
 \hline
 \end{tabular}
 \caption{\label{tab:Fixed_Points_XYZ} Critical points that arise at $y \rightarrow \infty$ from compactification of the full system in Eqs. \eqref{eqn:X'} -- \eqref {eqn:Y'}, with $\epsilon = -1$. $dS_{\pm}$ denotes an expanding (+) or contracting (-) de Sitter universe ($X' = Z' = 0$), and $S_{\pm}$ denotes a singularity with infinite expansion (+) or contraction (-).}
\end{table}

\begin{table}
\centering
 \begin{tabular}{||c | c ||} 
 \hline
 
 Name & Color \\ [0.5ex] 
 
\hline\hline

 $E$ & Cyan \\

 $dS_{1\pm}$ & Orange \\

 $dS_{2\pm}$ & Magenta \\

 $dS_{3\pm}$ & Dark blue \\

 $dS_{4\pm}$ & Orange \\

 $dS_{5\pm}$ & Magenta \\

 $dS_{6\pm}$ & Dark blue \\

 $S_{2\pm}$ & Black \\

 \hline
 \end{tabular}
     \caption{\label{tab:Fixed_Points_XYZ_Colour} Colors of the fixed points and critical points in the $X$-$Y$-$Z$ phase spaces. $E$ denotes a static Einstein universe and $dS_{\pm}$ an expanding (+) or contracting (-) de Sitter universe ($X' = Z' = 0$). $S_{\pm}$ denotes a singularity with infinite expansion (+) or contraction (-).}
\end{table}

\begin{table}
\centering
 \begin{tabular}{|| c | c ||} 
 \hline
 
 Name & Color \\ [0.5ex] 
 
\hline\hline

Closed Friedmann separatrix & Black curve \\

Zero acceleration surface ($a'' = 0)$ & Red surface \\

Closed trajectories &  Purple \\

Flat trajectories & Green \\

Open trajectories & Dark blue \\

High energy analytic approximation ($q > 3$ phase spaces only) & Orange section of the trajectories \\
 
 \hline
 \end{tabular}
 \caption{Colors of different features in the three-dimensional $X$-$Y$-$Z$ phase spaces.}
 \label{tab:3-D_features}
\end{table}

In this subsection, we focus our analysis on the set of trajectories that evolve to the right of the separatrices in Fig. \ref{fig:SR_qLT3_Cross}. The fixed points of the full system are shown in Table \ref{tab:Fixed_Points_xyz} and the critical points are shown in Table \ref{tab:Fixed_Points_XYZ}. Their color in the three-dimensional plots are shown in Table \ref{tab:Fixed_Points_XYZ_Colour}, which correlates to the two-dimensional phase spaces. The color of the different features in the three-dimensional phase spaces are given in Table \ref{tab:3-D_features}. We plot on a constant $Z_0$ and $X_0$ surface in three dimensions, meaning the $Z$-$X$ plane in three dimensions at any $Y$ looks the same as the two-dimensional phase space.

\begin{figure}
    \centering
    \includegraphics[width=\linewidth]{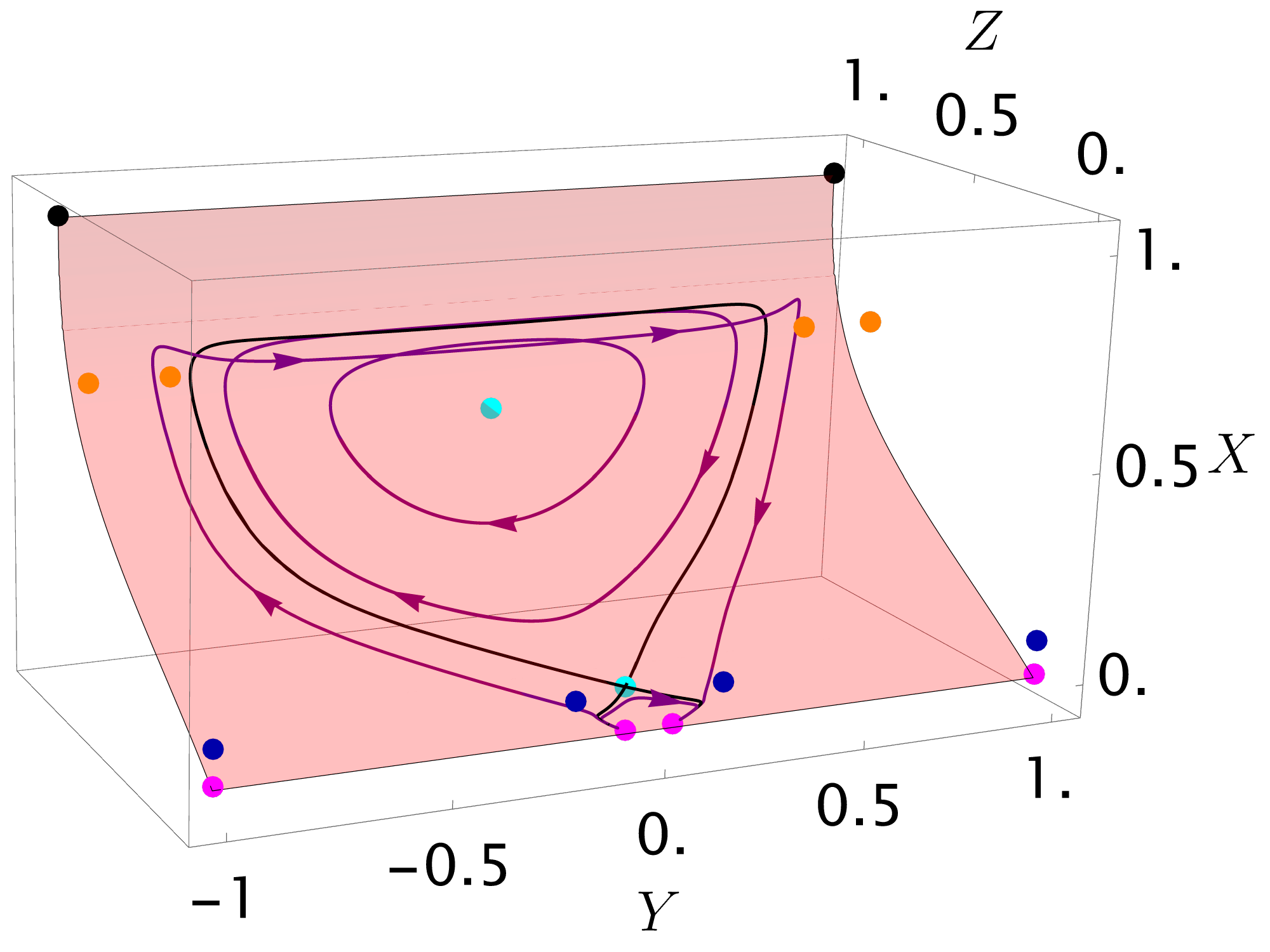}
    \caption{$X$-$Y$-$Z$ phase space corresponding to Fig. \ref{fig:SR_qLT3_Cross}, where $w_x=-0.9$ and $q=1.5$. The trajectories themselves are plotted on a first integral surface where $X_0 = 0.1$ and $Z_0 = 0.5$, which corresponds to a trajectory in Fig. \ref{fig:SR_qLT3_Cross} that evolves to the right of the separatrices, expanding to $dS_{2+}$. In this figure, only the trajectories with positive spatial curvature are plotted. Two Einstein fixed points (cyan) exist at $Y = 0$ on the zero acceleration surface. Trajectories inside the CFS either bounce once, contracting from $dS_{2-}$ and expanding to $dS_{2+}$, or are cyclic around an Einstein fixed point. The bouncing models inside the CFS always accelerate, and the cyclic models cross the zero acceleration surface once during expansion, meaning they have a decelerated phase but no late-time acceleration. Trajectories outside the CFS bounce once between $dS_{2-}$ and $dS_{2+}$, and during expansion cross the zero acceleration surface twice, meaning they have a decelerated phase and a late-time acceleration.}
    \label{fig:SR_XYZ_Cross_2EPts_OS_Closed}
\end{figure}

\begin{figure}
    \centering
    \includegraphics[width=\linewidth]{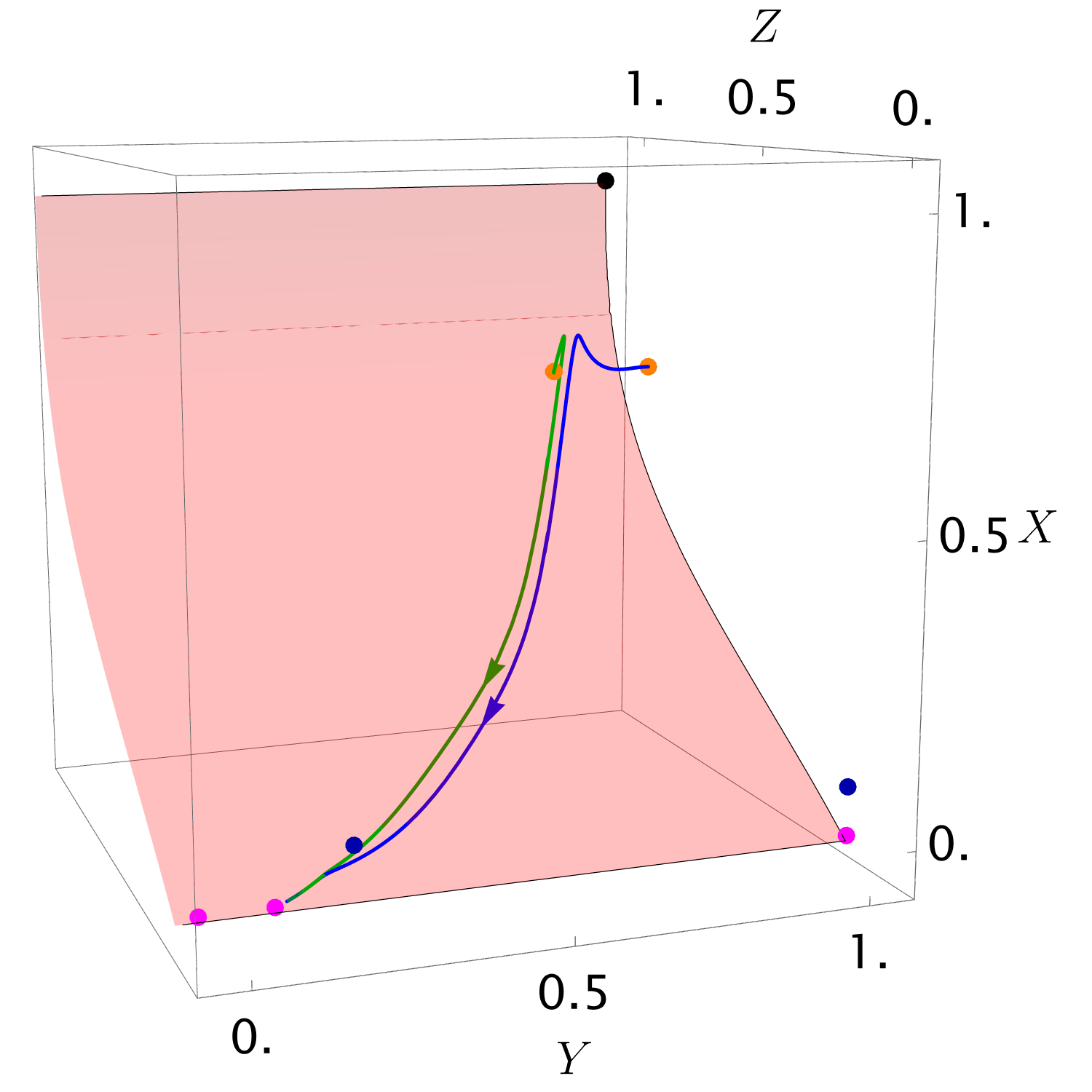}
    \caption{Expanding flat (green) and open (blue) trajectories in the $X$-$Y$-$Z$ phase space corresponding to Fig. \ref{fig:SR_qLT3_Cross}, where $w_x=-0.9$ and $q=1.5$. The trajectories themselves are plotted on a first integral surface where $X_0 = 0.1$ and $Z_0 = 0.5$, which corresponds to a trajectory in Fig. \ref{fig:SR_qLT3_Cross} that evolves to the right of the separatrices, expanding to $dS_{2+}$. The flat and open trajectories emerge from $dS_{1+}$ and $dS_{4+}$, respectively, and both cross the zero acceleration surface twice, meaning they initially accelerate, then have a period of deceleration, then have an accelerated phase as they expand toward $dS_{2+}$ at low energy.}
    \label{fig:SR_XYZ_Cross_2EPts_OS_FlatOpen}
\end{figure}

Figures \ref{fig:SR_XYZ_Cross_2EPts_OS_Closed} and \ref{fig:SR_XYZ_Cross_2EPts_OS_FlatOpen} show a first integral surface in the $X$-$Y$-$Z$ phase space with two Einstein fixed points, which corresponds to a trajectory in Fig. \ref{fig:SR_qLT3_Cross} that evolves to the right of the separatrices and intersects the $a'' = 0$ curve twice. In this case, one Einstein fixed point is a saddle and one is a center. A separatrix exists through the saddle Einstein fixed point, which we call the closed Friedmann separatrix (CFS) \cite{Ananda2006}, which separates different types of trajectories in the three-dimensional phase space. Figure \ref{fig:SR_XYZ_Cross_2EPts_OS_Closed} shows the trajectories with positive spatial curvature only, and Fig. \ref{fig:SR_XYZ_Cross_2EPts_OS_FlatOpen} shows the trajectories corresponding to expanding flat and open models.

The trajectories in Fig. \ref{fig:SR_XYZ_Cross_2EPts_OS_Closed} all have a bounce. 
The trajectories that evolve outside the CFS, as well as a subset of those that evolve inside the CFS, bounce once. These trajectories contract from $dS_{2-}$, go through a nonsingular bounce, and then expand towards $dS_{2+}$. There is also a subset of cyclic models within the CFS. These cycle around the Einstein fixed point which is a center, and repeatedly expand, reach a turnaround, contract and bounce. The trajectories that evolve inside the CFS are not of physical interest, as qualitatively they do not match the observed Universe. The cyclic models accelerate through the bounce, and then have a decelerated expanding period; however, they never have a late-time accelerated expansion \footnote{For an example of cyclic models that have a decelerated phase followed by a late-time accelerated expansion, see Ref. \cite{Bruni2022}.}. The bouncing models that evolve inside the CFS always accelerate. The trajectories that bounce once outside of the CFS are the models of interest, as they have an accelerated era evolving toward a high energy quasi-de Sitter phase with a bounce, becoming flatter as they start to expand, then have a decelerated expanding phase followed by a final period of acceleration as they evolve toward the low energy cosmological constant represented by $dS_{2+}$.
 
The expanding open and flat trajectories in Fig. \ref{fig:SR_XYZ_Cross_2EPts_OS_FlatOpen} are all qualitatively interesting. These models emerge from $dS_{1+}$ and $dS_{4+}$, respectively, and initially accelerate. They then have a decelerated period as they cross below the $a'' = 0$ surface, and have a final late-time acceleration as they asymptotically tend toward $dS_{2+}$. We note that although the system can violate the NEC, only closed models can have a bounce as we only consider the region of phase space where we have positive energy densities ($0<X<1$, $0<Z<1$). Therefore, $H \neq 0$ when $k \leq 0$.

\section{Dynamics with $q > 3$} \label{sec:qGT3}

Although nonsingular models exist in the $q < 3$ phase spaces, our main result lies in the $q > 3$ case, as all trajectories are nonsingular for any initial condition. As before, we show the $Z$-$X$ phase spaces, and take $Y > 0$ ($H > 0$) so the trajectories are expanding. Again, we set $\scriptR = 0.01$ to produce readable phase spaces. As in the $q < 3$ cases, there are three possible types of stability character for the $dS_{1+}$ fixed point: a repelling node, an improper node or a spiral repellor. In the two-dimensional phase spaces, only one separatrix exists, which is the repellor-saddle separatrix that joins the $dS_{1+}$ repellor and the $dS_{3+}$ saddle (see Fig. \ref{fig:FPs_Only}). When $q > 3$, all the critical points representing singularities have generalized saddle stability, which means all trajectories avoid the singularities, and expand from $dS_{1+}$ and asymptotically tend towards $dS_{2+}$ at late times. The zero acceleration curve either crosses the $Z = 0$ axes below the $dS_{2+}$ attractor, or does not cross it at all, therefore all trajectories have late-time acceleration as they approach a cosmological constant. The behavior of the trajectories at low energy is as in Fig. \ref{fig:SR_No_Cross}(\subref{subfig:ZX_low_energy}).

As before, the repellor-saddle separatrix may or may not intersect the zero acceleration ($a'' = 0$) curve. When $dS_{1+}$ is a spiral repellor, three subcases exist for the two-dimensional phase space: the repellor-saddle separatrix and $a'' = 0$ curve can intersect twice, touch or do not intersect at all depending on the specific value of $q$. However, when $dS_{1+}$ is a repelling or an improper node, the only subcase that exists is when the separatrix and zero acceleration curve do not intersect. We do not focus on the cases where $dS_{1+}$ is an improper or repelling node here; however, we explore them in the Appendix and highlight the physically interesting models. 

The subcase where $dS_{1+}$ is a spiral repellor, and the repellor-saddle separatrix intersects the $a'' = 0$ curve twice is particularly important, as all trajectories in this phase space are qualitatively interesting. In this case, trajectories can intersect the zero acceleration curve two, three or four times, which correspond to first integral surfaces in three dimensions which have two, three or four Einstein fixed points, respectively. In the following, we present the $Z$-$X$ phase spaces where $dS_{1+}$ is a spiral repellor, so we can show all possible models. The color scheme of the fixed points and critical points, and the color scheme of the curves are given in Tables \ref{tab:ZX_FPs_Stability_Colour} and \ref{tab:2-D_features}, respectively.

\subsection{$Z$-$X$ phase spaces}

\subsubsection{The repellor-saddle separatrix and $a'' = 0$ curve do not intersect} \label{sec:qGT3_CDNC}

\begin{figure}
    \centering
    \includegraphics[width=\linewidth]{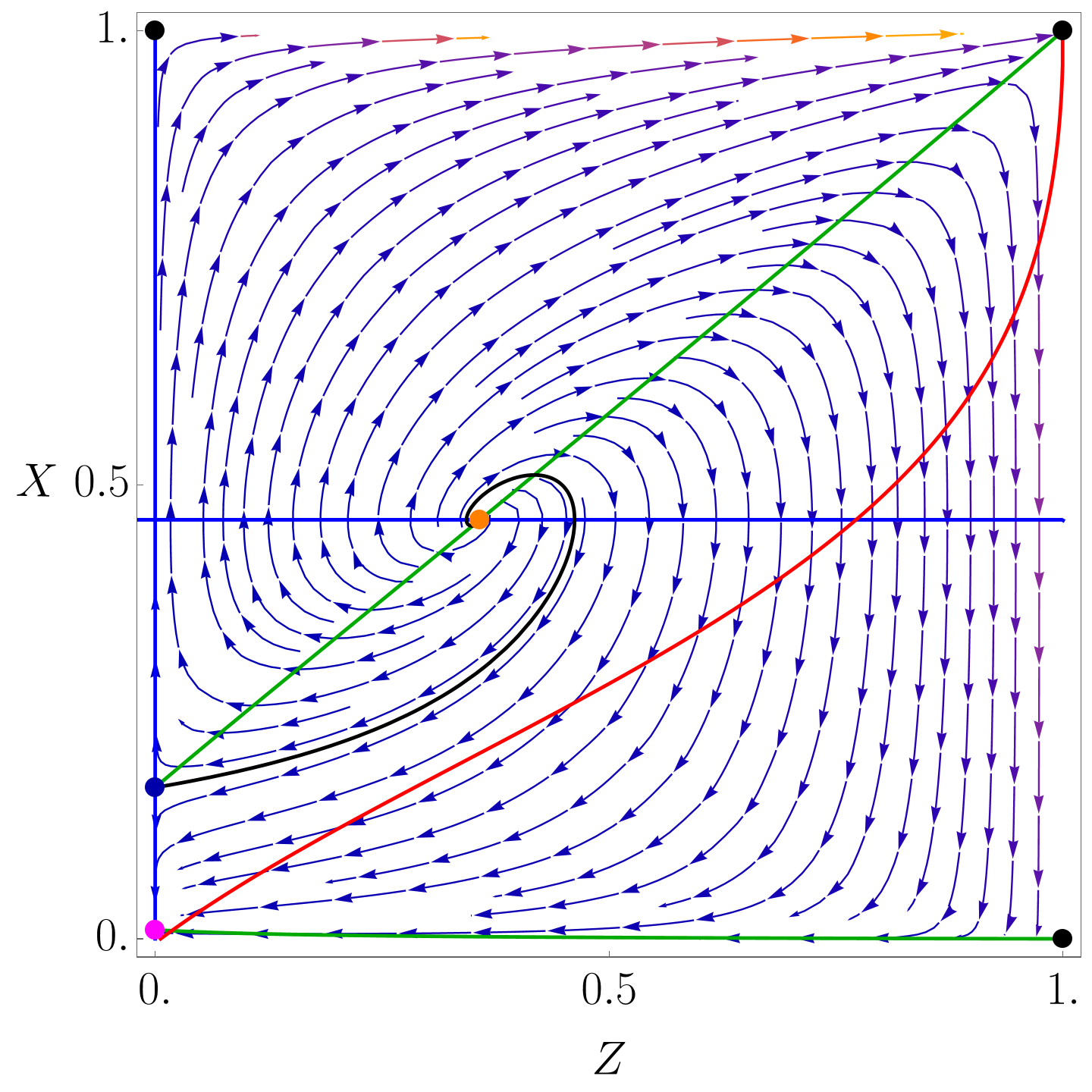}
    \caption{$Z$-$X$ phase space referred to in Sec. \ref{sec:qGT3_CDNC}. Here, the parameters are set to $w_x=-0.8$ and $q=3.5$. All trajectories expand from $dS_{1+}$ (orange) and evolve to $dS_{2+}$ (magenta), except for the separatrix which expands to $dS_{3+}$ (dark blue). Therefore, all models avoid the singularities. In this case; however,, the separatrix does not intersect the zero acceleration curve. Consequently, there are trajectories that always accelerate, and there are trajectories which intersect the $a'' = 0$ curve twice and have a decelerated period. There is also a trajectory that touches the zero acceleration curve, but otherwise accelerates.}
    \label{fig:qGT3_SR_NoCross}
\end{figure}

The first subcase is when the repellor-saddle separatrix and $a'' = 0$ curve do not intersect, which is shown in Fig. \ref{fig:qGT3_SR_NoCross}. Not all trajectories intersect the $a'' = 0$ curve, and those trajectories correspond to first integral surfaces in three dimensions that have no Einstein fixed points. These trajectories never have a decelerated period, and so are not of interest. A trajectory exists that touches, but does not intersect the $a'' = 0$ curve. This trajectory corresponds to a surface in three dimensions that has one Einstein fixed point, and has a point where there is no acceleration, but never decelerates so is not of interest. Finally, there are trajectories that intersect the $a'' = 0$ curve twice. These correspond to surfaces in the three-dimensional phase space with two Einstein fixed points. Initially these trajectories accelerate, then decelerate when they cross below the red $a'' = 0$ curve, and then cross back above the $a'' = 0$ curve and accelerate as they asymptotically approach $dS_{2+}$. 

\subsubsection{The repellor-saddle separatrix and $a'' = 0$ curve touch} \label{sec:qGT3_CT}

\begin{figure}[h!]
    \centering
    \includegraphics[width=\linewidth]{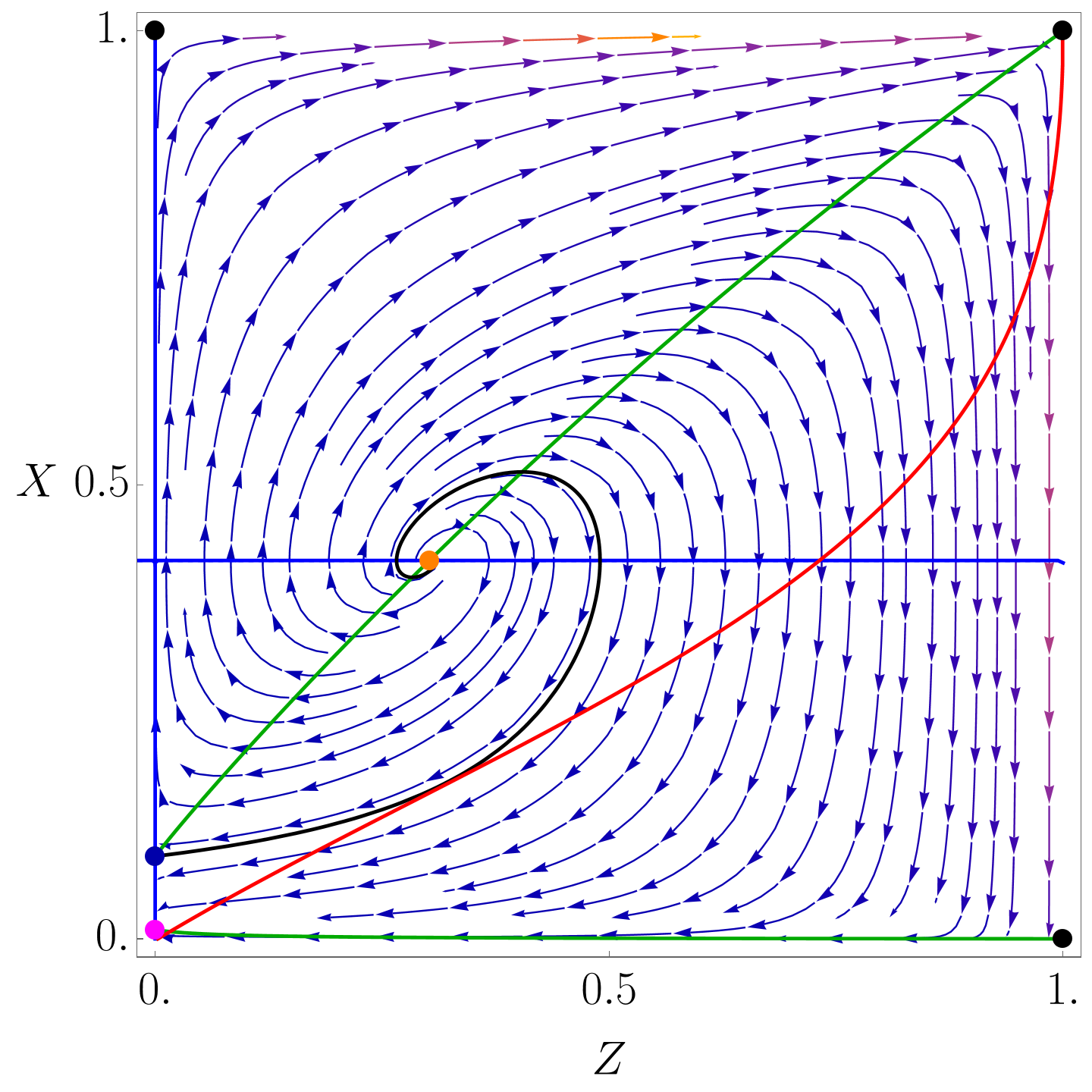}
    \caption{$Z$-$X$ phase space referred to in Sec. \ref{sec:qGT3_CT}. Here, the parameters are set to $w_x=-0.9$ and $q=4.2$. In this case, the separatrix touches the zero acceleration curve, therefore the separatrix itself is a trajectory that reaches a point of zero acceleration, but otherwise accelerates. All other trajectories initially accelerate as they expand from $dS_{1+}$, then cross below the red $a'' = 0$ curve and have a decelerated period, and finally cross above the $a '' = 0$ curve again, accelerating as they evolve to $dS_{2+}$.}
    \label{fig:qGT3_SR_Touch}
\end{figure}

Figure \ref{fig:qGT3_SR_Touch} shows the subcase where the separatrix and $a'' = 0$ curve touch. In this case, the separatrix itself is a trajectory which corresponds to a surface in the $X$-$Y$-$Z$ phase space with one Einstein fixed point. This trajectory reaches a point with zero acceleration; however, it never decelerates. All other trajectories intersect the $a'' = 0$ curve twice, initially accelerating, then decelerating when they cross below the $a'' = 0$ curve, and accelerating again at late times as they asymptotically expand toward $dS_{2+}$. These trajectories have corresponding first integral surfaces in three dimensions that have two Einstein fixed points.

\subsubsection{The repellor-saddle separatrix and $a'' = 0$ curve intersect twice} \label{sec:qGT3_CC}


\begin{figure}
\begin{subfigure}[h]{0.47\linewidth}
\includegraphics[width=\linewidth]{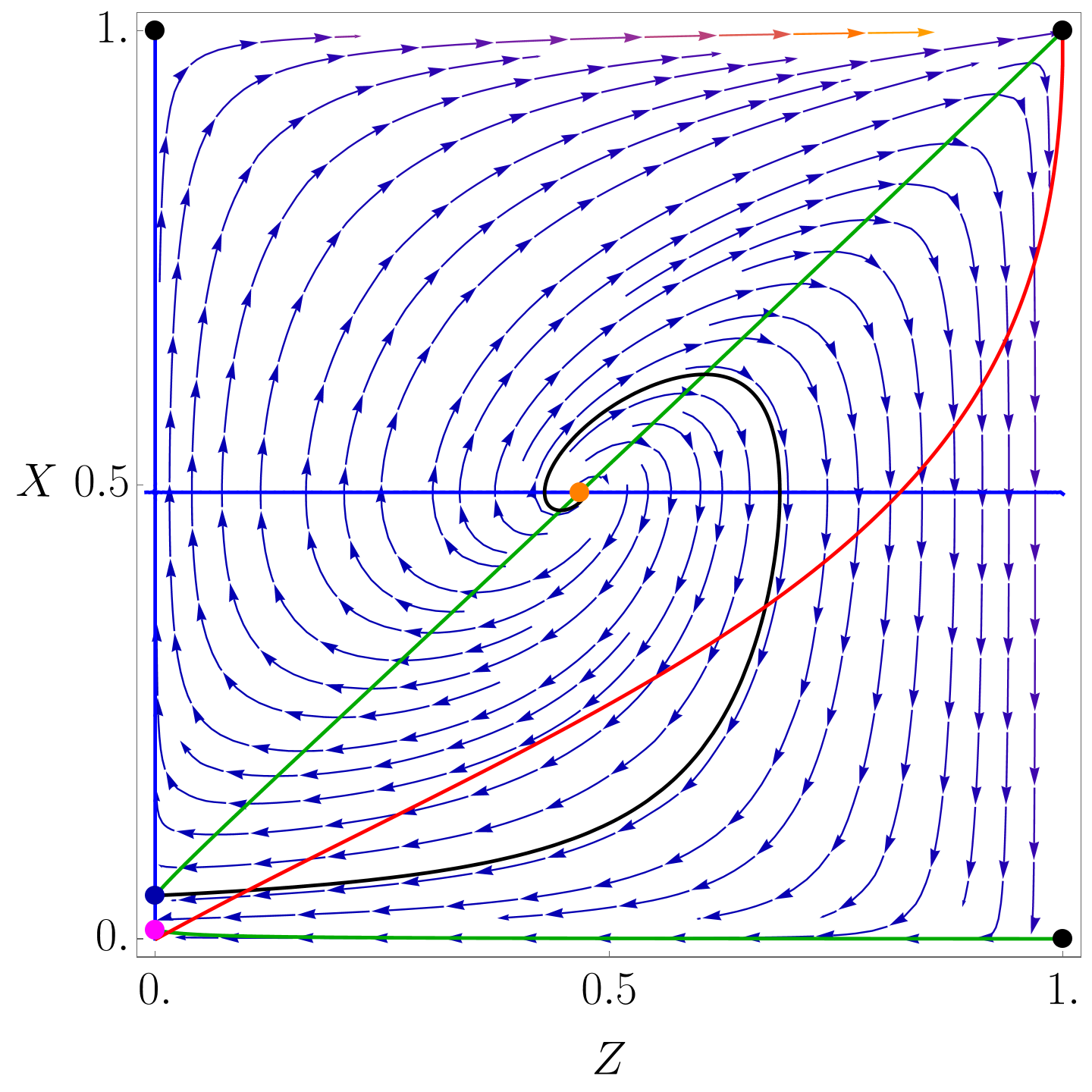}
\caption{Full phase space.}
\label{subfig:ZX_qGT3_CC}
\end{subfigure}
\hfill
\begin{subfigure}[h]{0.47\linewidth}
\includegraphics[width=\linewidth]{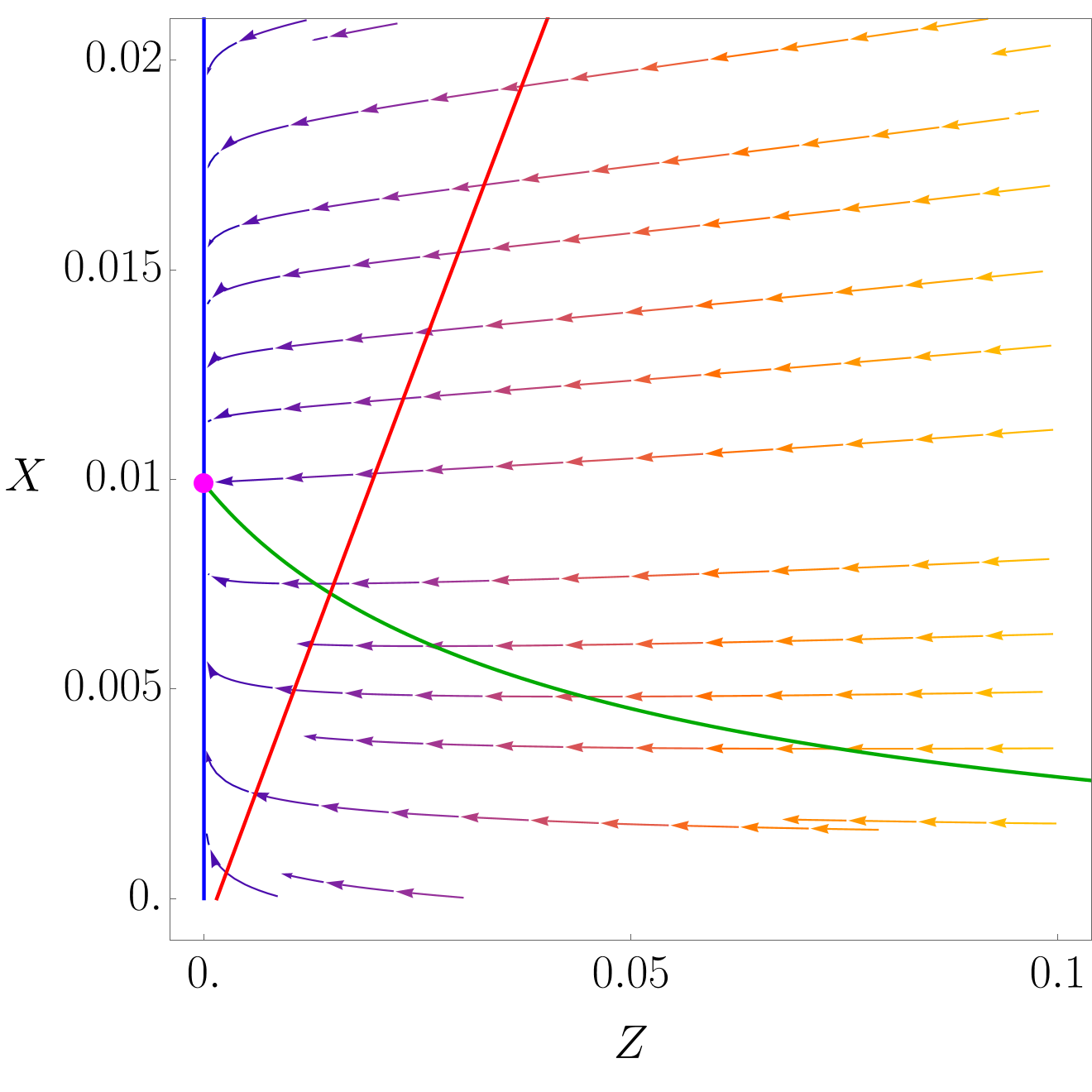}
\caption{The behavior of the trajectories as they approach $dS_{2+}$.}
\label{subfig:ZX_low_energy_qGT3}
\end{subfigure}
\caption{$Z$-$X$ phase space referred to in Sec. \ref{sec:qGT3_CC}. Here, the parameters are set to $w_x=-0.95$ and $q=3.1$. (\subref{subfig:ZX_qGT3_CC}) shows the full phase space, where all trajectories expand from $dS_{1+}$ (orange) and evolve to $dS_{2+}$ (magenta), except for the saddle-repellor separatrix which expands to $dS_{3+}$ (dark blue). The separatrix intersects the zero acceleration curve twice, which means all trajectories have a decelerated period. Trajectories which intersect the zero acceleration curve four times have two periods of deceleration, and those that intersect it twice have one decelerated phase. There is a trajectory that expands from the repellor and touches the zero acceleration curve and then intersects it twice. All trajectories in this case are of interest, as all are nonsingular and have a decelerated phase where large-scale structure could form. (\subref{subfig:ZX_low_energy_qGT3}) shows the behavior of the trajectories as they approach $dS_{2+}$. Some trajectories have decreasing dark energy density and evolve towards $dS_{2+}$ from above the green $X' = 0$ curve, and some cross below the $X' = 0$ curve and exhibit phantom behavior with increasing dark energy density.}
\label{fig:qGT3_ZX_SR_Cross}
\end{figure}

The final subcase is where the separatrix and zero acceleration curve intersect twice, which is shown in Fig. \ref{fig:qGT3_ZX_SR_Cross}. In this case all trajectories have at least one period of deceleration, and are therefore all of interest. Some trajectories intersect the $a'' = 0$ curve twice, which correspond to surfaces in the $X$-$Y$-$Z$ phase space that have two Einstein fixed points. These trajectories initially accelerate, then have one period of deceleration, and accelerate again at late times. One trajectory in the phase space will touch the $a'' = 0$ curve, and then intersect it twice; this trajectory also has one decelerated period, and corresponds to a surface in three dimensions that has three Einstein fixed points. Finally, there are trajectories which intersect the $a'' = 0$ curve four times, which correspond to first integral surfaces in three dimensions that have four Einstein fixed points. These trajectories have two periods of deceleration when they cross below the $a'' = 0$ curve, and have three acceleration periods when they are above it. All of the trajectories here are of interest, therefore we concentrate our analysis in three dimensions on this subcase.

\vspace{0.2cm}
\subsection{Three-dimensional phase spaces}

We will focus our analysis in three dimensions on the subcase shown in Fig. \ref{fig:qGT3_ZX_SR_Cross}, as all of these trajectories qualitatively match observations, as they have a decelerated matter dominated era followed by a late-time acceleration. There are three cases we present in the following sections, as the trajectories in Fig. \ref{fig:qGT3_ZX_SR_Cross} can correspond to first integral surfaces in three dimensions that have two, three or four Einstein fixed points. Some of the trajectories plotted in Figs. \ref{fig:SR_qGT3_XYZ_3EPts_Closed} -- \ref{fig:qGT3_XYZ_4EPts_Closed_Side} have orange sections. The numerical solutions in \texttt{Mathematica} break down close to $S_{2\pm}$, so it is not possible to plot some trajectories completely numerically. In order to plot these trajectories fully, we patch together the numeric solutions with analytic approximations. We solve the equations numerically until the solution breaks down. We then take initial conditions from the numeric solution, and solve the high energy analytic approximation in Eq. \eqref{eqn:x{z}_HE} around the singularity, which we plot in orange. We then take initial conditions from the high energy analytic approximation to plot the rest of the trajectory numerically. The color scheme of the fixed points and critical points in the three-dimensional phase spaces are given in Table \ref{tab:Fixed_Points_XYZ_Colour}, and the color scheme of the curves and surfaces are given in Table \ref{tab:3-D_features}.

\subsubsection{Two Einstein fixed points}

\begin{figure}[h!]
    \centering
    \includegraphics[width=\linewidth]{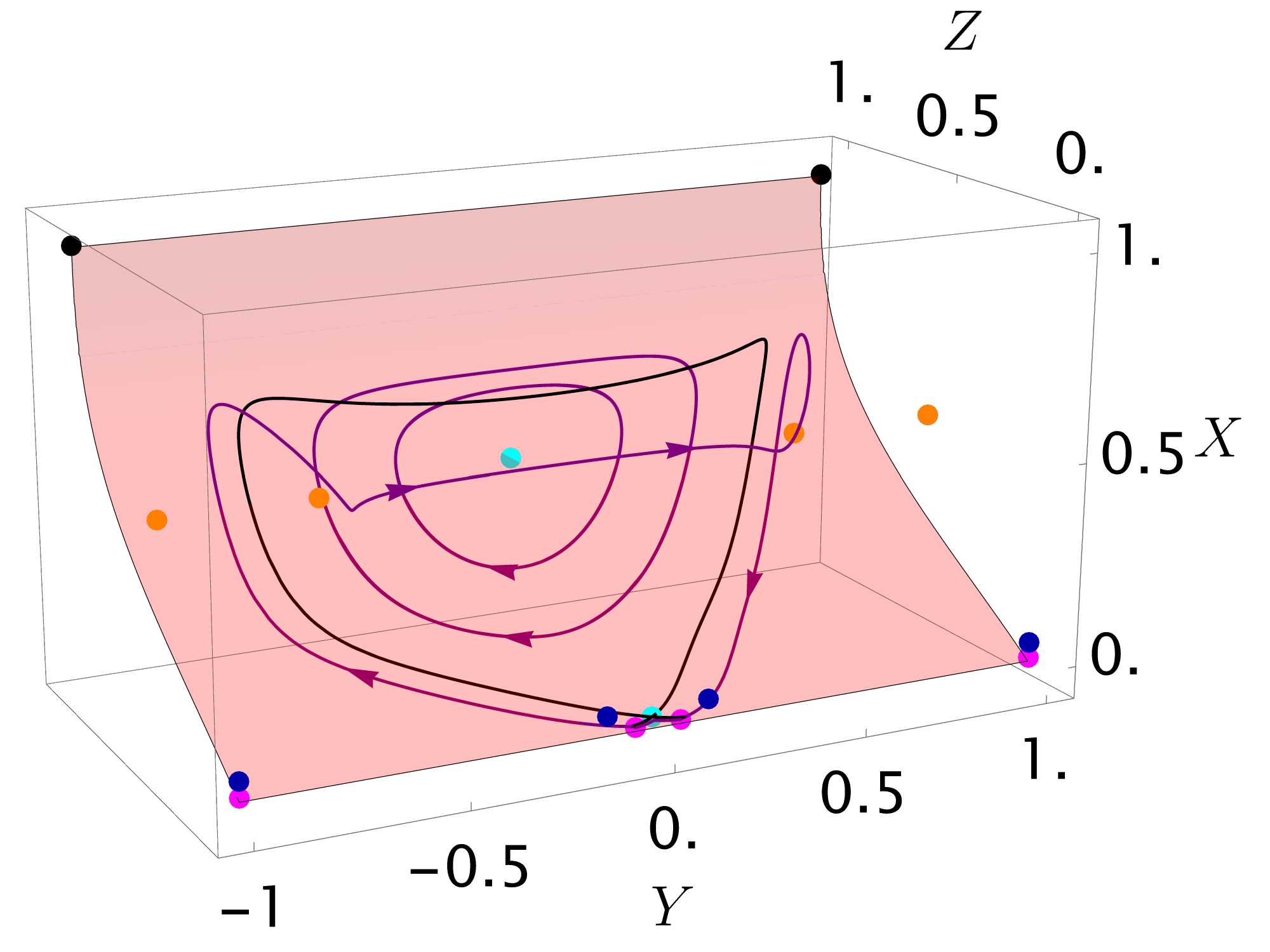}
    \caption{$X$-$Y$-$Z$ phase space corresponding to Fig. \ref{fig:qGT3_ZX_SR_Cross}, where $w_x=-0.95$ and $q=3.1$. The trajectories themselves are plotted on a first integral surface where $X_0 = 0.02$ and $Z_0 = 0.2$, which corresponds to a trajectory in Fig. \ref{fig:qGT3_ZX_SR_Cross} that intersects the red zero acceleration curve twice. In this figure, only the trajectories with positive spatial curvature are plotted. Two Einstein fixed points (cyan) exist at $Y = 0$ on the zero acceleration surface. Trajectories inside the CFS either bounce once between $dS_{2-}$ and $dS_{2+}$ and always accelerate, or are cyclic around an Einstein fixed point and cross the zero acceleration surface once during expansion, meaning they have a decelerated phase but no late-time acceleration. Trajectories outside the CFS bounce once, contracting from $dS_{2-}$ and expanding to $dS_{2+}$. During expansion, these trajectories cross the zero acceleration surface twice, meaning they have a decelerated phase and a late-time acceleration.}
    \label{fig:SR_qGT3_XYZ_2EPts_Closed}
\end{figure}

\begin{figure}[h!]
    \centering
    \includegraphics[width=\linewidth]{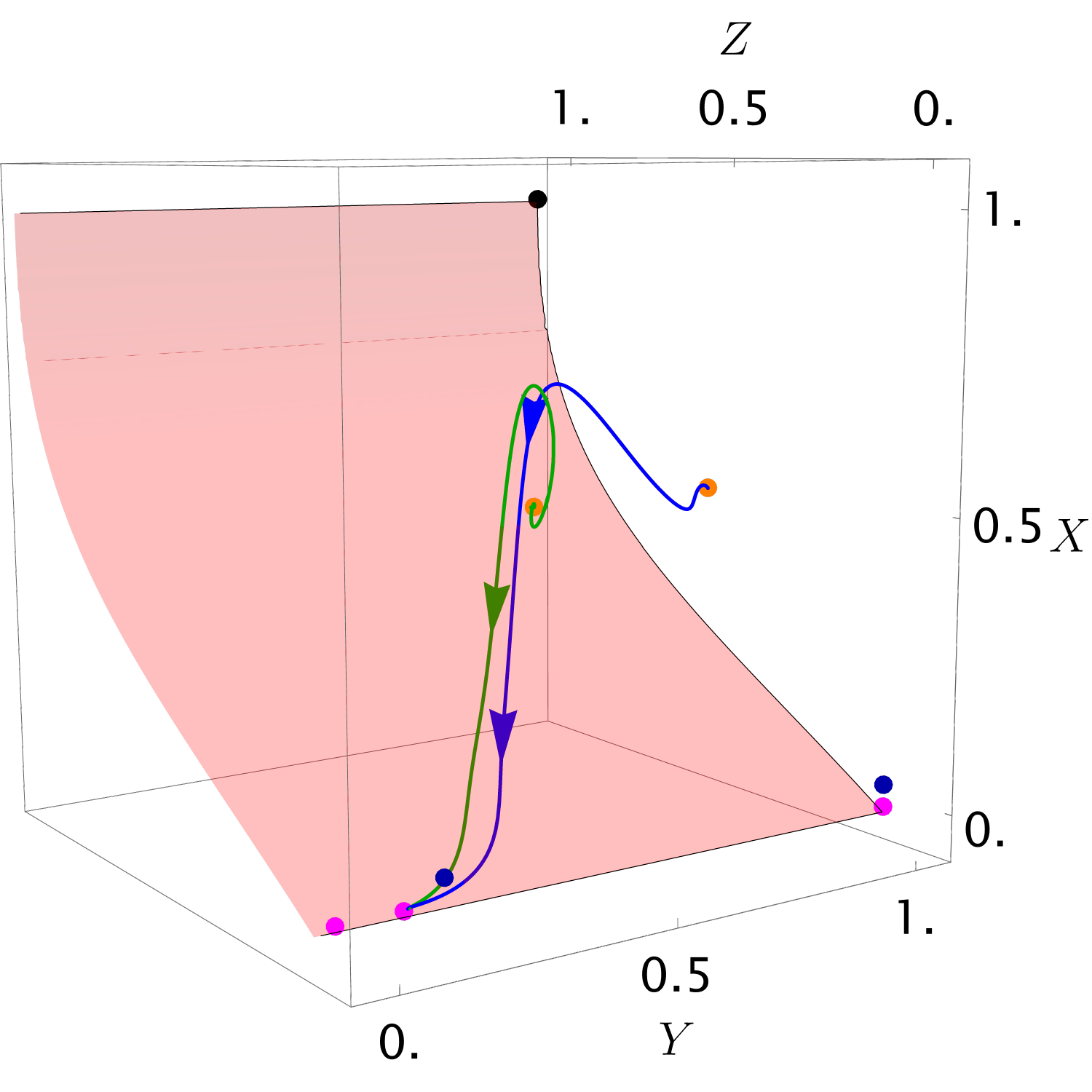}
    \caption{Expanding flat (green) and open (blue) trajectories in the $X$-$Y$-$Z$ phase space corresponding to Fig. \ref{fig:qGT3_ZX_SR_Cross}, where $w_x=-0.95$ and and $q=3.1$. The trajectories themselves are plotted on a first integral surface where $X_0 = 0.02$ and $Z_0 = 0.2$, which corresponds to a trajectory in Fig. \ref{fig:qGT3_ZX_SR_Cross} that intersects the red zero acceleration curve twice. The flat and open trajectories emerge from $dS_{1+}$ and $dS_{4+}$, respectively, and cross the zero acceleration surface twice, meaning they initially accelerate, then have a period of deceleration, and then have a final accelerated phase as they expand toward $dS_{2+}$ at low energy.}
    \label{fig:SR_qGT3_XYZ_2EPts_FlatOpen}
\end{figure}

Figures \ref{fig:SR_qGT3_XYZ_2EPts_Closed} and \ref{fig:SR_qGT3_XYZ_2EPts_FlatOpen} show a first integral surface in the $X$-$Y$-$Z$ phase space with two Einstein fixed points, which corresponds to a trajectory in Fig. \ref{fig:qGT3_ZX_SR_Cross} that intersects the $a'' = 0$ curve twice. One Einstein point is a center, and the other is a saddle which the CFS passes through. Figure \ref{fig:SR_qGT3_XYZ_2EPts_Closed} shows only the trajectories with positive spatial curvature, which all have a bounce. Trajectories that evolve outside of the CFS, and a subset that evolve inside the CFS, bounce once and contract from $dS_{2-}$ and expand to $dS_{2+}$. There is also a subset of trajectories that are cyclic around the center Einstein fixed point. These repeatedly expand, turnaround, contract and bounce. The trajectories inside the CFS are not of interest, as the bouncing models always accelerate, and the cyclic models have an initial accelerated expansion followed by a decelerated period, but they turn around and contract during the decelerated period so there is no late-time accelerated expansion. As for the $q < 3$ case, the trajectories that bounce once outside of the CFS are the models of interest. They have an accelerated era in which they evolve toward a high energy quasi-de Sitter phase when they bounce, and then become flatter as they start to expand; they then have a decelerated expanding phase followed by a final period of acceleration as they evolve toward the low energy cosmological constant represented by $dS_{2+}$.


Figure \ref{fig:SR_qGT3_XYZ_2EPts_FlatOpen} shows the expanding flat and open trajectories which are qualitatively interesting. These trajectories emerge from the de Sitter fixed points $dS_{1+}$ and $dS_{4+}$, respectively, and  cross the zero acceleration curve twice, so have a decelerated period followed by a final late-time acceleration as they asymptotically approach $dS_{2+}$.

\subsubsection{Three Einstein fixed points}

\begin{figure}[h!]
    \centering
    \includegraphics[width=\linewidth]{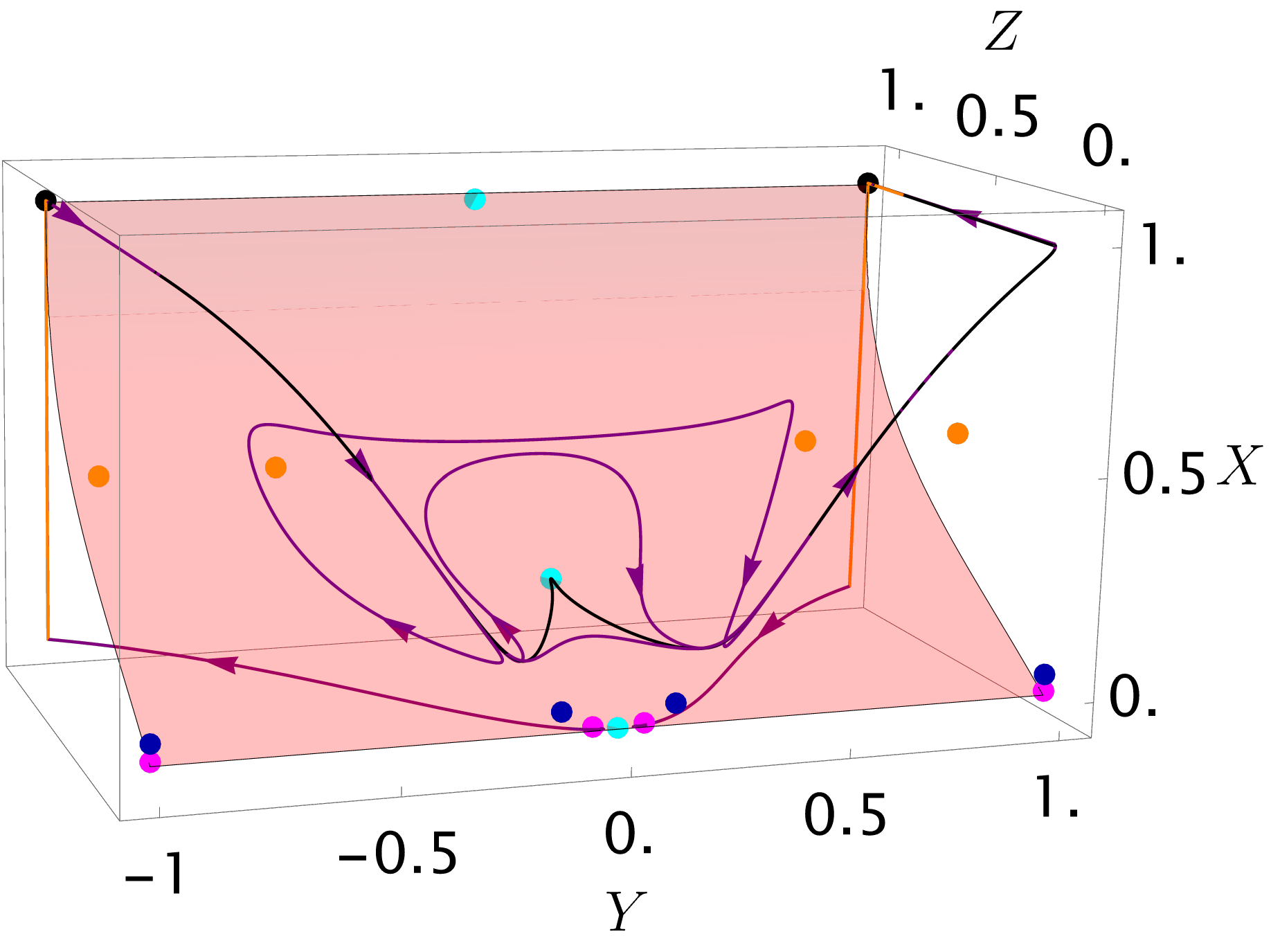}
    \caption{$X$-$Y$-$Z$ phase space corresponding to Fig. \ref{fig:qGT3_ZX_SR_Cross}, where $w_x=-0.95$ and $q=3.1$. The trajectories themselves are plotted on a first integral surface where $X_0 = 0.145$ and $Z_0 = 0.1$, which corresponds to a trajectory in Fig. \ref{fig:qGT3_ZX_SR_Cross} that first touches the red zero acceleration curve, and then intersects it twice. In this figure, only the trajectories with positive spatial curvature are plotted. Three Einstein fixed points (cyan) exist at $Y = 0$ on the zero acceleration surface. All trajectories bounce once, contracting from $dS_{2-}$ and expanding to $dS_{2+}$, and all cross the red zero acceleration surface twice, so have a decelerated phase followed by a late-time acceleration.}
    \label{fig:SR_qGT3_XYZ_3EPts_Closed}
\end{figure}

\begin{figure}[h!]
    \centering
    \includegraphics[width=\linewidth]{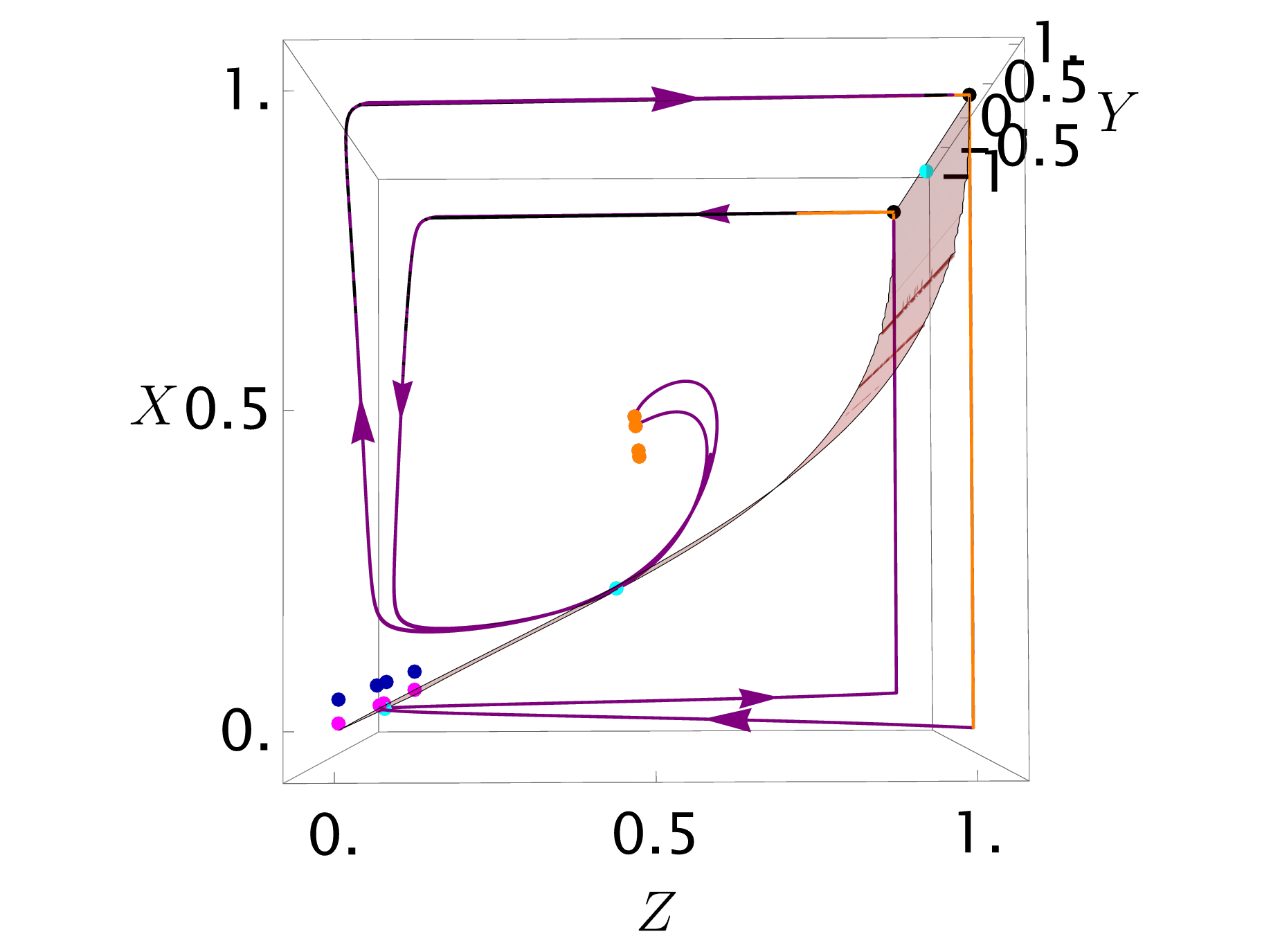}
    \caption{Side view of Fig. \ref{fig:SR_qGT3_XYZ_3EPts_Closed}. All trajectories intersect the red zero acceleration surface twice during expansion, accelerating when they are above the surface and decelerating when they are below it.}
    \label{fig:SR_qGT3_XYZ_3EPts_Closed_Side}
\end{figure}

\begin{figure}[h!]
    \centering
    \includegraphics[width=\linewidth]{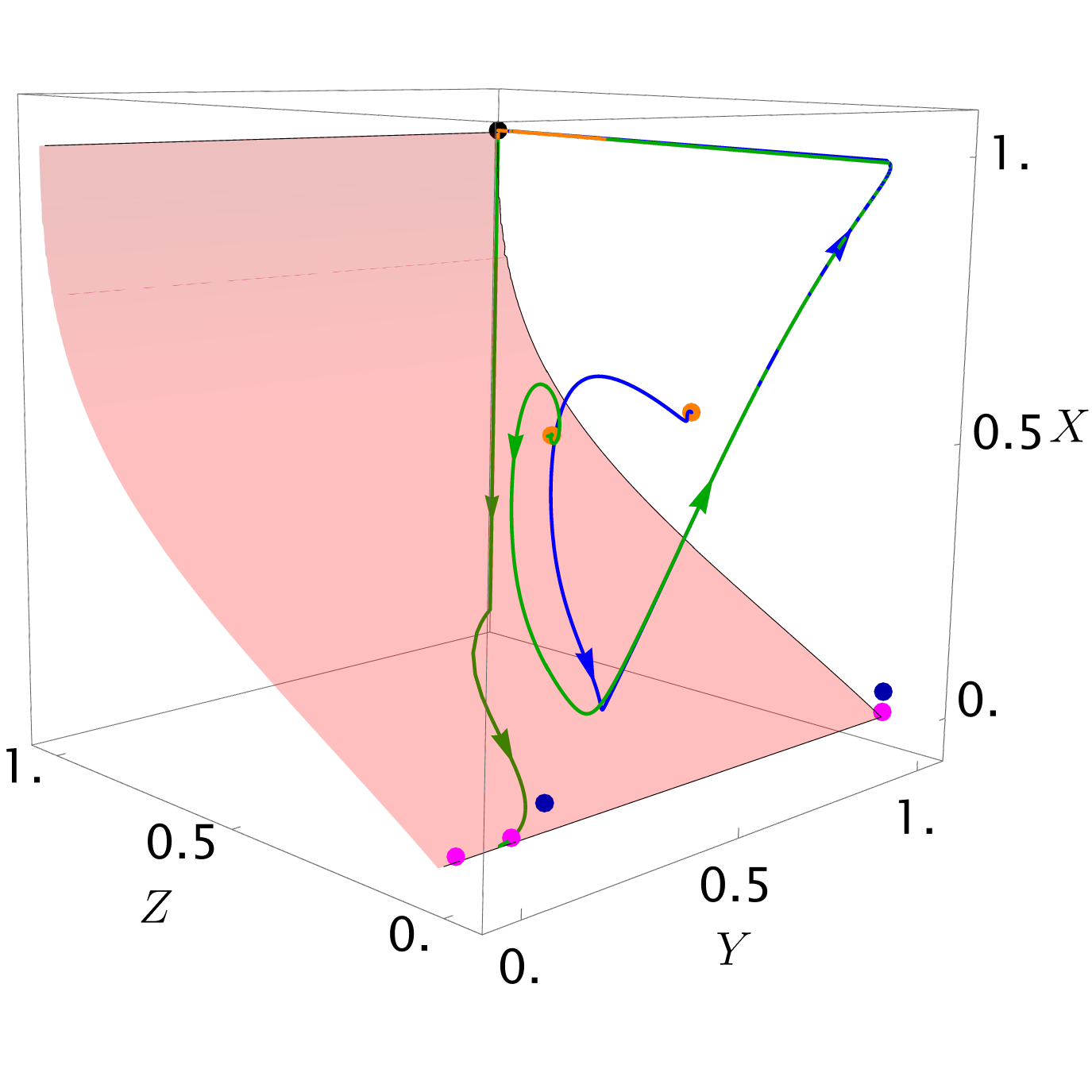}
    \caption{Expanding flat (green) and open (blue) trajectories in the $X$-$Y$-$Z$ phase space corresponding to Fig. \ref{fig:qGT3_ZX_SR_Cross}, where $w_x=-0.95$ and $q=3.1$. The trajectories themselves are plotted on a first integral surface where $X_0 = 0.145$ and $Z_0 = 0.1$, which corresponds to a trajectory in Fig. \ref{fig:qGT3_ZX_SR_Cross} that first touches the red zero acceleration curve, and then intersects it twice. The flat and open trajectories emerge from $dS_{1+}$ and $dS_{4+}$, respectively, and then touch the red zero acceleration surface before crossing it twice, eventually expanding toward $dS_{2+}$ (magenta) at low energy.}
    \label{fig:SR_qGT3_XYZ_3EPts_FlatOpen}
\end{figure}

Figures \ref{fig:SR_qGT3_XYZ_3EPts_Closed} -- \ref{fig:SR_qGT3_XYZ_3EPts_FlatOpen} show a first integral surface in three dimensions with three Einstein fixed points, which corresponds to the trajectory in Fig. \ref{fig:qGT3_ZX_SR_Cross} which touches the $a'' = 0$ curve, and then intersects it twice. Figure \ref{fig:SR_qGT3_XYZ_3EPts_Closed} shows the trajectories with positive spatial curvature. A CFS exists through the cusp Einstein fixed point in the middle, which is plotted in the figure. The trajectories above and below this CFS all contract from $dS_{2-}$, then accelerate through a bounce and become flatter as they start to expand; the trajectories above the CFS bounce during a quasi-de Sitter phase. The bouncing models then have a decelerated expanding period followed by a late-time acceleration as they evolve towards $dS_{2+}$. We expect another CFS to exist through the Einstein fixed point close to $X = 0$, which is a saddle, and loop around the Einstein fixed point close to $X = 1$, which is a center (similar to the CFS in Fig. \ref{fig:SR_qGT3_XYZ_2EPts_Closed}). We have not plotted this separatrix due to the numerics breaking down at high energy; however, we do not expect trajectories within this CFS to be qualitatively interesting; we expect there are bouncing trajectories that always accelerate, and cyclic trajectories which have no late-time acceleration. Figure \ref{fig:SR_qGT3_XYZ_3EPts_Closed_Side} shows the side view of the three-dimensional phase space through the $Z$-$X$ plane, which more clearly shows the corresponding trajectory in Fig. \ref{fig:qGT3_ZX_SR_Cross}.

Figure \ref{fig:SR_qGT3_XYZ_3EPts_FlatOpen} shows the expanding flat and open trajectories, which emerge from the de Sitter fixed points $dS_{1+}$ and $dS_{4+}$, respectively, and asymptotically approach $dS_{2+}$ at late times. Both flat and open models intersect the $a'' = 0$ curve twice, so have a decelerated period followed by a final late-time acceleration.

\subsubsection{Four Einstein fixed points}

\begin{figure}[h!]
    \centering
    \includegraphics[width=\linewidth]{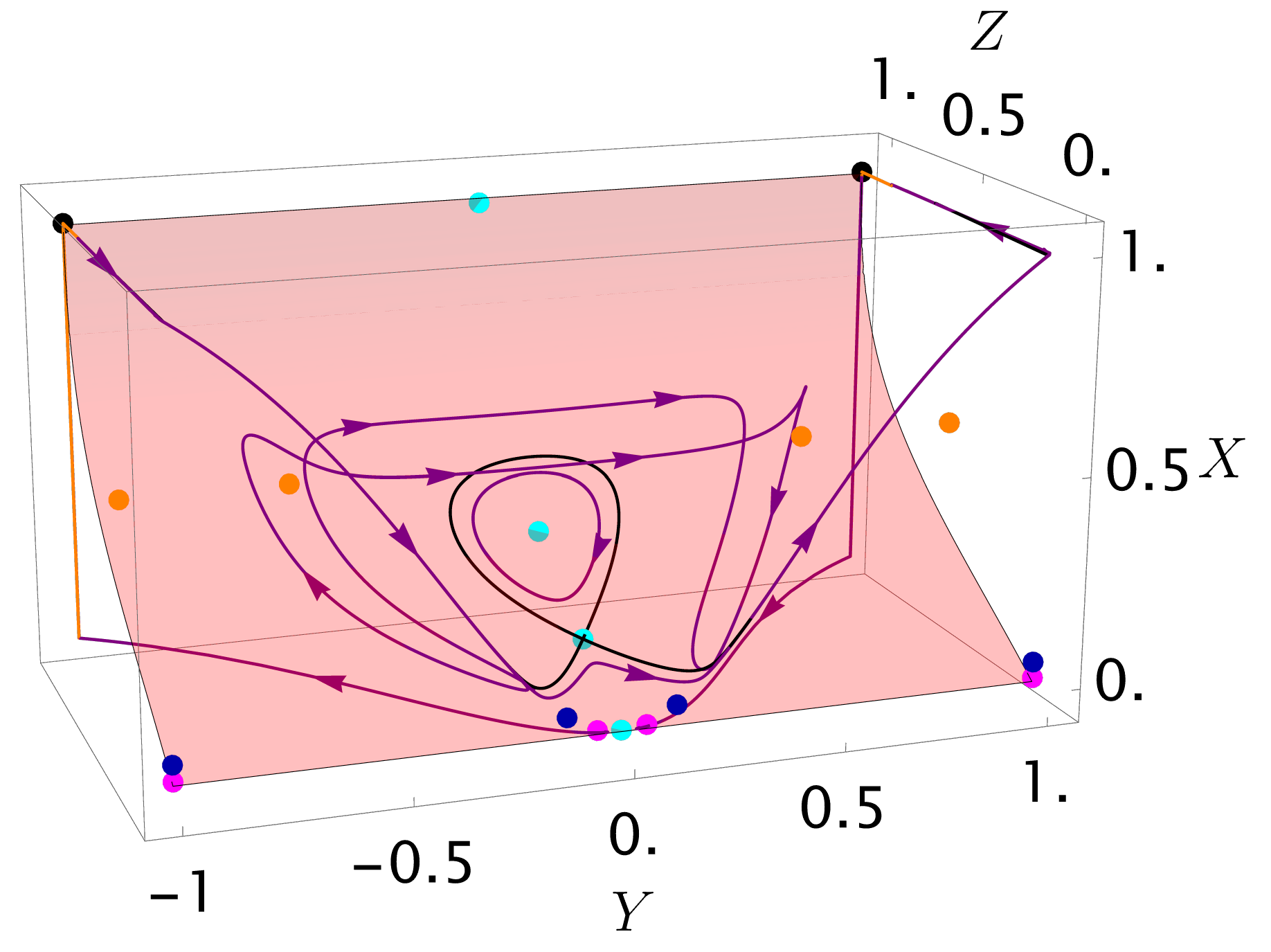}
    \caption{$X$-$Y$-$Z$ phase space corresponding to Fig. \ref{fig:qGT3_ZX_SR_Cross}, where $w_x=-0.95$ and $q=3.1$. The trajectories themselves are plotted on a first integral surface where $X_0 = 0.14$ and $Z_0 = 0.3$, which corresponds to a trajectory in Fig. \ref{fig:qGT3_ZX_SR_Cross} that intersects the red zero acceleration curve four times. In this figure, only the trajectories with positive spatial curvature are plotted. Four Einstein fixed points (cyan) exist at $Y = 0$ on the zero acceleration surface. The only trajectories that are not viable are the cyclic trajectories within the CFS, as during expansion there is no late-time acceleration. All other trajectories bounce once and contract from $dS_{2-}$, then expand towards $dS_{2+}$. Those within the CFS have one decelerated period followed by late-time acceleration, and those outside have two decelerated phases with a late-time acceleration.}
    \label{fig:qGT3_XYZ_4EPts_Closed}
\end{figure}

\begin{figure}[h!]
    \centering
    \includegraphics[width=\linewidth]{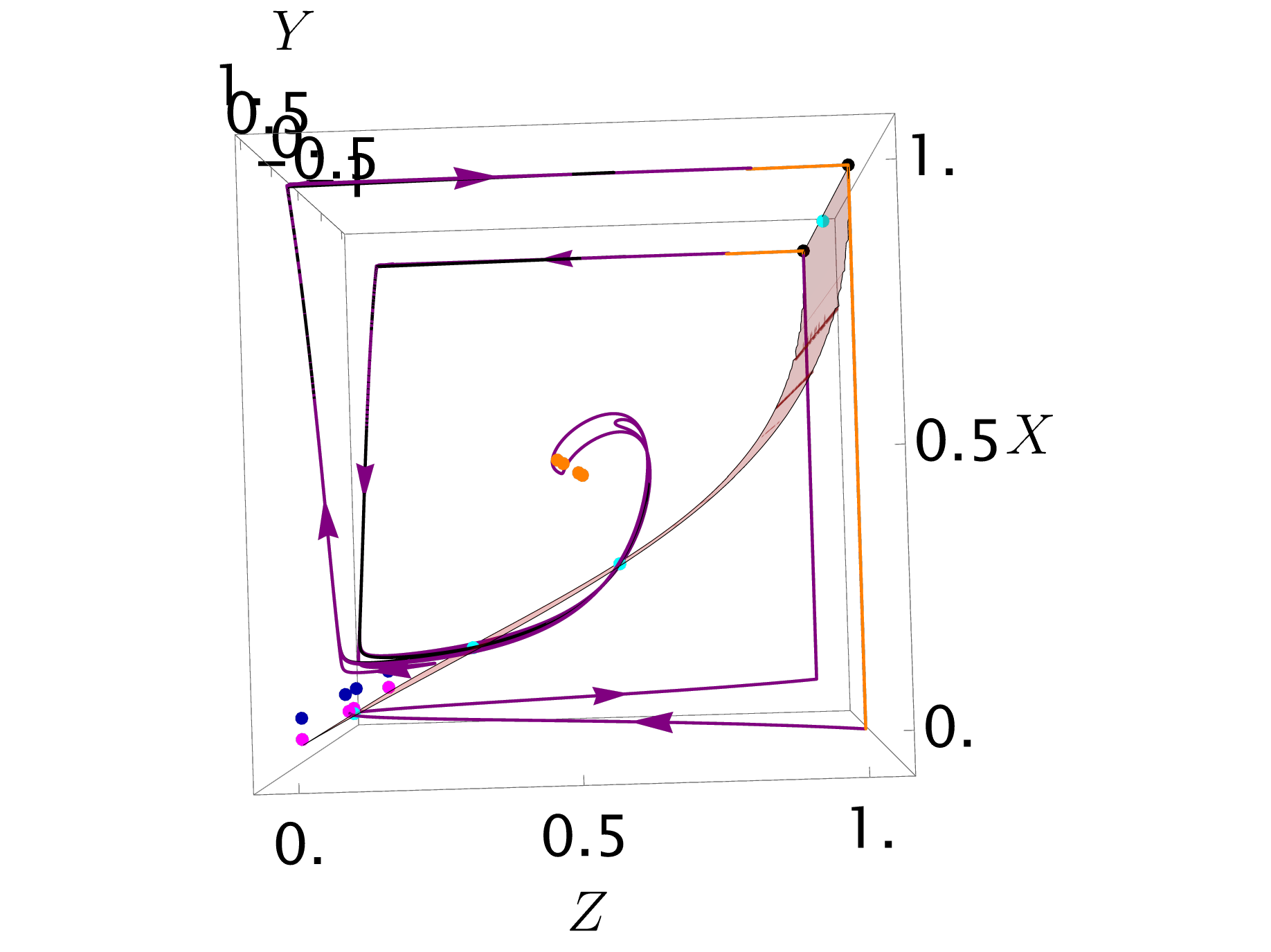}
    \caption{Side view of Fig. \ref{fig:qGT3_XYZ_4EPts_Closed}. The cyclic trajectories cross the red zero acceleration surface once during expansion. All other trajectories intersect this surface two or four times during expansion, accelerating when they are above the surface and decelerating when they are below it.}
    \label{fig:qGT3_XYZ_4EPts_Closed_Side}
\end{figure}

\begin{figure}[h!]
    \includegraphics[width=\linewidth]{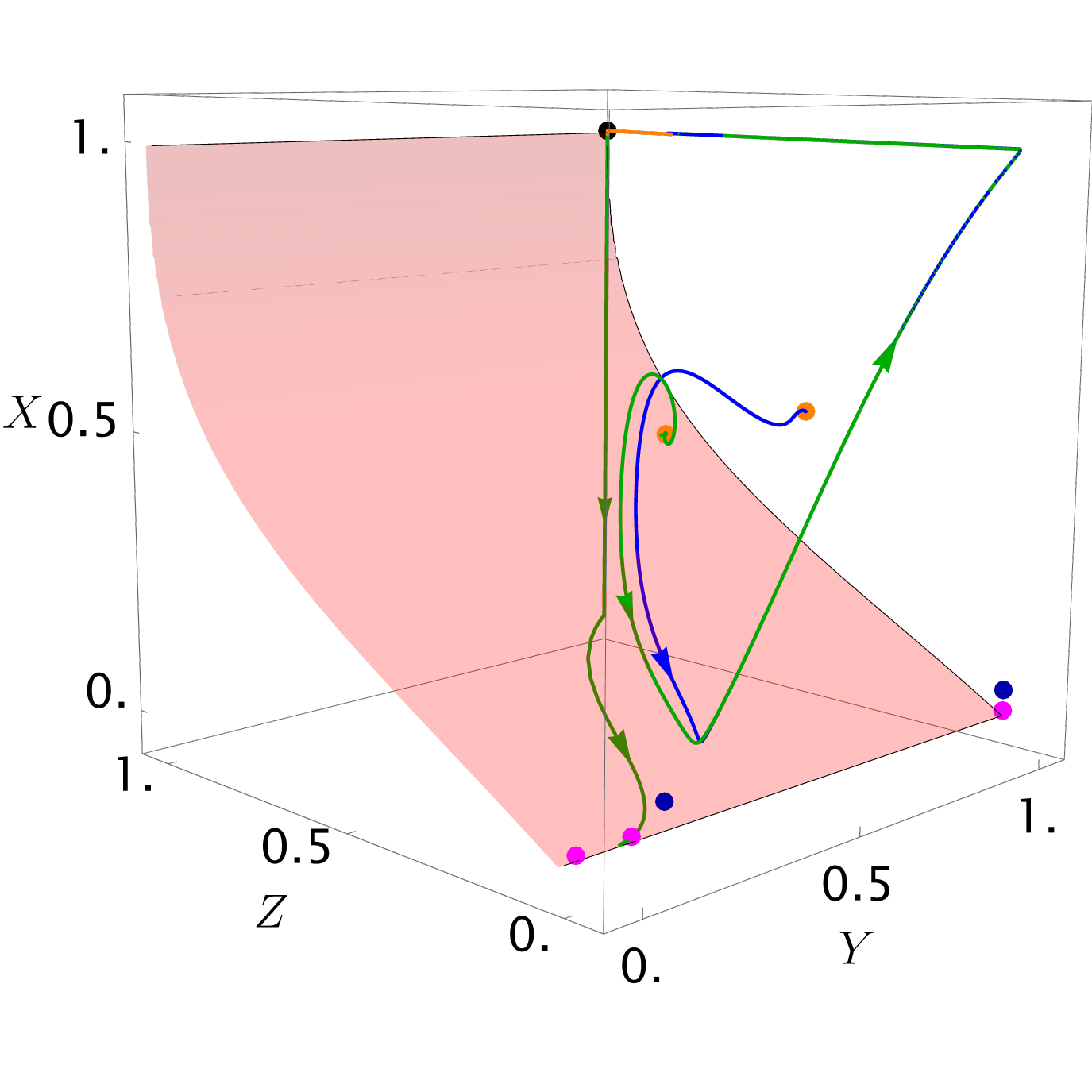}
    \caption{Expanding flat (green) and open (blue) trajectories in the $X$-$Y$-$Z$ phase space corresponding to Fig. \ref{fig:qGT3_ZX_SR_Cross}, where $w_x=-0.95$ and $q=3.1$. The trajectories themselves are plotted on a first integral surface where $X_0 = 0.14$ and $Z_0 = 0.3$, which corresponds to a trajectory in Fig. \ref{fig:qGT3_ZX_SR_Cross} which intersects the red zero acceleration curve four times. The flat and open trajectories emerge from $dS_{1+}$ and $dS_{4+}$, respectively, and cross the red zero acceleration surface four times, eventually expanding toward $dS_{2+}$ at low energy.}
    \label{fig:qGT3_XYZ_4EPts_FlatOpen}
\end{figure}

Figures \ref{fig:qGT3_XYZ_4EPts_Closed} -- \ref{fig:qGT3_XYZ_4EPts_FlatOpen} show a first integral surface in the three-dimensional phase space, corresponding to a trajectory in Fig. \ref{fig:qGT3_ZX_SR_Cross} which intersects the $a'' = 0$ curve four times. Figure \ref{fig:qGT3_XYZ_4EPts_Closed} shows trajectories with positive spatial curvature, which all have a bounce. One CFS is plotted, which passes through one of the Einstein fixed points, which is a saddle, and loops round the Einstein fixed point above it, which is a center. A subset of trajectories within this CFS are cyclic, and have an initial accelerated expansion followed by a decelerated period, but no late-time acceleration.
Bouncing trajectories within this CFS have one decelerated period, and those outside the CFS have two decelerated periods, both with a late-time acceleration. These models all contract from $dS_{2-}$, then accelerate through a bounce and become flatter as they start to expand, with those outside the CFS bouncing during a quasi-de Sitter phase; they then have one or two decelerated expanding phases, followed by a late-time acceleration as they evolve to $dS_{2+}$, therefore these bouncing models can all qualitatively match the observed Universe. The Einstein fixed point close to $X = 1$ is a center, and the Einstein fixed point close to $X=0$ is a saddle. We expect this saddle fixed point to form part of another CFS, which would loop around the center Einstein fixed point close to $X=1$; however, we do not plot this separatrix as the numerics breaks down at high energy. We do not expect trajectories within this CFS to be qualitatively interesting; we expect bouncing trajectories to always accelerate, and cyclic trajectories to have no late-time acceleration. Figure \ref{fig:qGT3_XYZ_4EPts_Closed_Side} shows the view of the three-dimensional phase space through the $Z$-$X$ plane, where it is clearer which trajectory this surface corresponds to in the two-dimensional phase space in Fig. \ref{fig:qGT3_ZX_SR_Cross}. 

Figure \ref{fig:qGT3_XYZ_4EPts_FlatOpen} shows the expanding flat and open trajectories. These emerge from $dS_{1+}$ and $dS_{4+}$, respectively, and intersect the $a'' = 0$ surface four times. This means flat and open models have two periods of deceleration and three accelerated phases, including a late-time acceleration as they expand towards $dS_{2+}$.

\section{Conclusions} \label{sec:Conclusions}

In this paper we have studied the dynamics of FLRW models with standard pressureless dark matter interacting with dark energy with a quadratic EoS \eqref{eqn:DE_EoS_interacting}. This EoS introduces two energy scales: $\rho_\Lambda$ turns out to be an effective cosmological constant, which plays the role of the asymptotic state of all models evolving toward low energies, and $\rho_*> \rho_\Lambda$ characterizes the quadratic term. The interaction is also quadratic and characterized by an energy scale $\rho_i$, and defined so that energy flows from the dark energy to the dark matter. Given this form of interaction, the dark matter energy density $\rho_m$ is always positive, while the dark energy density $\rho_x$ remains positive if it was always positive in the past. The quadratic terms are only relevant at very high energies, such that singularities can be avoided. 

This work is an extension of Paper I \cite{Burkmar2023}, where we studied the dynamics of FLRW models containing dark matter, radiation and dark energy with a quadratic EoS, without any interaction term. In Paper I \cite{Burkmar2023}, we found nonsingular models that had an expanding decelerated phase and a late-time acceleration; however, the decelerated period was lost when the effective cosmological constants were set to realistic energy scales, i.e. when the effective low energy cosmological constant $\rho_\Lambda$ is close to the observed value of $\Lambda$, and when the characteristic energy scale of the dark energy $\rho_*$ is between Planck and inflationary scales. Here, our aim was twofold. First, to see if including an interaction between the dark components allows for nonsingular models, and second, to see if these nonsingular models could evolve over realistic energy scales, with a decelerated period and late-time accelerated expansion. 

In Sec. \ref{sec:Equations} we presented the system of equations both in two dimensions, where we take $H > 0$ and have the equations for dark energy and dark matter, and in three dimensions where we include the equation for the Hubble function. We also defined dimensionless variables. In Sec. \ref{sec:param_space} we explored the parameter space in order to obtain a phase space topology in two dimensions with a nonsingular high energy repellor that trajectories can expand from, and a late-time attractor such that trajectories asymptotically tend towards a cosmological constant at late times. We also defined compact variables to be able to study the dynamics at infinity. We found a high energy repellor and low energy attractor when $\epsilon = -1$, $0 < \scriptR < 1$, $w_x > -1 +\scriptR$ and $q=\rho_*/\rho_i < 3/\scriptR$. We then set $\epsilon = -1$ and $\scriptR = 0.01$, such that the system only depended on two dimensionless parameters, $w_x$ and $q$. In reality, we would expect $10^{-120} < \scriptR < 10^{-111}$ if $\rho_\Lambda$ is close to the current observed dark energy density \cite{Planck2018VI,Prat2022} and $\rho_*$ is between inflationary and Planck scales \cite{Bass2015}. However, the dynamics are unaffected by setting $\scriptR$ to a larger value, and this helps to have readable phase space plots. As $\scriptR \rightarrow 0$, the attractor $dS_{2+}$ representing the low energy cosmological constant just moves closer to the origin.

In general, two cases for the topology exist: one when $q < 3$, and the other when $q > 3$. When $q < 3$, two separatrices in the two-dimensional phase space exist, which we present in Sec. \ref{sec:qLT3}. Some trajectories expand from the high energy repellor and tend toward the late-time low energy attractor  $dS_{2+}$; however, some trajectories expand from the high energy repellor and evolve toward a critical point representing a singularity at late times. We have not explicitly analyzed this; however, we expect it to be a big-rip singularity as $\rho \rightarrow \infty$ and $H \rightarrow \infty$, which is caused by the phantom behavior. There are qualitatively interesting expanding models when $q < 3$, as there are trajectories that have a decelerated expansion phase followed by a late-time acceleration, asymptotic to the late-time cosmological constant represented by  $dS_{2+}$. We then presented the corresponding three-dimensional dynamics for these qualitatively interesting trajectories. In this case, expanding open and flat models emerge from the high energy repellor, a fixed point in phase space representing a de Sitter model, and evolve toward the low energy cosmological constant represented by  $dS_{2+}$. These all have a decelerated phase followed by a late-time acceleration. 

Positively curved models are either cyclic, which repeatedly contract and expand through bounces and turnarounds, or they bounce once, contracting from a low energy de Sitter state and expanding toward $dS_{2+}$. The positively curved models of qualitative interest are the subset which go through a quasi-de Sitter bounce at high energy, as they have a decelerated expansion phase followed by a late-time acceleration as they evolve towards $dS_{2+}$.

In Sec. \ref{sec:qGT3}, we show the phase spaces for $q > 3$. In this case, only one separatrix between the high energy repellor and low energy attractor exists in the two-dimensional phase spaces, and therefore all trajectories avoid a singularity. For certain parameter values, there is a case where the separatrix and zero acceleration curve intersect twice. In this case, all trajectories in the two-dimensional phase space have a decelerated matter dominated phase during expansion, followed by a late-time acceleration dominated by dark energy, eventually evolving toward the cosmological constant represented by $dS_{2+}$. Therefore in this subcase, all trajectories represent models which qualitatively correspond to the observed Universe, for which we then presented the three-dimensional dynamics. The expanding open and flat models are qualitatively interesting, emerging from the high energy de Sitter fixed point and expanding to the late-time cosmological constant represented by $dS_{2+}$, with either one or two decelerated phases and a late-time accelerated expansion. Some positively curved models are cyclic, and some are asymptotic in the past to a contracting low energy de Sitter state, have a bounce, then expand, asymptotically evolving toward the low energy de Sitter fixed point $dS_{2+}$. The bouncing trajectories which are qualitatively interesting have one or two decelerated phases during expansion followed by a late-time acceleration, some with a quasi-de Sitter transition phase at high energy.

Overall, our analysis shows that for our system of interacting dark energy and dark matter, a parameter range exists such that all trajectories represent singularity-free cosmological models, and are qualitatively consistent with our observed universe. In particular, trajectories expand from a high energy nonsingular fixed point, have a decelerated expansion phase where large scale structure could form, and have a late-time acceleration where trajectories asymptotically tend toward a cosmological constant. 

In future work, we will include perturbations and test this model against recent observational data. To test the model at high energy, we would need to constrain it with the current bounds on primordial gravitational waves. Models with a linear interaction term have been constrained with supernovae data \cite{Quercellini2008}, and if we linearize the equations at low energy, then we obtain a model with affine EoS which has been tested against, and is consistent with, observational data from WMAP, SDSS and the Hubble Space Telescope \cite{Balbi2007,Pietrobon2008}; these studies found that models with affine EoS $P = P_0 + \alpha\rho$ fit the data just as well as the $\Lambda$CDM model if flatness is not assumed. We therefore expect our model to fit this observational data at least as well as the $\Lambda$CDM model; however, it is beyond the scope of this paper to perform such tests. We reiterate that the aim of this paper is to understand whether singularities can be avoided in an FLRW cosmology, and we therefore do not necessarily seek to improve a phenomenological fit compared to the $\Lambda$CDM model. This is a qualitative analysis, and one may tweak the EoS of the dark energy and the interaction term without necessarily changing the dynamics; however, this may have implications for the fit to data. Naturally, the next step will be to test the specific model in this paper at low energy with more recent cosmic microwave background (CMB), baryon acoustic oscillations (BAO) and type Ia supernova (SNIa) observations. In particular, the recent DESI results favor a dynamical dark energy over a cosmological constant at late times \cite{DESIDR2II}, and we are eager to investigate whether suitable modifications of the model presented here can be made to accommodate the DESI data at low energy, while preserving the desirable qualitative features we found in this paper. It will also be interesting to quantitatively determine whether the decelerated periods are long enough for large scale structure to form, and to include radiation in the setup to analyze if there is any effect on the overall dynamical behavior; however, we do not expect it to affect the dynamics if it is produced through a reheating phase, as we found in Paper I \cite{Burkmar2023}.



\section*{ACKNOWLEDGEMENTS}

This work has been supported by UK STFC Grants No. ST/W507738/1 and No. ST/S000550/1.

\section*{DATA AVAILABILITY}

The data that support the findings of this article are openly available \cite{Planck2018I,BurkmarIntDEnb}.

\newpage

\bibliographystyle{apsrev4-2}
\bibliography{export}

\newpage

\appendix

\section{$q < 3$}

Here, we show examples of the phase spaces for $q < 3$, where $dS_{1+}$ is a repelling node and an improper node. We fix $\epsilon = -1$ and $\scriptR = 0.01$ as before, and change $w_x$ and $q$. The color scheme of the fixed points and critical points in the two-dimensional $Z$-$X$ phase spaces are given in Table \ref{tab:ZX_FPs_Stability_Colour}, and the color scheme of the curves are given in Table \ref{tab:2-D_features}. For the three-dimensional phase spaces, the color scheme of the fixed points and critical points are given in Table \ref{tab:Fixed_Points_XYZ_Colour}, and the color scheme of the curves and surfaces are given in Table \ref{tab:3-D_features}.

There are three subcases each for the phase spaces where $dS_{1+}$ is a repelling node and when it is an improper node, which are the same as the subcases for the spiral repellor in Sec. \ref{sec:qLT3}. In each case, the zero acceleration curve and the repellor-saddle separatrix between $dS_{1+}$ and $dS_{3+}$ can intersect twice, not intersect or touch depending on the specific value of $q$. In the following, we present the subcases where the repellor-saddle separatrix and zero acceleration curve intersect twice for both the repelling node and improper node case, as these are the subcases where all trajectories that evolve to the right of the separatrices are qualitatively interesting.

\subsection{Repelling node} \label{sec:qLT3_R}

\begin{figure}
    \includegraphics[width=\linewidth]{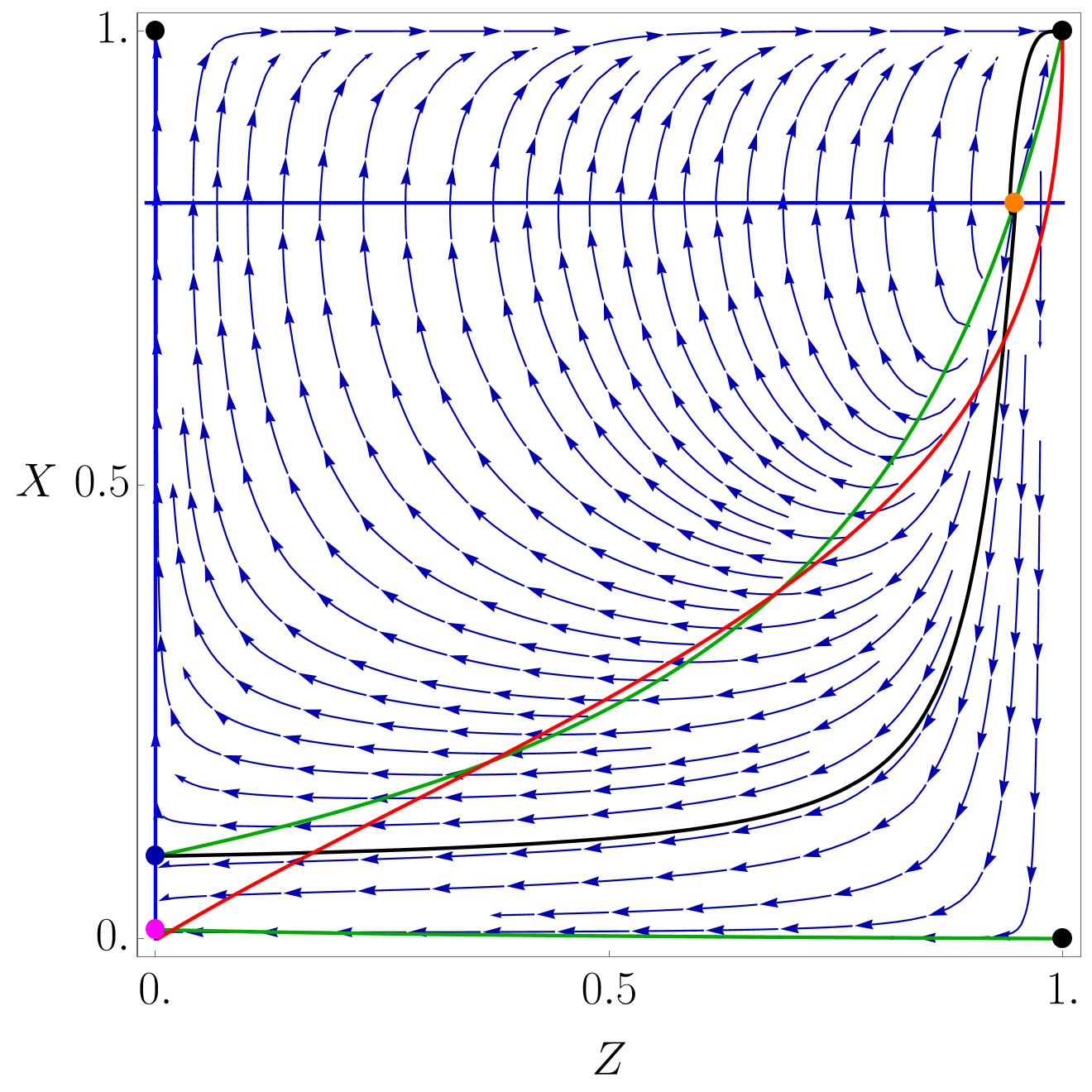}
    \caption{$Z$-$X$ phase space referred to in Appendix \ref{sec:qLT3_R}. Here, the parameters are set to $w_x=-0.9$ and $q=0.7$. In this case, $dS_{1+}$ is a repelling node and the zero acceleration curve intersects the repellor-saddle separatrix twice. Trajectories to the left of the separatrices expand to $S_{2+}$ which represents a singularity, some with and some without a decelerated period, and those to the right of the separatrices all have a decelerated period, then accelerate as they expand towards $dS_{2+}$. Therefore, the trajectories that evolve to the right of the separatrices are qualitatively of interest.}
    \label{fig:ZX_R_qLT3}
\end{figure}

Figure \ref{fig:ZX_R_qLT3} shows an example of the $Z$-$X$ phase space where $q < 3$ and $dS_{1+}$ is a repelling node. In this case, the repellor-saddle separatrix and the zero acceleration curve intersect twice. Trajectories that evolve to the left of the two separatrices expand from $dS_{1+}$ toward $S_{2+}$. These trajectories either always accelerate if they never touch the zero acceleration curve, or have a deceleration period if they intersect this curve twice. There is also a limiting case where a trajectory touches the red curve and has a point of zero acceleration, but otherwise accelerates. Trajectories that evolve to the right of the separatrices expand from $dS_{1+}$, and asymptotically tend towards a late-time cosmological constant as they approach $dS_{2+}$. The trajectories that evolve to the right of the separatrices are all of interest, as they all cross the red zero acceleration curve twice and therefore have a decelerated period, followed by a late-time accelerated expansion.

\begin{figure}
    \centering
    \includegraphics[width=\linewidth]{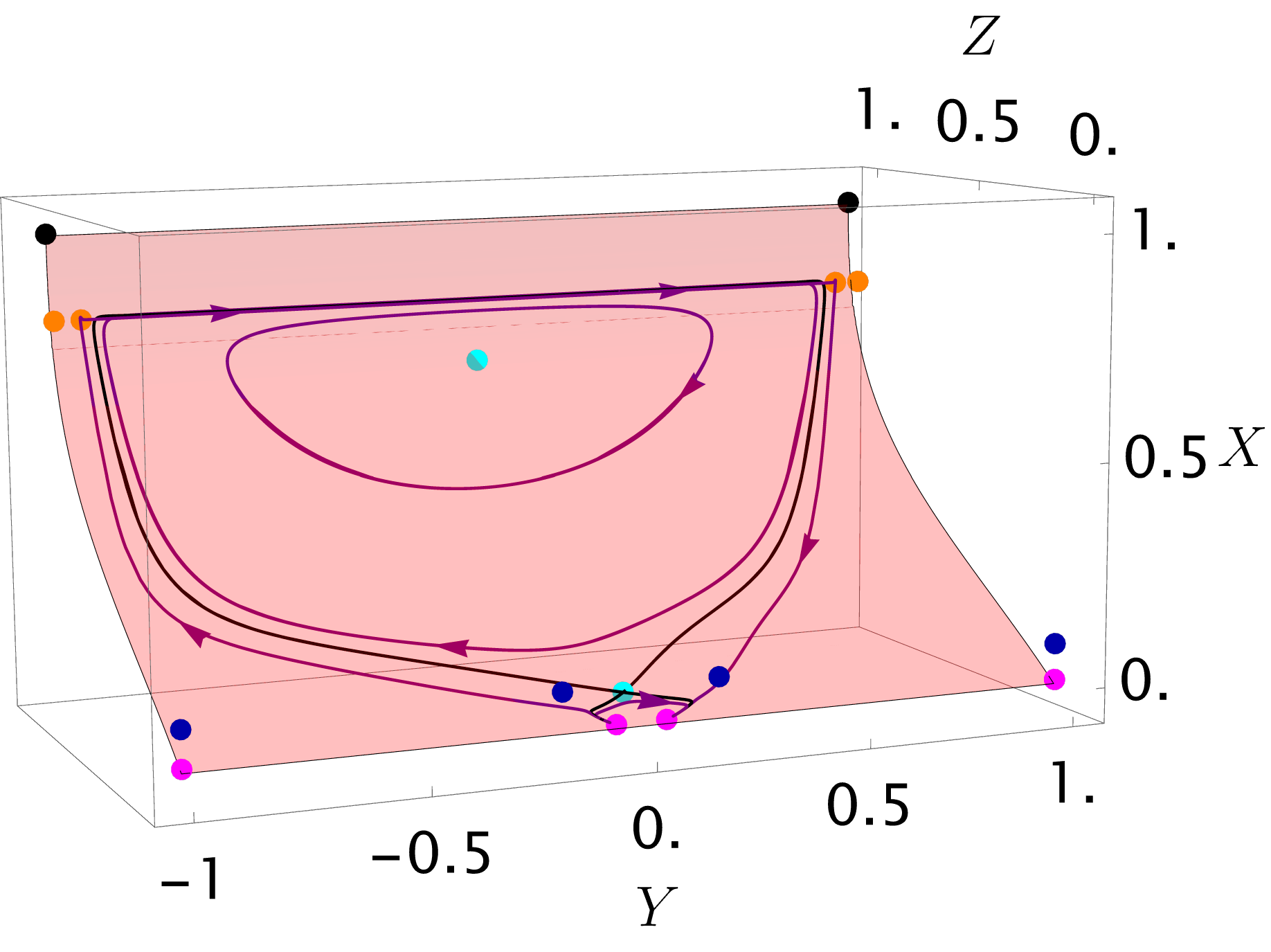}
    \caption{$X$-$Y$-$Z$ phase space corresponding to Fig. \ref{fig:ZX_R_qLT3}, where $w_x=-0.9$ and $q=0.7$. The trajectories are plotted on a first integral surface where $X_0 = 0.07$ and $Z_0 = 0.5$, which corresponds to a trajectory in Fig. \ref{fig:ZX_R_qLT3} that evolves to the right of the separatrices. In this figure, only the trajectories with positive spatial curvature are plotted. Two Einstein fixed points (cyan) exist at $Y = 0$ on the zero acceleration surface. Trajectories inside the CFS either bounce once, contracting from $dS_{2-}$ and expanding towards $dS_{2+}$, or are cyclic around an Einstein fixed point. The bouncing models inside the CFS always accelerate, and the cyclic models cross the zero acceleration surface once during expansion, meaning they have a decelerated phase but no late-time acceleration. Trajectories outside the CFS bounce once between $dS_{2-}$ and $dS_{2+}$, and during expansion intersect the zero acceleration surface twice, meaning they have a decelerated phase and a late-time acceleration.}
    \label{fig:XYZ_R_qLT3}
\end{figure}

\begin{figure}
    \centering
    \includegraphics[width=\linewidth]{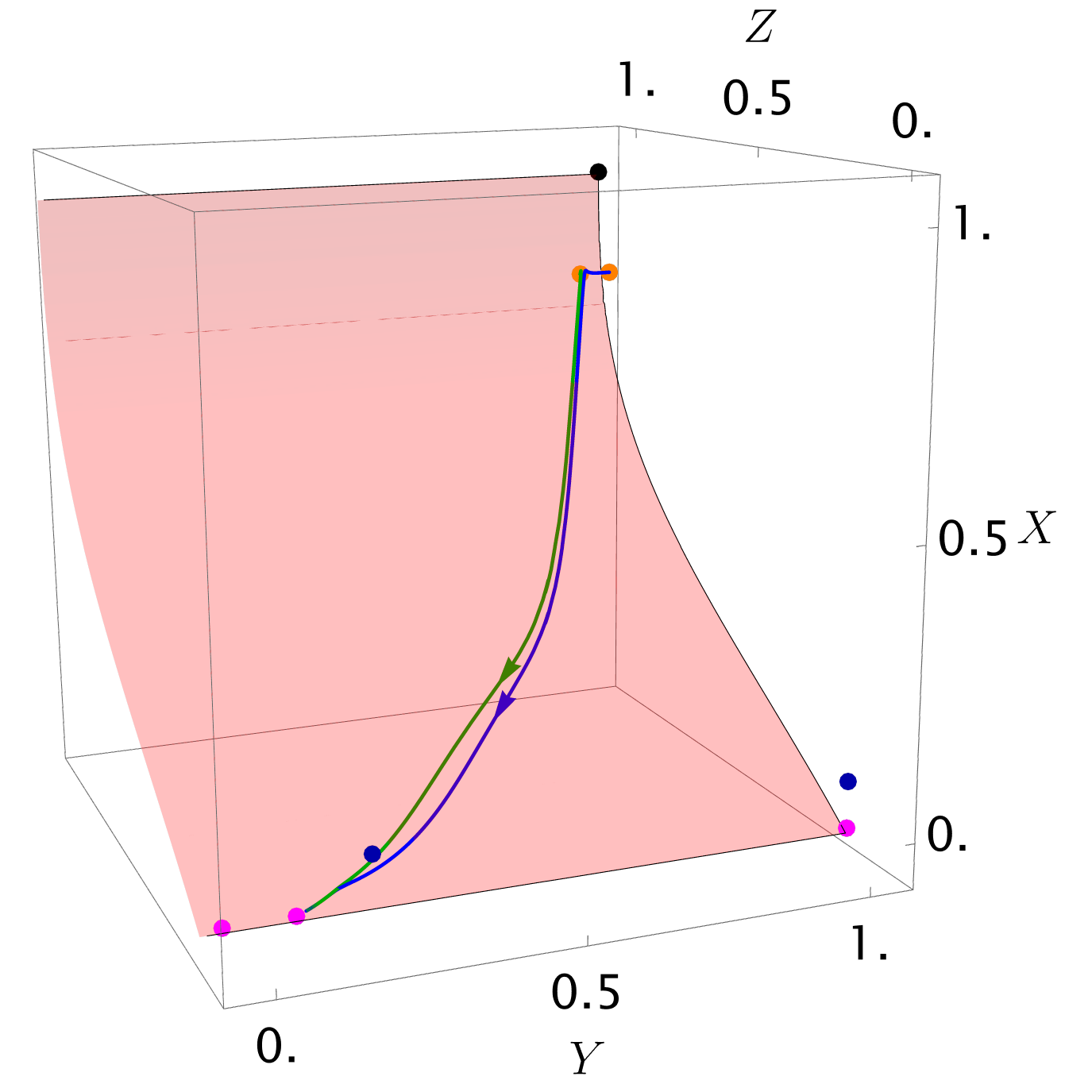}
    \caption{Expanding flat (green) and open (blue) trajectories in the $X$-$Y$-$Z$ phase space corresponding to Fig. \ref{fig:ZX_R_qLT3}, where $w_x=-0.9$ and $q=0.7$. The trajectories themselves are plotted on a first integral surface where $X_0 = 0.07$ and $Z_0 = 0.5$, which corresponds to a trajectory in Fig. \ref{fig:ZX_R_qLT3} that evolves to the right of the separatrices, expanding to $dS_{2+}$. Both the flat and open trajectories intersect the zero acceleration surface twice, meaning they initially accelerate as they expand from $dS_{1+}$ and $dS_{4+}$, respectively, then have a period of deceleration, and then have a final accelerated phase as they expand toward $dS_{2+}$ at low energy.}
    \label{fig:XYZ_R_qLT3_FlatOpen}
\end{figure}

Figures \ref{fig:XYZ_R_qLT3} and \ref{fig:XYZ_R_qLT3_FlatOpen} show trajectories on a first integral surface in the $X$-$Y$-$Z$ phase space generated by fixing initial conditions $X_0$ and $Z_0$, corresponding to a single trajectory in Fig. \ref{fig:ZX_R_qLT3} that evolves to the right of the separatrices and intersects the zero acceleration curve twice. Figure \ref{fig:XYZ_R_qLT3} shows the trajectories with positive spatial curvature, and Fig. \ref{fig:XYZ_R_qLT3_FlatOpen} shows expanding flat and open trajectories. As in the main part of the paper, the three-dimensional phase spaces in the appendixes all include a red zero acceleration surface, which corresponds to the zero acceleration curve in the two-dimensional phase spaces. The trajectories with positive spatial curvature in Fig. \ref{fig:XYZ_R_qLT3} all have a bounce. Two Einstein 
fixed points exist in the phase space; one has saddle stability which the CFS passes through, and the other is a center. Trajectories inside the CFS either bounce once between $dS_{2-}$ and $dS_{2+}$, or are cyclic around the center Einstein fixed point. These trajectories are not of interest, as the bouncing trajectories always accelerate, and the cyclic trajectories have a decelerated phase during expansion, but no late-time acceleration. The trajectories that evolve outside the CFS bounce once and contract from $dS_{2-}$, go through a quasi-de Sitter bounce, then expand towards $dS_{2+}$. These trajectories intersect the zero acceleration surface twice during expansion, meaning they accelerate though the bounce, becoming flatter as they start to expand, then have a decelerated phase where large scale structure could form, and then have a final late-time accelerated expansion as they approach a cosmological constant. These trajectories that evolve outside the CFS are therefore the models of interest.

The expanding open and flat trajectories in Fig. \ref{fig:XYZ_R_qLT3_FlatOpen} are all of interest. Flat and open trajectories emerge from $dS_{1+}$ and $dS_{4+}$, respectively, and tend toward $dS_{2+}$ at low energy. They also intersect the red zero acceleration surface twice, meaning they initially accelerate, then have a decelerated phase, and finally have a late-time acceleration as they asymptotically approach a cosmological constant.

\subsection{Improper node} \label{sec:qLT3_IN}

\begin{figure}
    \centering
    \includegraphics[width=\linewidth]{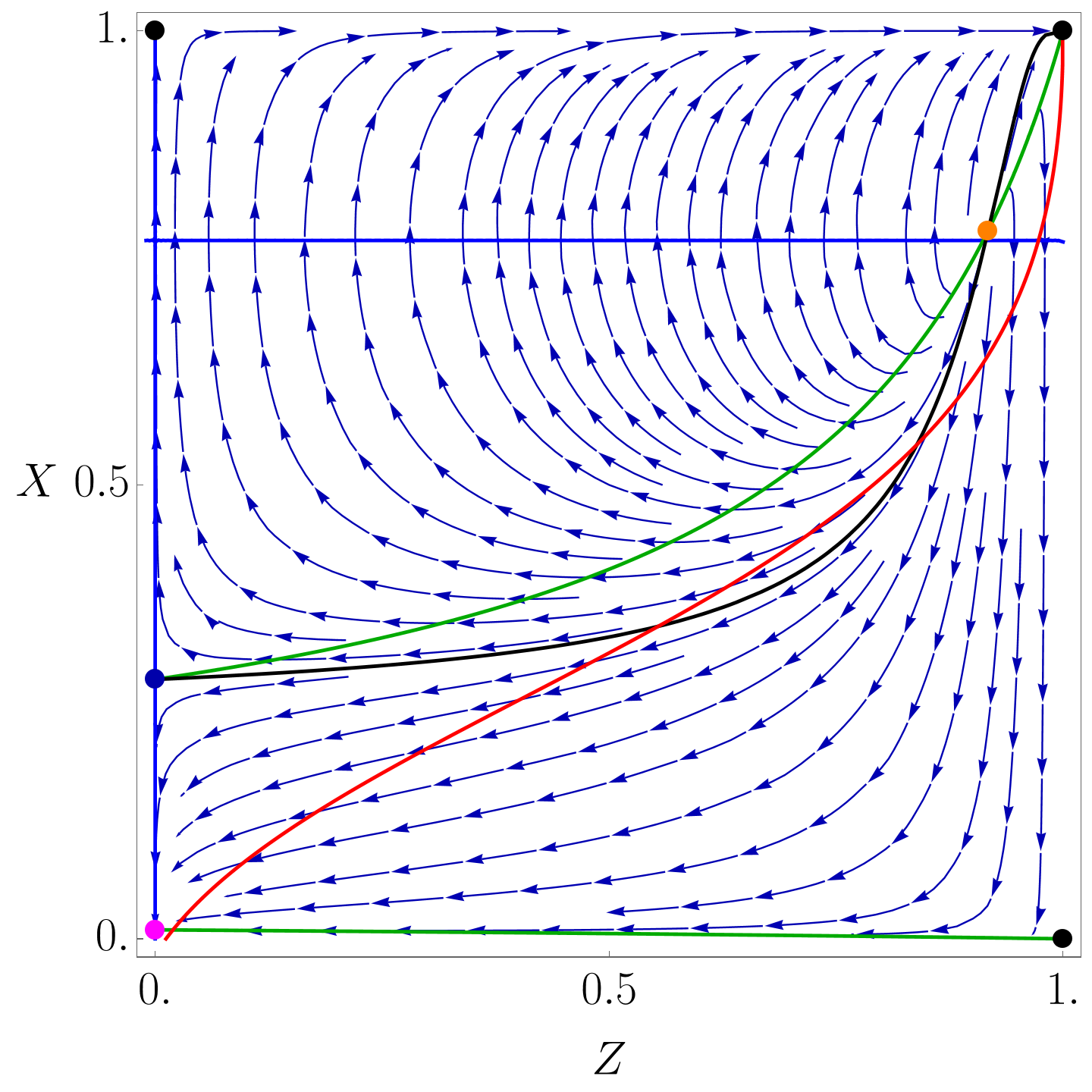}
    \caption{$Z$-$X$ phase space referred to in Appendix \ref{sec:qLT3_IN}. Here, the parameters are set to $w_x=-0.6$ and $q=0.847386$. In this case, the stability of $dS_{1+}$ (orange) is an improper node, and the zero acceleration curve intersects the repellor-saddle separatrix twice. Trajectories to the left of the separatrices expand to a singularity, some with and some without a decelerated period. Those to the right of the separatrices all have a decelerated period and accelerate as they asymptotically expand towards  $dS_{2+}$ (magenta). Therefore, the trajectories that evolve to the right of the separatrices are qualitatively of interest.}
    \label{fig:ZX_IN_qLT3}
\end{figure}

An example of the $Z$-$X$ phase space where $q < 3$ and the high energy repellor is an improper node is shown in Fig. \ref{fig:ZX_IN_qLT3}. Qualitatively, this case is similar to the repelling node case in Fig. \ref{fig:ZX_R_qLT3}. The trajectories of interest are those that evolve to the right of the separatrices, as they accelerate as they expand from $dS_{1+}$, then have a decelerated phase, and finally have a late-time acceleration as they approach $dS_{2+}$. 

\begin{figure}
    \centering
    \includegraphics[width=\linewidth]{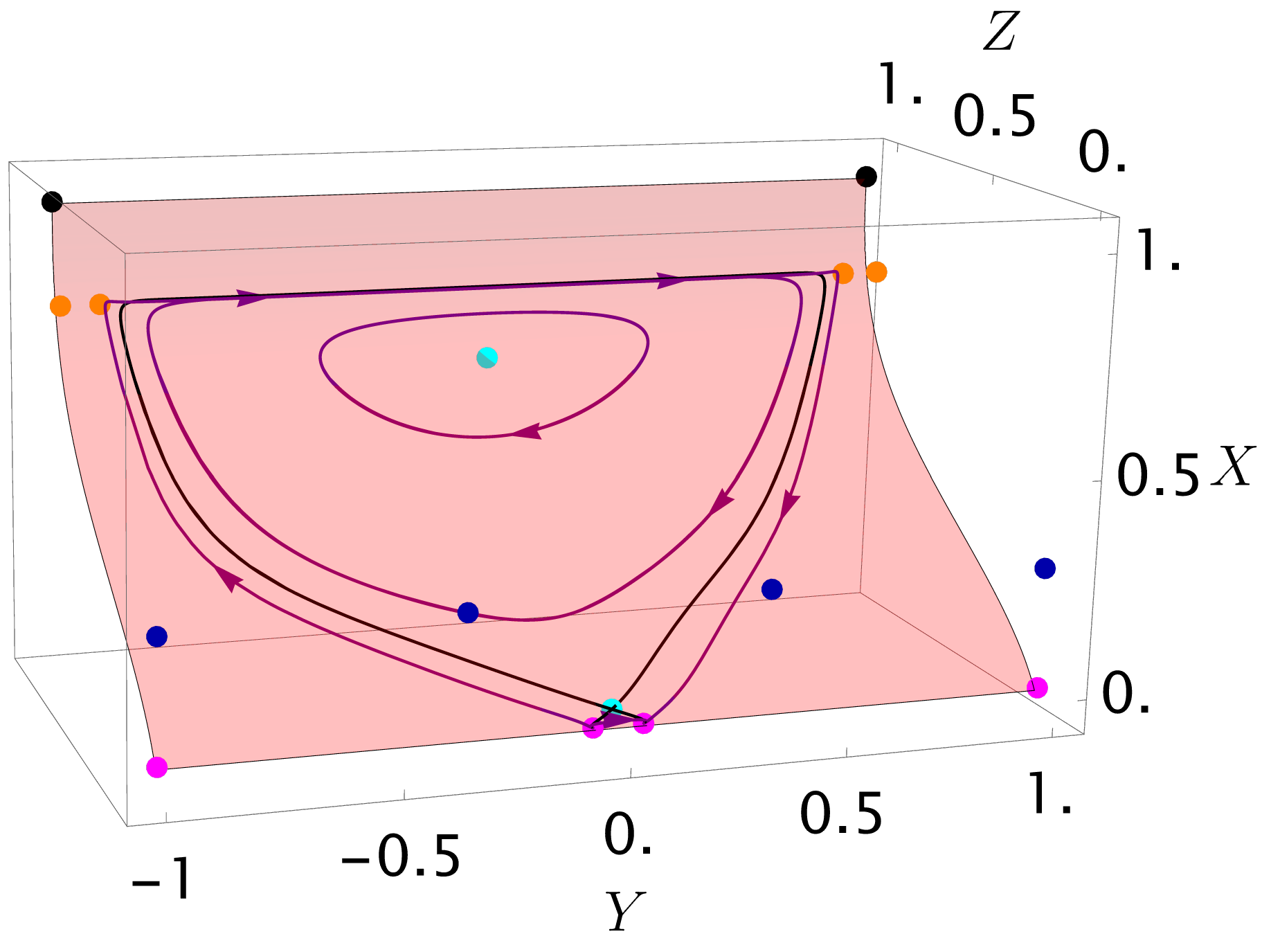}
    \caption{$X$-$Y$-$Z$ phase space corresponding to Fig. \ref{fig:ZX_IN_qLT3}, where $w_x=-0.6$ and $q=0.847386$. The trajectories themselves are plotted on a first integral surface where $X_0 = 0.1$ and $Z_0 = 0.5$, which corresponds to a trajectory in Fig. \ref{fig:ZX_IN_qLT3} that evolves to the right of the separatrices. In this figure, only the trajectories with positive spatial curvature are plotted. Two Einstein fixed points (cyan) exist at $Y = 0$ on the zero acceleration surface. Trajectories inside the CFS either bounce once between $dS_{2-}$ and $dS_{2+}$ and always accelerate, or are cyclic around an Einstein fixed point and cross the zero acceleration surface once during expansion, meaning they have a decelerated phase but no late-time acceleration. Trajectories outside the CFS bounce once between between $dS_{2-}$ and $dS_{2+}$, and intersect the zero acceleration surface twice during expansion, meaning they have a decelerated phase and a late-time acceleration.}
    \label{fig:XYZ_IN_qLT3}
\end{figure}

\begin{figure}
    \centering
    \includegraphics[width=\linewidth]{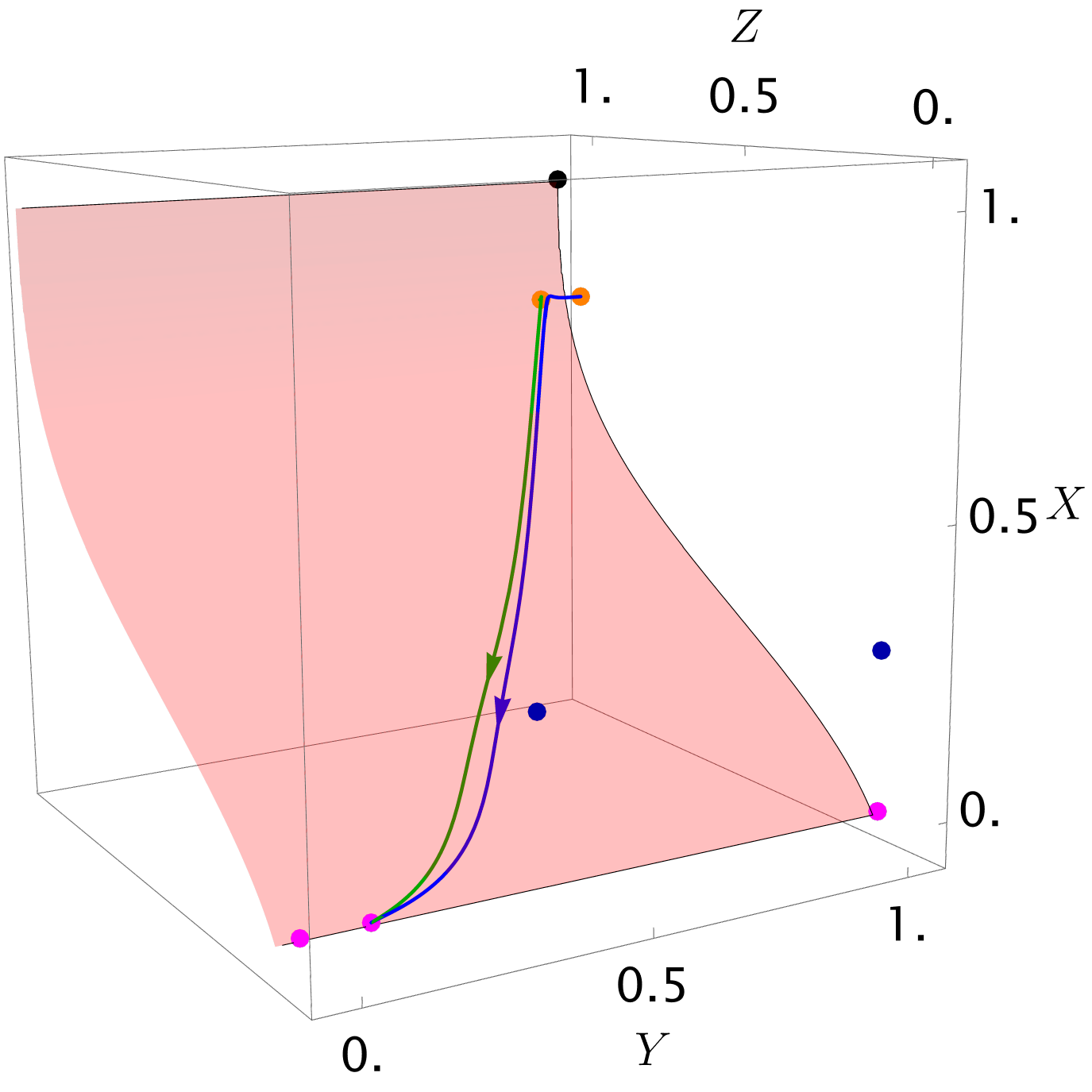}
    \caption{Expanding flat (green) and open (blue) trajectories in the $X$-$Y$-$Z$ phase space corresponding to Fig. \ref{fig:ZX_IN_qLT3}, where $w_x=-0.6$ and $q=0.847386$. The trajectories themselves are plotted on a first integral surface where $X_0 = 0.1$ and $Z_0 = 0.5$, which corresponds to a trajectory in Fig. \ref{fig:ZX_IN_qLT3} that evolves to the right of the separatrices, expanding to $dS_{2+}$. Both the flat and open trajectories intersect the zero acceleration surface twice, meaning they initially accelerate as they expand from $dS_{1+}$ and $dS_{4+}$, respectively, then have a period of deceleration, and finally have an accelerated phase as they expand toward $dS_{2+}$ at low energy.}
    \label{fig:XYZ_IN_qLT3_FlatOpen}
\end{figure}

Figures \ref{fig:XYZ_IN_qLT3} and \ref{fig:XYZ_IN_qLT3_FlatOpen} show the three-dimensional $X$-$Y$-$Z$ phase spaces corresponding to a trajectory that evolves to the right of the separatrices in Fig. \ref{fig:ZX_IN_qLT3}. Qualitatively, these phase spaces are similar to that of Figs. \ref{fig:XYZ_R_qLT3} and \ref{fig:XYZ_R_qLT3_FlatOpen}. Trajectories with positive spatial curvature are shown in Fig. \ref{fig:XYZ_IN_qLT3}. Two Einstein points exist in the phase space, one of which is a center, and the other is a saddle which forms part of the CFS. The trajectories that evolve outside the CFS are qualitatively interesting, as they contract from $dS_{2-}$, go through a quasi-de Sitter bounce, and then expand towards $dS_{2+}$, intersecting the zero acceleration surface twice during expansion. Expanding flat and open trajectories are shown in Fig. \ref{fig:XYZ_IN_qLT3_FlatOpen} and are all of interest, as they emerge from $dS_{1+}$ and $dS_{4+}$, respectively, and intersect the zero acceleration surface twice. This means they have a decelerated phase, followed by a late-time acceleration as they expand towards $dS_{2+}$.

\section{$q > 3$}

In this Appendix, we analyze the dynamics when $q > 3$ and $dS_{1+}$ is either a repelling node or an improper node. We fix $\epsilon = -1$ and $\scriptR = 0.01$ as before, and change $w_x$ and $q$. As shown in Sec. \ref{sec:qGT3}, when $dS_{1+}$ is a spiral repellor three subcases exist: {\it i)}  the repellor-saddle separatrix and the zero acceleration curve do not intersect at all; {\it ii)} they just touch;  {\it ii)} they intersect twice. However, when $dS_{1+}$ is a repelling node or an improper node, we only find one subcase where the repellor-saddle separatrix and zero acceleration curve do not intersect. This is because of the dependence from $q$ of the repellor fixed point, as well as how the vector field behaves when $dS_{1+}$ is not a spiral.

The $z$-coordinate of $dS_{1+}$ is proportional to $1/q^2$ and the $x$-coordinate is proportional to $1/q$. When $q < 3$, $dS_{1+}$ will therefore have larger $z$- and $x$-values than when $q > 3$. In particular, when $q > 3$, $dS_{1+}$ has an $x$-value of $x < 1$. This means the repellor-saddle separatrix will be shorter when $q > 3$, as the repelling fixed point is closer to the saddle $dS_{3+}$. We find that when $dS_{1+}$ is a spiral repellor, the separatrix and zero acceleration curve can intersect as the vector field spirals out from $dS_{1+}$, which means the separatrix also spirals out from $dS_{1+}$. However, when $dS_{1+}$ is a repelling or an improper node, the separatrix does not spiral out, and so does not intersect the zero acceleration curve. In the following, we present the two-dimensional and three-dimensional phase spaces where $dS_{1+}$ is a repelling node and an improper node. The color scheme of the fixed points and critical points in the two-dimensional $Z$-$X$ phase spaces is as in Table \ref{tab:ZX_FPs_Stability_Colour}, and the color scheme of the curves is given in Table \ref{tab:2-D_features}. The color scheme of the fixed points and critical points in the three-dimensional phase spaces is given in Table \ref{tab:Fixed_Points_XYZ_Colour}, and the color scheme of the curves and surfaces is as in Table \ref{tab:3-D_features}.

\subsection{Repelling node} \label{sec:qGT3_R}

\begin{figure}
    \centering
    \includegraphics[width=\linewidth]{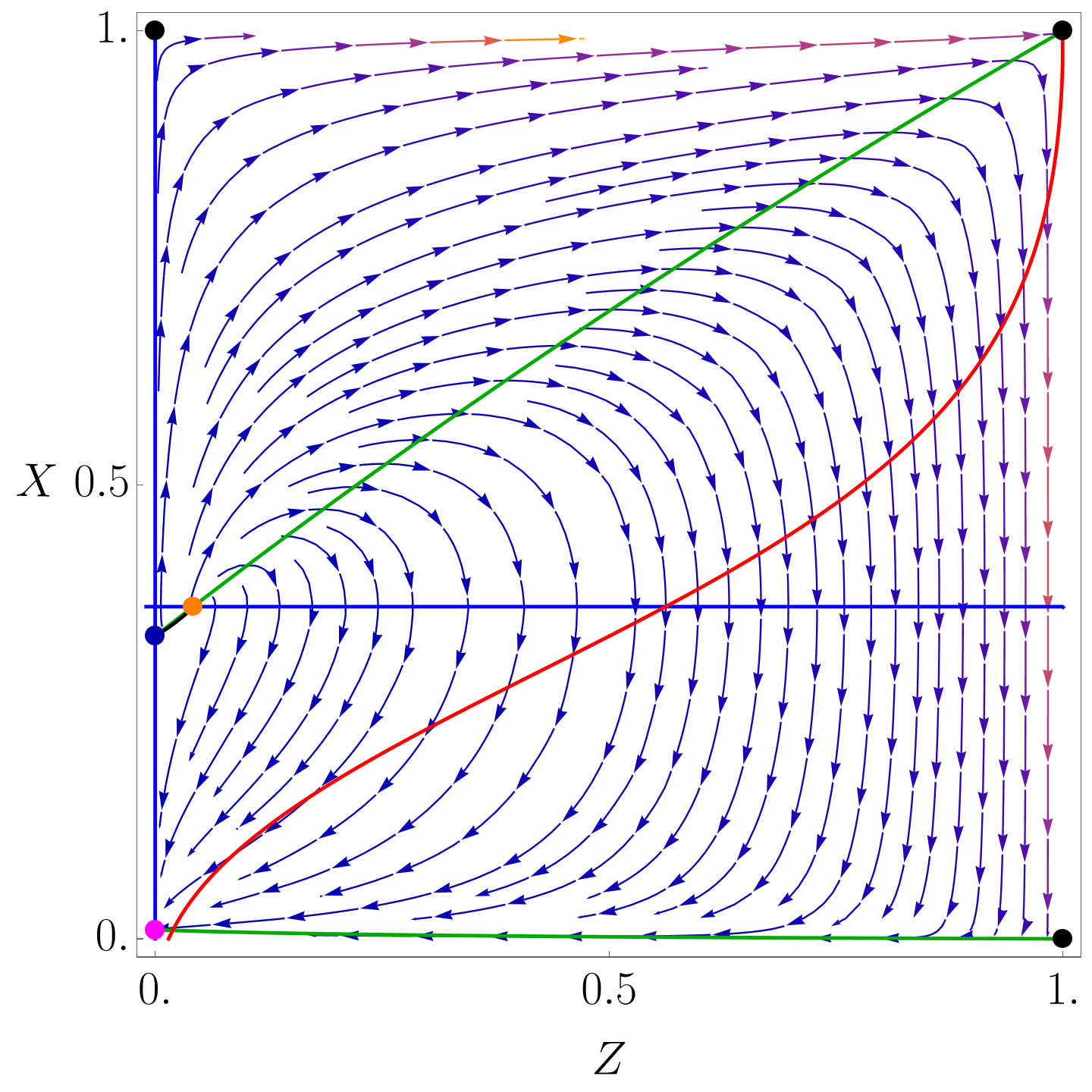}
    \caption{$Z$-$X$ phase space referred to in Appendix \ref{sec:qGT3_R}. Here, the parameters are set to $w_x=-0.5$ and $q=5.2$. The stability of $dS_{1+}$ (orange) is a repelling node, and the zero acceleration curve does not intersect the repellor-saddle separatrix. Apart from the separatrix itself which evolves to $dS_{3+}$ (dark blue), all other trajectories expand towards $dS_{2+}$ (magenta) at late times. Trajectories which do not intersect the $a'' = 0$ curve always accelerate, and those that intersect it twice have a decelerated period. One trajectory in the phase space will touch the zero-acceleration curve, but will otherwise accelerate.}
    \label{fig:ZX_R_qGT3}
\end{figure}

Figure \ref{fig:ZX_R_qGT3} shows the $Z$-$X$ phase space when $q > 3$ and $dS_{1+}$ is a repelling node. All trajectories are nonsingular, and expand from $dS_{1+}$ at high energy, and tend towards $dS_{2+}$ at late times. The trajectories that do not intersect the red zero acceleration curve always accelerate, and those that cross the red curve twice have a period of deceleration, followed by a late-time acceleration. There is also a limiting case where a trajectory touches the red curve and has a point of zero acceleration, but otherwise accelerates. The trajectories that intersect the zero acceleration curve twice are the models of interest.

\begin{figure}
    \centering
    \includegraphics[width=\linewidth]{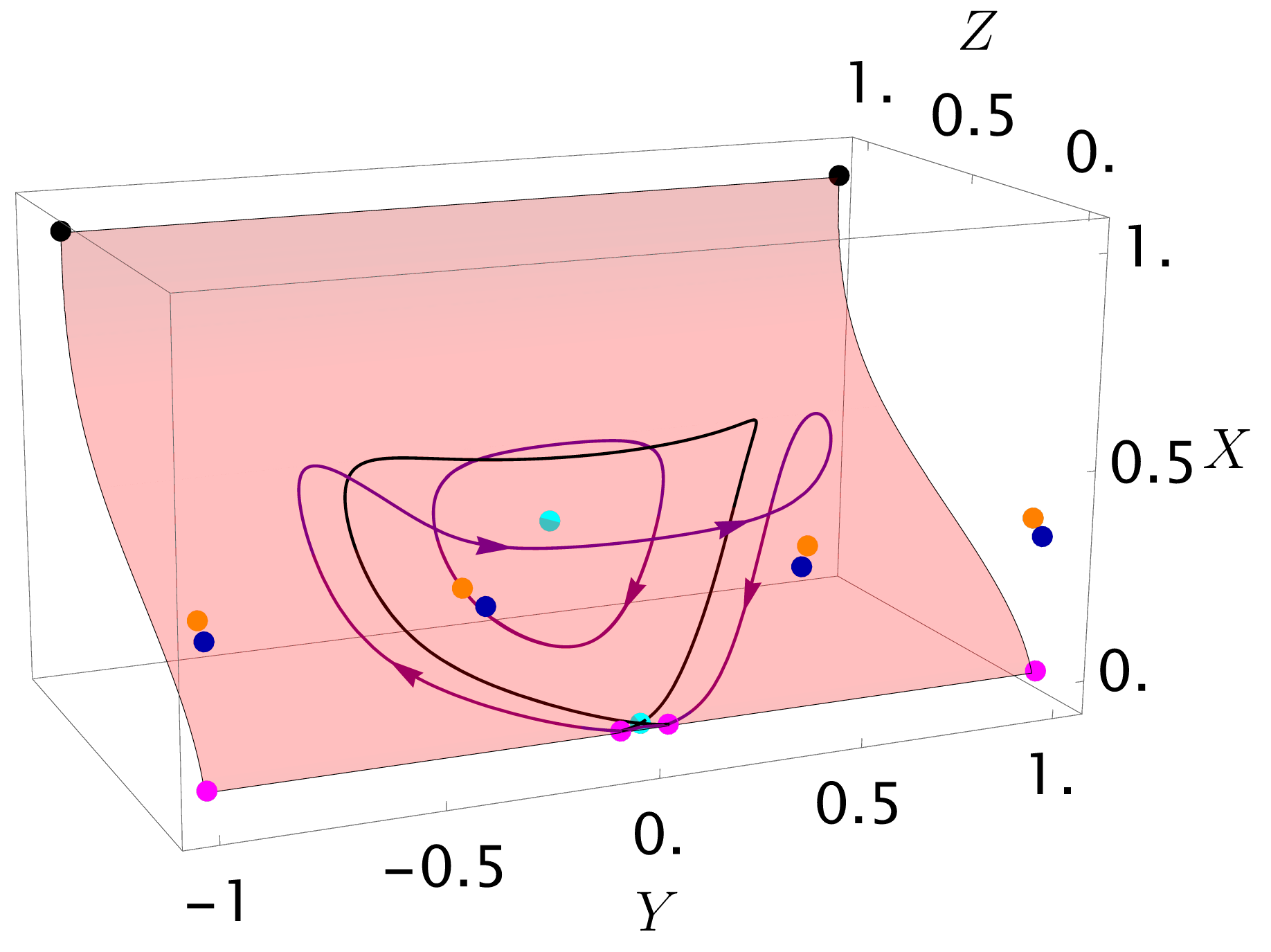}
    \caption{$X$-$Y$-$Z$ phase space corresponding to Fig. \ref{fig:ZX_R_qGT3}, where $w_x=-0.5$ and $q=5.2$. The trajectories themselves are plotted on a first integral surface where $X_0 = 0.1$ and $Z_0 = 0.4$, which corresponds to a trajectory in Fig. \ref{fig:ZX_R_qGT3} that intersects the red zero acceleration curve twice. In this figure, only the trajectories with positive spatial curvature are plotted. Two Einstein fixed points (cyan) exist at $Y = 0$ on the zero acceleration surface. Trajectories inside the CFS either bounce once between $dS_{2-}$ and $dS_{2+}$, and always accelerate, or are cyclic around an Einstein fixed point and cross the zero acceleration surface once during expansion, meaning they have a decelerated phase but no late-time acceleration. Trajectories outside the CFS bounce once between $dS_{2-}$ and $dS_{2+}$, and intersect the zero acceleration surface twice during expansion, meaning they have a decelerated phase followed by a late-time acceleration.}
    \label{fig:XYZ_R_qGT3}
\end{figure}

\begin{figure}
    \centering
    \includegraphics[width=\linewidth]{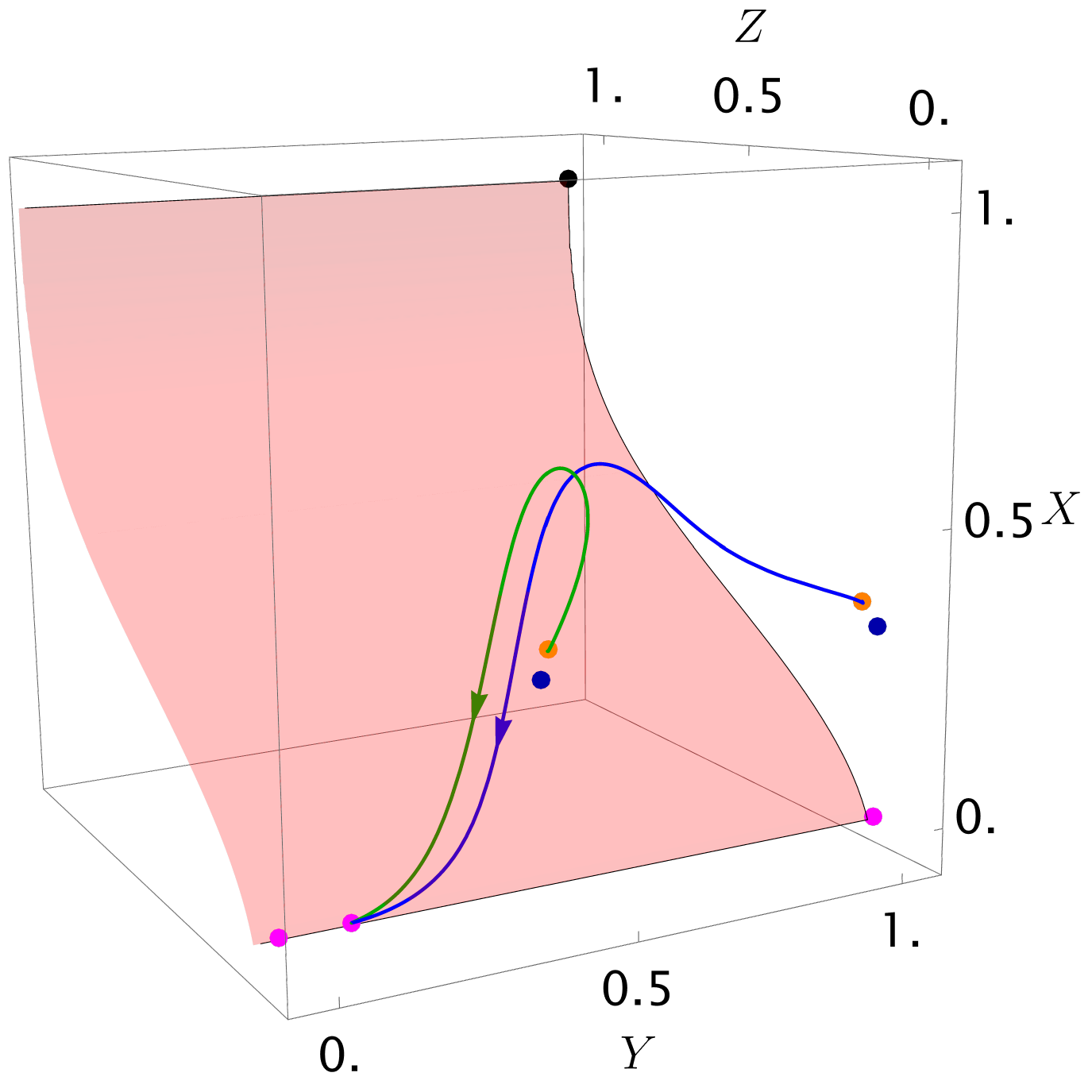}
    \caption{Expanding flat (green) and open (blue) trajectories in the $X$-$Y$-$Z$ phase space corresponding to Fig. \ref{fig:ZX_R_qGT3}, where $w_x=-0.5$ and $q=5.2$. The trajectories themselves are plotted on a first integral surface where $X_0 = 0.1$ and $Z_0 = 0.4$, which corresponds to a trajectory in Fig. \ref{fig:ZX_R_qGT3} that intersects the red zero acceleration surface twice. Both the flat and open trajectories cross the zero acceleration surface twice, meaning they initially accelerate as they expand from $dS_{1+}$ and $dS_{4+}$, respectively, then have a period of deceleration, and then have a final accelerated phase as they expand toward $dS_{2+}$ at low energy.}
    \label{fig:XYZ_R_qGT3_FlatOpen}
\end{figure}

Figures \ref{fig:XYZ_R_qGT3} and \ref{fig:XYZ_R_qGT3_FlatOpen} show a first integral surface in the $X$-$Y$-$Z$ phase space corresponding to a trajectory that intersects the zero acceleration curve twice in Fig. \ref{fig:ZX_R_qGT3}. Figure \ref{fig:XYZ_R_qGT3} shows the trajectories with positive spatial curvature, and Fig. \ref{fig:XYZ_R_qGT3_FlatOpen} shows expanding flat and open trajectories. All the trajectories in Fig. \ref{fig:XYZ_R_qGT3} bounce at least once. Two Einstein points exist at $Y = 0$, one of which is a saddle which the CFS passes through, and the other is a center. Trajectories inside the CFS either bounce once between $dS_{2-}$ and $dS_{2+}$, or are cyclic around the center Einstein fixed point. These trajectories are not of interest, as the bouncing trajectories always accelerate, and the cyclic trajectories have a decelerated phase during expansion, but no late-time acceleration. The trajectories that evolve outside the CFS bounce once during a quasi-de Sitter phase, contracting from $dS_{2-}$ then expanding towards $dS_{2+}$. These trajectories intersect the zero acceleration surface twice during expansion, meaning they accelerate though the bounce, becoming flatter as they start to expand, then have a decelerated phase where large scale structure could form, and then have a final late-time accelerated expansion as they approach a late-time cosmological constant represented by $dS_{2+}$. The trajectories outside the CFS are therefore of interest.

The expanding flat and open trajectories in Fig. \ref{fig:XYZ_R_qGT3_FlatOpen} are all qualitatively interesting, as they emerge from $dS_{1+}$ and $dS_{4+}$, respectively, and tend toward $dS_{2+}$ at low energy. They cross the red zero acceleration surface twice, meaning they initially accelerate, then have a decelerated phase, and finally have a late-time acceleration as they approach $dS_{2+}$.

\subsection{Improper node} \label{sec:qGT3_IN}

\begin{figure}
    \centering
    \includegraphics[width=\linewidth]{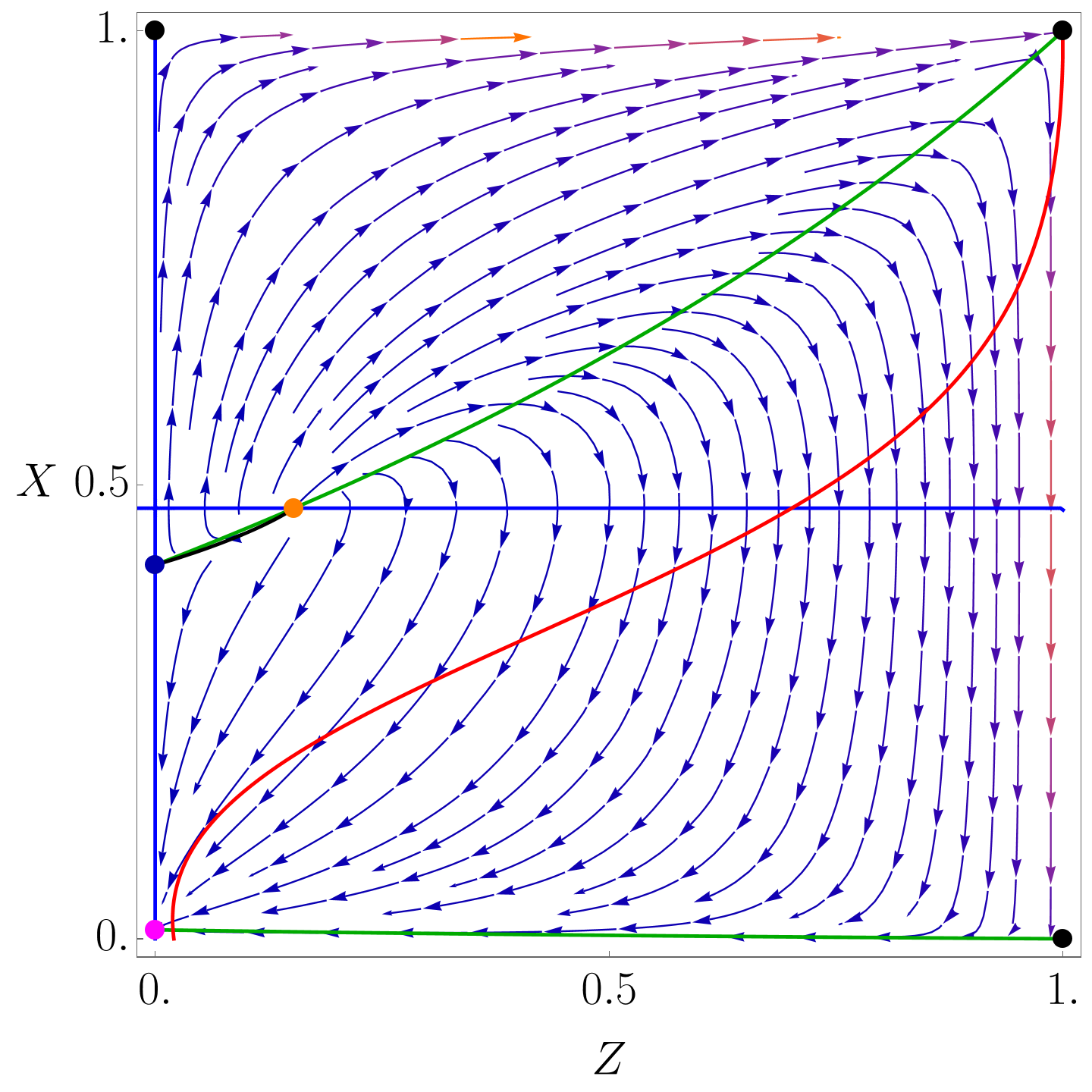}
    \caption{$Z$-$X$ phase space referred to in Appendix \ref{sec:qGT3_IN}. Here, the parameters are set to $w_x=-0.3$ and $q=3.32488$. Here, the stability of $dS_{1+}$ (orange) is an improper node, and the zero acceleration curve does not intersect the repellor-saddle separatrix. Except for the separatrix which expands to $dS_{3+}$ (dark blue), all other trajectories expand towards $dS_{2+}$ (magenta) at late times. Trajectories which do not intersect the $a'' = 0$ curve always accelerate, and those that intersect it twice have a decelerated phase. One trajectory in the phase space will touch the zero-acceleration curve, but will otherwise accelerate.}
    \label{fig:ZX_IN_qGT3}
\end{figure}

The $Z$-$X$ phase space where $q > 3$ and $dS_{1+}$ is an improper node is shown in Fig. \ref{fig:ZX_IN_qGT3}. Qualitatively, this case is similar to the repelling node case in Fig. \ref{fig:ZX_R_qGT3}. The trajectories of interest are those that intersect the red zero acceleration curve twice, as they have a decelerated phase where large scale structure could form, and a late-time acceleration as they approach $dS_{2+}$. 

\begin{figure}
    \centering
    \includegraphics[width=\linewidth]{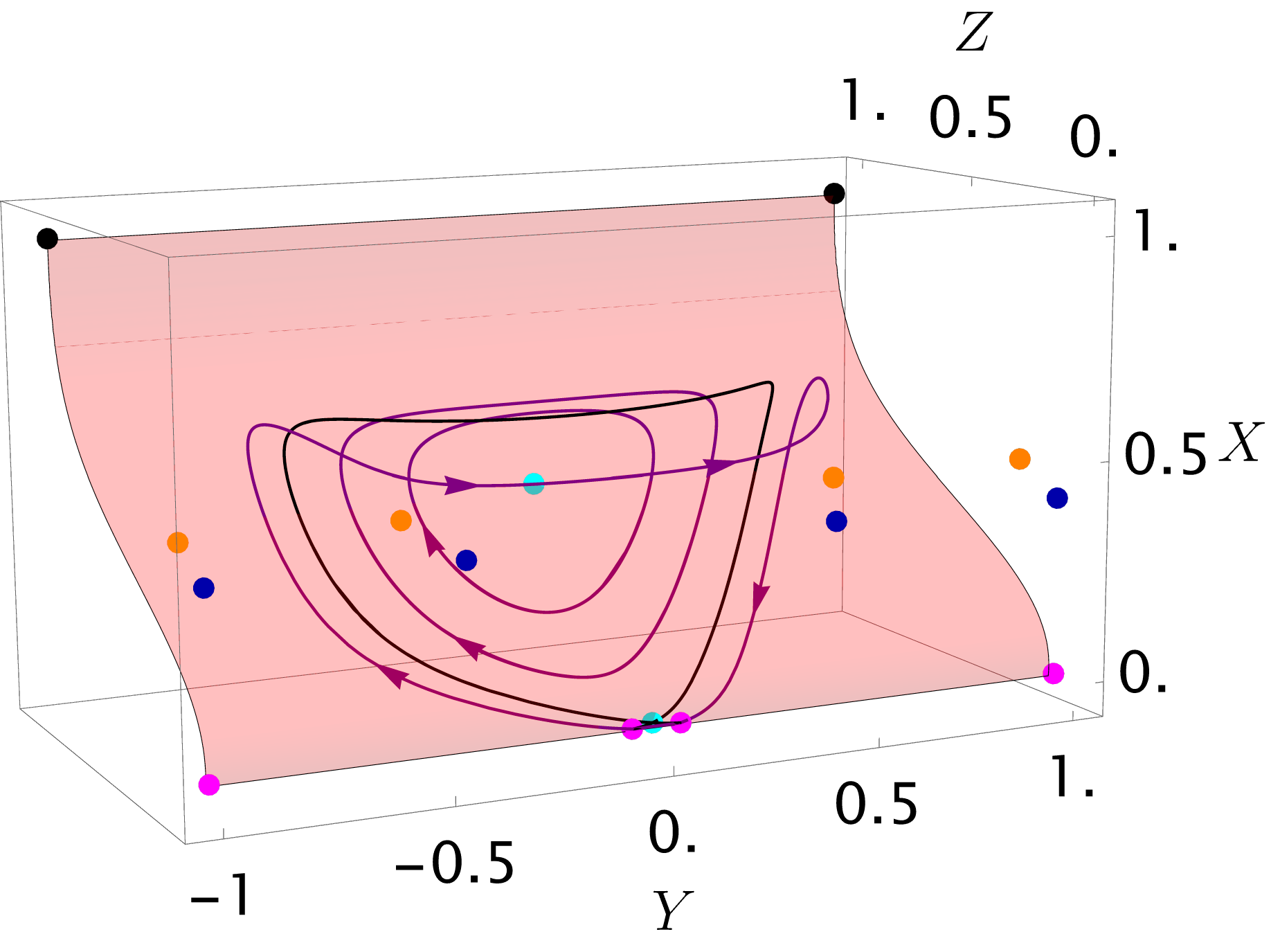}
    \caption{$X$-$Y$-$Z$ phase space corresponding to Fig. \ref{fig:ZX_IN_qGT3}, where $w_x=-0.3$ and $q=3.32488$. The trajectories themselves are plotted on a first integral surface where $X_0 = 0.05$ and $Z_0 = 0.3$, which corresponds to a trajectory in Fig. \ref{fig:ZX_IN_qGT3} that intersects the red zero acceleration curve twice. In this figure, only the trajectories with positive spatial curvature are plotted. Two Einstein fixed points (cyan) exist at $Y = 0$ on the zero acceleration surface. Trajectories inside the CFS either bounce once between $dS_{2-}$ and $dS_{2+}$, and always accelerate, or are cyclic around an Einstein fixed point and cross the zero acceleration surface once during expansion, meaning they have a decelerated phase but no late-time acceleration. Trajectories outside the CFS bounce once between $dS_{2-}$ and $dS_{2+}$, and intersect the zero acceleration surface twice during expansion, meaning they have a decelerated phase followed by a late-time acceleration.}
    \label{fig:XYZ_IN_qGT3}
\end{figure}

\begin{figure}
    \centering
    \includegraphics[width=\linewidth]{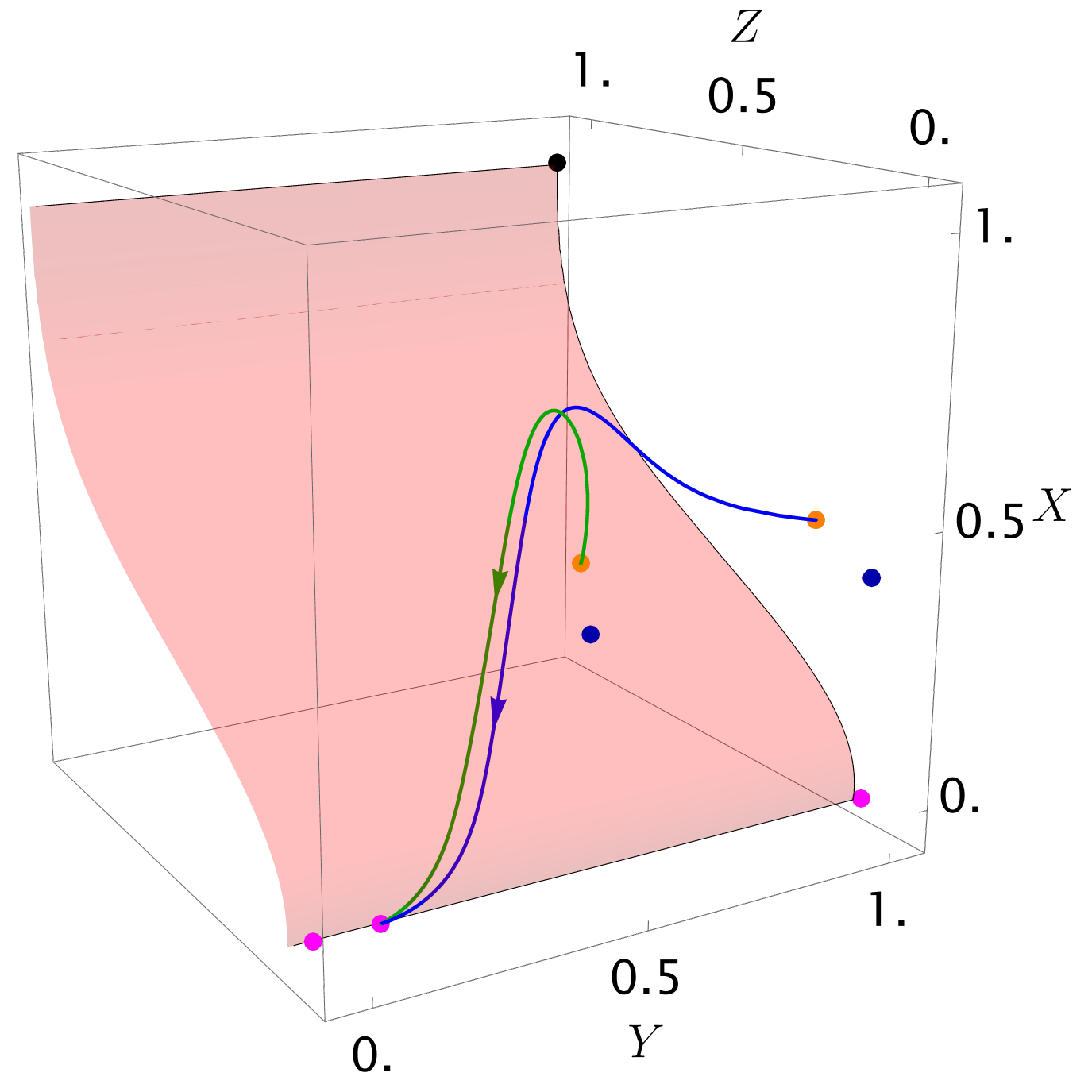}
    \caption{Expanding flat (green) and open (blue) trajectories in the $X$-$Y$-$Z$ phase space corresponding to Fig. \ref{fig:ZX_IN_qGT3}, where $w_x=-0.3$ and $q=3.32488$. The trajectories are plotted on a first integral surface where $X_0 = 0.05$ and $Z_0 = 0.3$, which corresponds to a trajectory in Fig. \ref{fig:ZX_IN_qGT3} that intersects the red zero acceleration curve twice. Both the flat and open trajectories intersect the zero acceleration surface twice, meaning they initially accelerate as they expand from $dS_{1+}$ and $dS_{4+}$, respectively, then have a period of deceleration, and then have a final accelerated phase as they expand toward $dS_{2+}$ at low energy.}
    \label{fig:XYZ_IN_qGT3_FlatOpen}
\end{figure}

Figures \ref{fig:XYZ_IN_qGT3} and \ref{fig:XYZ_IN_qGT3_FlatOpen} show the three-dimensional $X$-$Y$-$Z$ phase spaces corresponding to a trajectory that intersects the zero acceleration curve twice in Fig. \ref{fig:ZX_IN_qGT3}. Qualitatively, the phase spaces are similar to that of Figs. \ref{fig:XYZ_R_qGT3} and \ref{fig:XYZ_R_qGT3_FlatOpen}. Figure \ref{fig:XYZ_IN_qGT3} shows the trajectories with positive spatial curvature. Two Einstein fixed points exist in the phase space, one of which is a center and the other is a saddle which forms part of the CFS. The trajectories that evolve outside the CFS  and bounce once are the models that are qualitatively interesting. These trajectories contract from $dS_{2-}$, then bounce during a quasi de Sitter phase, and finally expand towards $dS_{2+}$. They cross the zero acceleration surface twice during expansion, meaning they accelerate through the bounce, becoming flatter as they start to expand, then have a decelerated period followed by a late-time accelerated expansion. Expanding flat and open trajectories are shown in Fig. \ref{fig:XYZ_IN_qGT3_FlatOpen}, and are all of interest. Flat and open trajectories emerge from $dS_{1+}$ and $dS_{4+}$, respectively, and intersect the $a'' = 0$ surface twice. This means they have a decelerated phase where large scale structure could form, and accelerate at late times as they expand towards $dS_{2+}$.

\end{document}